\documentclass[11pt]{article}
\pdfoutput=1
\usepackage{jheppub}
\usepackage{graphicx}
\usepackage{amssymb}
\usepackage{amsmath}
\usepackage{amsfonts}
\usepackage{latexsym}
\usepackage{hyperref}
\usepackage{physics}
\usepackage{bbm}
\usepackage{empheq}
\usepackage{tikz}
\usepackage{cancel}
\usepackage{hhline}
\usepackage{caption}
\usepackage{soul}
\usepackage{float}
\usepackage{multirow}
\usepackage{hhline}
\usepackage{algorithm}
\usepackage{algpseudocode}

\usepackage[T1]{fontenc}

\usetikzlibrary{arrows.meta}
\usetikzlibrary{decorations.markings}
\tikzset{->-/.style={decoration={
			markings,
			mark=at position #1 with {\arrow{>}}},postaction={decorate}}}

\newcommand\nn{\nonumber}
\newcommand\fft[2]{\frac{#1}{#2}}

\newcommand\tDelta{\widetilde{\Delta}}

\newcommand\bxi{\boldsymbol{\xi}}
\newcommand\mA{\mathcal{A}}
\newcommand\mB{\mathcal{B}}
\newcommand\mC{\mathcal{C}}
\newcommand\tpsi{\widetilde{\psi}}
\newcommand\mfe{\mathfrak{e}}
\newcommand\mJ{\mathcal{J}}

\newcommand\mn{\mathfrak{n}}

\newcommand\mV{\mathcal{V}}

\newcommand\mN{\mathcal{N}}
\def\ri{{\rm i}}
\newcommand\mR{\mathcal{R}}

\newcommand\mf{\mathfrak{f}}
\newcommand\mO{\mathcal{O}}

\newcommand\mZ{\mathcal{Z}}

\newcommand\bk{\boldsymbol{k}}
\newcommand\bDelta{\boldsymbol{\Delta}}

\newcommand\btDelta{\boldsymbol{\widetilde{\Delta}}}
\newcommand\mD{\mathcal{D}}
\newcommand\mfc{\mathfrak{c}}

\newcommand\hg{\hat{g}}
\newcommand\hf{\hat{f}}
\newcommand{\sign}{{\rm sign}}

\makeatletter
\newcommand*{\rom}[1]{\expandafter\@slowromancap\romannumeral #1@}
\makeatother

\begin{document}
	
\title{An Airy Tale at Large $N$}
	
\author[a,b]{Nikolay Bobev,}
\author[a]{Pieter-Jan De Smet,}
\author[c]{Junho Hong,}
\author[d,e]{Valentin Reys,}
\author[a,f]{Xuao Zhang}

\affiliation[a]{Institute for Theoretical Physics, KU Leuven\,,\\ Celestijnenlaan 200D, B-3001 Leuven, Belgium}
\affiliation[b]{Leuven Gravity Institute, KU Leuven\,,\\ Celestijnenlaan 200D, B-3001 Leuven, Belgium}
\affiliation[c]{Department of Physics \& Center for Quantum Spacetime, Sogang University\,,\\ 35 Baekbeom-ro, Mapo-gu, Seoul 04107, Republic of Korea}
\affiliation[d]{Université Paris-Saclay, CNRS, CEA, \\	Institut de physique théorique, 91191, Gif-sur-Yvette, France}
\affiliation[e]{Laboratoire de Physique de l’Ecole normale supérieure,\\
CNRS, PSL Research University and Sorbonne Universités,
24 rue Lhomond, 75005 Paris, France}
\affiliation[f]{Kavli Institute for Theoretical Sciences (KITS)\,,\\
University of Chinese Academy of Sciences, Beijing 100190, China}

\emailAdd{nikolay.bobev@kuleuven.be}
\emailAdd{pieterjan.desmet@kuleuven.be}
\emailAdd{junhohong@sogang.ac.kr}
\emailAdd{valentin.reys@phys.ens.fr}
\emailAdd{zhangxuao@ucas.ac.cn}


\abstract{We study the sphere partition function of 3d $\mathcal{N}=2$ holographic SCFTs arising on the worldvolume of $N$ coincident M2-branes in the presence of squashing and real mass deformations. We argue that the all-order large $N$ perturbative expansion of the partition function resums into a compact expression in terms of an Airy function. We provide ample evidence for this conjecture using numerical methods, saddle point analysis in the 't~Hooft limit, and the relation of the squashed sphere partition function to the Cardy-like limit of the superconformal index. We also discuss the relation of our results to topological string theory and the statistical mechanics of Fermi gases. Our conjecture has a bearing on the AdS/CFT correspondence and the structure of perturbative M-theory corrections to 11d supergravity which we discuss.}
	
\maketitle \flushbottom

\section{Introduction}
\label{sec:intro}

The large $N$ limit of matrix models has an illustrious history of providing seminal insights into quantum field theory and mathematical physics \cite{tHooft:1973alw,Brezin:1977sv}. The development of supersymmetric localization methods for QFT on compact Euclidean manifolds, complemented with advances in the AdS/CFT correspondence, has led to the incorporation of the powerful large $N$ matrix model techniques into holography and string theory, see \cite{Pestun:2016zxk} for a review.

A poster child for the successful application of these methods to holography is provided by the ABJM theory which describes the low-energy dynamics of coincident M2-branes on $\mathbb{C}^4/\mathbb{Z}_k$ \cite{Aharony:2008ug}. The $S^3$ partition function of this $\mathcal{N}=6$ ${\rm U}(N)_{k}\times {\rm U}(N)_{-k}$ Chern-Simons-matter theory in the presence of real mass deformations and for certain supersymmetric squashing deformations can be computed by supersymmetric localization \cite{Kapustin:2009kz,Jafferis:2010un,Hama:2010av,Hama:2011ea,Imamura:2011wg}. The resulting matrix model is amenable to an explicit analysis by employing a variety of large $N$ techniques and the results can be successfully compared with calculations in the holographically dual 11d supergravity. These results have been generalized and can be extended to a variety of quiver 3d $\mathcal{N}=2$ CS-matter theories that admit a holographic description in terms of AdS$_4$ vacua of string and M-theory, see~\cite{Marino:2016new} for a review and further references.

It was realized in \cite{Fuji:2011km,Marino:2011eh} that the large $N$ expansion of the $S^3$ partition function of the ABJM SCFT at fixed $k$ can be resummed in a compact form in terms of an Airy function. This remarkable result was later generalized in various ways by turning on deformations of the ABJM theory, like real masses and squashing of the round $S^3$ metric, that break the conformal invariance but preserve part of the supersymmetry. In particular, it was shown that the large $N$ partition function can still be expressed in terms of an Airy function in the presence of two  specific real mass parameters \cite{Nosaka:2015iiw}, as well as for the particular value for the $S^3$ squashing parameter $b^2=3$ \cite{Hatsuda:2016uqa}. The same form of the large $N$ sphere partition function was also derived in other holographic SCFTs, see \cite{Marino:2011eh,Marino:2012az,Mezei:2013gqa,Grassi:2014vwa}. These results, together with advances in the holographic calculation of subleading corrections to the large $N$ expansion of the sphere free energy, lead to the conjecture in \cite{Bobev:2022jte,Bobev:2022eus,Bobev:2023lkx} (see also \cite{Hristov:2022lcw}) that the Airy form of the $S^3$ partition function is more general and applies to general values of the ABJM deformation parameters as well as to a number of other holographic SCFTs. This Airy conjecture can be formulated simply by stating that the large $N$ $S^3$ partition function of a large class of CS-matter 3d $\mathcal{N}=2$ SCFTs with a holographic dual takes the form
\begin{equation}\label{eq:Airyintro}
	Z(N,b,\bxi)=\mC^{-1/3}e^{\mA}\text{Ai}\Big[\mC^{-1/3}(N-\mB)\Big]\Big(1+\mO(e^{-\#\sqrt{N}})\Big)\,.
\end{equation}
Here the quantities $(\mA,\mB,\mC)$ do not depend on the rank of the gauge group $N$ but depend on the real mass parameters available in a given theory, denoted collectively by $\bxi$ in \eqref{eq:Airyintro}, and on the real parameter $b$ that determines the geometric squashing of $S^3$.\footnote{The collective variable $\bxi$ also encodes discrete data of the 3d gauge theory like the Chern-Simons level $k$ or the number of the flavor multiplets $N_f$.} Importantly the expression in \eqref{eq:Airyintro} pertains to the perturbative part in the $1/N$ expansion of the partition function. There are also non-perturbative corrections, indicated with $\mO(e^{-\#\sqrt{N}})$ in \eqref{eq:Airyintro}, which are hard to compute and in general do not admit a known closed-form expression. The main goal of this work is to summarize the status of the Airy conjecture and to provide substantial new evidence for its validity for five classes of 3d SCFTs, summarized in Table~\ref{table:abc} below, arising on the worldvolume of M2-branes probing conical singularities in M-theory.\footnote{We also present an Airy conjecture for two close cousins of the ABJM theory, namely the ABJ and mABJM SCFTs, which we discuss further in Section~\ref{sec:AiryTale:ex:ABJM}.} Our results underscore the broad validity of the simple form of the partition function in \eqref{eq:Airyintro} and strongly suggest that similar seemingly mysterious yet-to-be uncovered simplifications should occur in the perturbative expansion of the holographically dual M-theory.

It is important to note that the Airy conjecture in \eqref{eq:Airyintro} has been rigorously established for some SCFTs and for specific values of the deformation parameters $(b,\bxi)$. There is also additional supporting evidence for its validity in various limits that have been obtained using holography or various dualities and numerical techniques. These previous developments are summarized in Section~\ref{sec:AiryTale}. To extend these results and provide strong new evidence in support of the Airy conjecture we employ three different methods - direct numerical integration of the matrix model; calculations of $Z(N,b,\bxi)$ in a saddle point approximation; and a relation between the superconformal index of 3d $\mathcal{N}=2$ in the Cardy-like limit and the $S^3$ partition function in the limit of large values of the squashing parameter $b$.

The direct numerical integration method we develop exploits certain simplifications of the matrix model integrand for particular choices of the squashing and real mass parameters.\footnote{See \cite{Hanada:2012si,Hatsuda:2012hm,Putrov:2012zi} for previous work on different numerical approaches to the calculation of the $S^3$ partition function of the ABJM theory.} In this case, studied also recently in \cite{Kubo:2024qhq}, and for theories with a single gauge node like the ADHM theory, we develop an algorithm, based on mathematical work by Bornemann, that allows for the efficient numerical calculation of $Z(N,b,\bxi)$ for values of $N$ up to around $N\sim 20$. Since the non-perturbative corrections denoted by $\mO(e^{-\#\sqrt{N}})$ in \eqref{eq:Airyintro} are very small for these values of $N$, we are able to use these results to test the Airy conjecture with excellent numerical accuracy.  

The Airy conjecture is formulated in the so-called M-theory limit where $N$ is taken large with all other parameters in the theory being fixed. It is interesting however, to also take a different large $N$ limit, which we refer to as the IIA or 't Hooft limit, in which one also scales a different parameter in the CS-matter theory like the CS level $k$ or the number of flavor multiplets $N_f$. In these cases the Airy conjecture can be reorganized in a way that makes the SCFT partition function $Z(N,b,\bxi)$ take the form familiar from the genus expansion of a string theory path integral. Consequently, the Airy conjecture leads to simple explicit expressions for the string theory free energy at arbitrary values of the genus. We exploit this fact and focus on the genus-0 string free energy $F_0$ in the IIA limit of the ABJM and ADHM theories to perform a careful saddle point evaluation of $F_0$, see also \cite{Geukens:2024zmt} for recent related work. We do this by a combination of analytic and numerical techniques and find very strong support of the analytic form of the Airy conjecture.

It was pointed out in \cite{Choi:2019dfu} that the partition function of 3d $\mathcal{N}=2$ SCFTs on $S^1\times_{\omega} S^2$, also known as the superconformal index, is related to the partition function of the theory on a squashed $S^3$. In particular the leading two terms in the Cardy-like limit of the index $\omega \to 0$, corresponding to the small radius limit of the $S^1$, are equivalent to the leading two terms in the $b \to \infty$ expansion of the squashed $S^3$ partition function.\footnote{The squashed $S^3$ partition function is invariant under $b \to 1/b$ and one can therefore equivalently relate its $b\to 0$ limit to the Cardy-like limit of the superconformal index.} Notably, this relation between the two partition functions does not rely on taking any large $N$ limit. Recently, in~\cite{Bobev:2022wem,Bobev:2024mqw} new methods to compute the $S^1\times_{\omega} S^2$ partition function in the Cardy-like limit were developed. They were then successfully applied to find compact expressions that resum the $1/N$ expansion for the leading terms in the small $\omega$ expansion of the superconformal index for a plethora of holographic SCFTs. We combine these two sets of known results, together with the large $b$ expansion of the expression in \eqref{eq:Airyintro}, and find strong new evidence in support of the Airy conjecture.

In addition to its importance for unravelling the structure of perturbative corrections to 11d supergravity and understanding and extending the AdS/CFT correspondence beyond the leading supergravity approximation, the Airy conjecture has implications for the quantum dynamics of the SCFTs at hand. In particular it encodes integrated correlation functions of the SCFT on the round $S^3$ which can be extracted by expanding $Z(N,b,\bxi)$ for small values of the deformation parameters $b-1$ and $\bxi$. We exploit these relations to derive a simple compact expression, valid to all orders in the $1/N$ expansion, for the coefficient $C_T$ that determines the two-point function of the stress-energy tensor of the CFTs that admit an Airy form \eqref{eq:Airyintro} of their $S^3$ partition function. This extends and complements previous results in the literature for the ABJM and ADHM theories \cite{Agmon:2017xes,Chester:2020jay}.

Before we continue with the presentation of our results we note that the Airy form of the $S^3$ partition function, for certain values of the parameters $(b,\bxi)$, can be related to the matrix models arising in topological string theory on non-compact Calabi-Yau manifolds as well as in the physics of Fermi gases. It is natural to wonder whether the more general form of the Airy conjecture we propose in this work can be derived using similar techniques. While our attempts to do this have not been successful, we summarize previous developments on these relations and discuss the limitations to their generalizations in Section~\ref{sec:TS-ST}.

We continue our exposition in the next section with the formulation of the Airy conjecture and a summary of various results in the literature that support it in special cases. In Section~\ref{sec:evi1} we provide new evidence in favor of the conjecture by employing a direct numerical integration algorithm. In Section~\ref{sec:evi2} we focus on the 't Hooft limit of the ABJM and ADHM theories and use a saddle point approximation to arrive at additional evidence supporting the Airy conjecture. In Section~\ref{sec:evi3} we exploit the relation between the squashed $S^3$ and $S^1\times_{\omega} S^2$ path integrals of 3d $\mathcal{N}=2$ SCFTs in the Cardy-like limit to provide additional support for the conjecture. In Section~\ref{sec:TS-ST} we discuss the relation between the Airy conjecture and the so-called topological string/spectral theory (TS/ST) correspondence. A collection of interesting open questions for future work that stem from our results is provided in Section~\ref{sec:discussion}. In the five appendices we present some of the technicalities pertinent to our analysis as well as more details on the numerical calculations we perform.

\section{An Airy tale}\label{sec:AiryTale}
In this section, we introduce the Airy conjecture and review the currently known evidence that supports it. We will first state the Airy conjecture in Section~\ref{sec:AiryTale:statement}, namely that the squashed $S^3$ partition function of a large class of $\mN=2$ Chern-Simons-matter quiver gauge theories is given by an Airy function which resums the $1/N$-perturbative expansion up to non-perturbative corrections. In Section~\ref{sec:AiryTale:ex} we review various examples for which the Airy conjecture is analytically proven or partially supported by analytic calculations.

\subsection{Airy conjecture}\label{sec:AiryTale:statement}

We consider $\mN=2$ Chern-Simons-matter quiver gauge theories with product unitary gauge group $G=\otimes_{r=1}^pU(N)_{k_r}$ for which the sum of the Chern-Simons (CS) levels~vanishes,~$\sum_{r=1}^pk_r=0$. This class of $\mN=2$ CS-matter theories contains various holographic SCFTs arising from the worldvolume theory of a stack of $N$ coincident M2-branes placed at the tip of a cone over a 7d Sasaki-Einstein manifold. We are interested in the squashed $S^3$ partition function of these quiver gauge theories which can be calculated using supersymmetric localization. The squashing preserves some supersymmetry and is parametrized by a positive real number $b>0$. The SU(2)$_L\times$SU(2)$_R$ isometry of the round $S^3$ is broken to its U(1)$\times$U(1) subgroup such that $S^3_b$ with the special value $b=1$ corresponds to the round $S^3$, see \cite{Hama:2011ea}. 

\medskip

The $S^3_b$ partition function of the $\mN=2$ CS-matter theories of interest can be written in terms of a matrix model via supersymmetric localization as \cite{Kapustin:2009kz,Kapustin:2010xq,Jafferis:2010un,Hama:2011ea,Imamura:2011wg}
\begin{equation}
\begin{split}
	Z(N,b,\bxi)&=\fft{1}{(N!)^p}\int\prod_{r=1}^p\Bigg[\prod_{i=1}^N\Big(\fft{d\mu_{r,i}}{2\pi}\,e^{\fft{\ri k_r}{4\pi}\mu_{r,i}^2-\Delta_{m,r}\mu_{r,i}}\Big)\prod_{i>j}^N\Big(2\sinh\fft{b\mu_{r,ij}}{2}\,2\sinh\fft{\mu_{r,ij}}{2b}\Big)\Bigg]\\
	&\quad\times\prod_\Psi\prod_{\rho_\Psi}s_b\bigg(\fft{\ri Q}{2}(1-\Delta_\Psi)-\fft{\rho_\Psi(\mu)}{2\pi}\bigg)\,,
\end{split}\label{Z}
\end{equation}
where $Q\equiv b+b^{-1}$. See Appendix \ref{app:double-sine} for the definition and various useful properties of the double sine function $s_b(x)$ governing the 1-loop contribution from the chiral multiplets collectively denoted by $\Psi$. We now discuss the various quantities appearing in the localization formula (\ref{Z}).
\begin{itemize}
	\item $\mu_{r,i}$ denotes the gauge zero modes coming from constant scalars in $\mN=2$ vector multiplets parametrizing the BPS locus relevant for supersymmetric localization. Their differences are written compactly as $\mu_{r,ij}=\mu_{r,i}-\mu_{r,j}$.
	
	\item $\Delta_{m,r}$ stands for bare monopole $R$ charges, see \cite{Jafferis:2011zi}.
	
	\item $\Delta_\Psi$ represents the $R$ charge of an $\mN=2$ chiral multiplet $\Psi$ and $\rho_\Psi$ runs over the weights of the representation $\mR_\Psi$ of the chiral multiplet $\Psi$ with respect to the gauge group $G$. We note that for supersymmetric QFTs on $S^3$ the $R$ charge parameters $\Delta_\Psi$ are related to real mass deformations of the theory and we will use this nomenclature interchangeably.
	
	\item $\bxi$ collectively represents the parameters describing a given theory including the CS levels $\bk=\{k_r\}$, the number of pairs of [anti-]fundamental chiral mutiplets $N_f$, and the $R$ charges of chiral multiplets $\bDelta=\{\Delta_\Psi\}$.
\end{itemize}

\medskip

Our main claim in this work is that the matrix integral (\ref{Z}) evaluated in the large $N$ limit can be written in terms of an Airy function up to exponentially suppressed non-perturbative corrections as
\setlength{\fboxsep}{7pt} 
\begin{empheq}[box=\fbox]{equation}
	Z(N,b,\bxi)=\mC^{-1/3}e^{\mA}\text{Ai}\Big[\mC^{-1/3}(N-\mB)\Big]\Big(1+\mO(e^{-\#\sqrt{N}})\Big)\label{Airy}
\end{empheq}
\setlength{\fboxsep}{3pt} 

\noindent This expression is valid for certain holographic $\mN=2$ CS-matter theories which flow to IR SCFTs arising from $N$ coincident M2-branes. Before we discuss explicit examples of such SCFTs we make some remarks on the Airy conjecture (\ref{Airy}).
\begin{itemize}
	\item The parameters $\mA,\mB,\mC$ in (\ref{Airy}) are independent of $N$ and are determined solely by the parameters $b$ and $\bxi$. Notably, we have found that the squashing parameter dependence of the coefficients $\mB$ and $\mC$ can be expressed more explicitly in terms of $Q\equiv b+b^{-1}$ as
	\begin{subequations}
		\begin{align}
			\mB(b,\bxi)&=\beta(\bxi)-\fft{4}{3Q^2}\gamma(\bxi)\,,\\
			\mC(b,\bxi)&=\left(\fft{8}{3\pi Q^2\alpha(\bxi)}\right)^2\,.
		\end{align}\label{mBmC:abc}%
	\end{subequations}
	The explicit form of the parameters $(\alpha,\beta,\gamma)$ for the various $\mN\geq2$ holographic SCFTs of interest in this work is given in Table~\ref{table:abc}. The physical meaning of the parameters collectively denoted by $\bxi$ for the different theories is discussed further in Section~\ref{sec:AiryTale:ex}.
	\begin{table}[t]
		\centering
		\footnotesize
		\renewcommand{\arraystretch}{1.4}
		\begin{tabular}{ |c||c|c|c| } 
			\hline
			& $\alpha(\bxi)$ & $\beta(\bxi)$ & $\gamma(\bxi)$ \\
			\hline\hline
			\text{ABJM} & $\fft{4\sqrt{2k\Delta_1\Delta_2\Delta_3\Delta_4}}{3}$ & $\fft{k}{24}-\fft{1}{12k}\sum_{a=1}^4\fft{1}{\Delta_a}$ & $-\fft{4-\sum_{a=1}^4\Delta_a^2}{16k\Delta_1\Delta_2\Delta_3\Delta_4}$ \\
			\hline
			\text{ADHM} & $\fft{4\sqrt{2N_f\tDelta_1\tDelta_2\tDelta_3\tDelta_4}}{3}$ & $\fft{N_f}{24}-\fft{N_f}{12}(\fft{1}{\tDelta_1}+\fft{1}{\tDelta_2})-\fft{1}{12N_f}(\fft{1}{\tDelta_3}+\fft{1}{\tDelta_4})$ & $-\fft{N_f}{8\tDelta_1\tDelta_2}+\fft{\tDelta_1^2+\tDelta_2^2-2(\tDelta_1+\tDelta_2)+\tDelta_1\tDelta_2}{8N_f\tDelta_1\tDelta_2\tDelta_3\tDelta_4}$ \\
			\hline
			$N^{0,1,0}$ & $\fft{2(k+N_f)}{3\sqrt{2k+N_f}}$ & $-\fft{7N_f-2k}{48}-\fft{2}{3(k+N_f)}$ & $-\fft{N_f}{4}-\fft{3k+2N_f}{(k+N_f)^2}$ \\ 
			\hline
			$V^{5,2}$ & $\fft{16\sqrt{N_f}}{27}$ & $-\fft{N_f}{6}-\fft{1}{4N_f}$ & $-\fft{9N_f}{16}-\fft{27}{16N_f}$ \\ 
			\hline
			$Q^{1,1,1}$ & $\fft{4\sqrt{N_f}}{3\sqrt{3}}$ & $-\fft{N_f}{6}$ & $-\fft{N_f}{4}-\fft{3}{4N_f}$ \\ 
			\hline
		\end{tabular}\caption{The parameters $(\alpha,\beta,\gamma)$ for the different $\mN\geq2$ SCFTs studied in this work.}\label{table:abc}%
	\end{table}

	It is remarkable that the dependence of the coefficients $\mB$ and $\mC$ on the squashing parameter $b$ takes the simple form presented in (\ref{mBmC:abc}). While it is not surprising that the dependence appears only through the combination $Q$, given that the matrix model (\ref{Z}) is invariant under the transformation $b\leftrightarrow b^{-1}$, the simplicity of the resulting expressions for $\mB$ and $\mC$ suggests some underlying structure that will be interesting to explore further.
	
	\item A closed-form expression for the parameter $\mA$ is not known in general. We will review the analytic expressions for $\mA$ that are known for certain special cases in Section~\ref{sec:AiryTale:ex}.
	
	\item The $S^3_b$ partition function (\ref{Airy}) can be expanded in the large $N$ limit, with the parameters $\mA,\mB,\mC$ held finite, as
	\begin{align}
		-\log Z(N,b,\bxi)=\fft23\mC^{-1/2}N^{3/2}-\mC^{-1/2}\mB N^{1/2}+\fft14\log N+\mO(N^0)\,,
	\end{align}
	where we have employed the asymptotic expansion of the Airy function
	\begin{equation}
		\begin{split}
			\text{Ai}[z] &= \frac{\text{exp}\bigl[-\frac23\,z^{3/2}\bigr]}{2\sqrt{\pi}\,z^{1/4}}\,\sum_{n = 0}^{\infty}\,\left(-\tfrac32\right)^n\,u_n\,z^{-3n/2}\,,\\
			u_n &= \frac{(6n-5)(6n-3)(6n-1)}{216(2n-1)n}\,u_{n-1}\qquad(n\geq1\,,~u_0=1)\,.
		\end{split}\label{Airy:asymp}
	\end{equation}
	The characteristic $1/4$ coefficient of the logarithmic term is universal, meaning its value remains unaffected by the specifics of the holographic SCFTs. This logarithmic coefficient has been successfully derived on the dual M-theory side through an 11d supergravity one-loop calculation for the ABJM theory with vanishing real masses and $b=1$ \cite{Bhattacharyya:2012ye}. It has also been derived for more general values of the squashing parameter using 4d Kaluza-Klein supergravity analysis \cite{Bobev:2023dwx}.
	
	\item The exponentially suppressed non-perturbative corrections to the partition function are captured by the $\mO(e^{-\#\sqrt{N}})$ term in \eqref{Airy} where the positive coefficient $\#$ depends on $b$ and $\bxi$. The general form of these non-perturbative corrections is not known and the precise value of the coefficient $\#$ has been determined only in special cases. For instance, in the ABJM theory at the superconformal point, $\#$ is given by $2\pi\sqrt{2/k}$ \cite{Drukker:2010nc,Hatsuda:2012dt}, and a partial generalization to include real mass deformations was recently presented in \cite{Nosaka:2024gle}.. The leading non-perturbative correction is expected to be determined by probe M2-branes in the holographically dual asymptotically AdS$_4$ 11d supergravity solution. This was indeed recently demonstrated for some specific examples in \cite{Gautason:2023igo,Beccaria:2023ujc}.
	
	\item It is not known at present for which large $N$ SCFTs the $S^3$ partition function takes the Airy form in \eqref{Airy}. All holographic SCFTs with $\mN\geq2$ supersymmetry that we study in this paper obey the Airy conjecture and are non-chiral, with matter content consisting of [anti-]fundamental, bi-fundamental, and adjoint chiral multiplets. This naturally raises the question of whether the Airy conjecture can be established under such general conditions, or perhaps with even fewer assumptions. We will speculate further about this in Section~\ref{sec:discussion}.
\end{itemize}
%

\subsection{Implications of the conjecture}\label{sec:AiryTale:implications}

The $S^3$ partition function in \eqref{Airy} encodes a host of important information about the underlying CFT which can be obtained by expanding the partition function in a power series for small values of the deformation parameters $b-1$ and $\bxi$. The coefficients in this expansion can be interpreted as integrated correlation functions of the CFT on $S^3$. Perhaps the simplest example of this is given by the two point function of the stress-energy tensor of the SCFT which is fixed by conformal symmetry up to one positive (for unitary CFTs) real number $C_T$. As shown in \cite{Closset:2012ru} for 3d $\mathcal{N}=2$ SCFTs this coefficient can be determined from the squashed $S^3$ partition function via the following relation (we use the same conventions as in \cite{Agmon:2017xes,Chester:2020jay})
\begin{equation}\label{eq:CTWard}
C_T =\; -\frac{32}{\pi^2}\frac{\partial^2 \log Z_{S^3_{b}}}{\partial b^2}\Big\vert_{b=1}\,.
\end{equation}
We can readily apply this to the Airy conjecture for the $S_b^3$ partition function in \eqref{Airy} to obtain explicit expressions for the coefficient $C_T$ of all SCFTs discussed in this work.\footnote{For the study of second and higher order derivatives of the free energy at $b=1$ employing a relation to the small $b$ expansion rather than the Airy formula, see \cite{Gang:2019jut}.} A straightforward calculation then leads to
\begin{align}
C_T(N,\bxi) =&\; -\frac{32}{9\pi^2}\Bigl\{\frac{\text{Ai}'\bigl[\mathcal{C}^{-1/3}(N-\mathcal{B})\bigr]}{\mathcal{C}^{7/3}\text{Ai}\bigl[\mathcal{C}^{-1/3}(N-\mathcal{B})\bigr]}\,\mathfrak{f}(N,b,\bxi) \nonumber\\ 
&\qquad\quad\;\; + \frac{\text{Ai}\bigl[\mathcal{C}^{-1/3}(N-\mathcal{B})\bigr]^2(N - \mathcal{B}) - \mathcal{C}^{1/3}\text{Ai}'\bigl[\mathcal{C}^{-1/3}(N-\mathcal{B})\bigr]^2}{\mathcal{C}^3\text{Ai}\bigl[\mathcal{C}^{-1/3}(N-\mathcal{B})\bigr]^2}\,\mathfrak{g}(N,b,\bxi)\Bigr\}\Big\vert_{b=1} \nonumber \\
&\; - \frac{128}{3\pi^2} - \frac{32}{\pi^2}\partial^2_b\mathcal{A}\big\vert_{b=1} + \mathcal{O}(e^{-\sqrt{N}}) \, ,
\end{align}
where we have defined
\begin{equation}
\begin{split}
\mathfrak{f}(N,b,\bxi) =&\; 4(N - \mathcal{B})(\partial_b\mathcal{C})^2 - 9\,\mathcal{C}^2 \partial^2_b \mathcal{B} + 3\,\mathcal{C}\bigl(2\,\partial_b\mathcal{C}\,\partial_b\mathcal{B} - (N - \mathcal{B})\partial^2_b\mathcal{C}\bigr) \, , \\
\mathfrak{g}(N,b,\bxi) =&\; \bigl[(N - \mathcal{B})\partial_b\mathcal{C} + 3\,\mathcal{C}\,\partial_b\mathcal{B}\bigr]^2 \, .
\end{split}
\end{equation}
Importantly, the expressions above should be evaluated for the superconformal configurations of all continuous parameters, i.e. we should set $b=1$ and all real mass parameters in $\bxi$ should vanish. Notice that $\mathcal{B}$ and $\mathcal{C}$ in \eqref{mBmC:abc} depend on $b$ only through the function $Q=b+1/b$. This implies that
\begin{equation}
\partial_b\mathcal{B}\big\vert_{b=1}=\partial_b\mathcal{C}\big\vert_{b=1}=0\,,
\end{equation}
and thus we find that for the holographic SCFTs of interest in this work $\mathfrak{g}(N,b,\bxi)=0$. Moreover, using the expressions in \eqref{mBmC:abc} we can rewrite $\mathfrak{f}(N,b,\bxi)$ at the superconformal configuration of parameters as\footnote{We will not write the argument $\bxi$ in the following expressions to avoid clutter in the formulae. It is assumed that all functions of $\bxi$ are evaluated for vanishing real masses.}
\begin{equation}
\mathfrak{f}(N,1) = \frac{32}{81\pi^4\alpha^4}\Bigl(6N-6\beta - \gamma\Bigr) \, .
\end{equation}
With this at hand we arrive at the following expression for $C_T$
\begin{equation}\label{eq:CT5holoex}
\begin{split}
C_T(N) =&\; -\frac{32}{3\pi^{4/3}}\Bigl(\frac{2\alpha^2}{3}\Bigr)^{1/3}\Bigl(6N - 6\beta - \gamma\Bigr)\frac{\text{Ai}'\bigl[\Bigl(\frac{3\pi\alpha}{2}\Bigr)^{2/3}(N - \beta + \frac{\gamma}{3})\bigr]}{\text{Ai}\bigl[\Bigl(\frac{3\pi\alpha}{2}\Bigr)^{2/3}(N - \beta + \frac{\gamma}{3})\bigr]} \\
&\; - \frac{128}{3\pi^2} - \frac{32}{\pi^2}\partial^2_b\mathcal{A}\big\vert_{b=1} + \mathcal{O}(e^{-\sqrt{N}}) \, .
\end{split}
\end{equation}
Using the values of $(\alpha,\beta,\gamma)$ from Table~\ref{table:abc} in the expression above leads to an explicit result for $C_T$ for the holographic SCFTs discussed in this work valid to all orders in the $1/N$ expansion. Since we do not have complete knowledge of the function $\mathcal{A}(b,\bxi)$, the $N^0$ term in \eqref{eq:CT5holoex} is not completely determined. It will be very interesting to find its value either analytically or numerically. Perhaps the method discussed in \cite{Chester:2020jay} can be used to do that.

For SCFTs with $\mathcal{N}=4$ or more supersymmetry the coefficient $C_T$ can also be determined by a different superconformal Ward identity from the one in \eqref{eq:CTWard}, that relates $C_T$ to the second derivative of the sphere partition function with respect to a real mass parameter. This was indeed exploited in \cite{Agmon:2017xes} and \cite{Chester:2020jay} to derive compact expressions for $C_T$ for the ABJM \cite{Agmon:2017xes} and ADHM \cite{Chester:2020jay} SCFTs in terms of an Airy function and its derivatives. The expression in \eqref{eq:CT5holoex}, together with \eqref{mBmC:abc} and the first two rows of Table~\ref{table:abc}, agrees with the results in~\cite{Agmon:2017xes} and~\cite{Chester:2020jay}.

We also note that the series expansion of the Airy function and its first derivative implies that there is no $\log N$ term in the large $N$ expansion of $C_T$, since
\begin{equation}
\frac{\text{Ai}'[a(N-b)]}{\text{Ai}[a(N-b)]} = -\sqrt{a}N^{1/2} + \frac{b\sqrt{a}}{2}N^{-1/2} + \ldots\,.
\end{equation}
This fact is compatible with general expectations about two-point functions in large $N$ 3d CFTs as well as with the holographic calculations of the $\log N$ terms in the partition function of 3d SCFTs arising from M2-branes in M-theory \cite{Bobev:2023dwx}. \\

The Airy conjecture as formulated in \eqref{Airy} applies to the large $N$ limit of holographic SCFTs where all parameters in $(b,\bxi)$ do not scale with $N$. As shown in \cite{Bobev:2022eus}, however, it is possible to use the Airy form of the $S^3$ partition function to reorganize the large $N$ expansion to look like a genus expansion in a manner expected from the general arguments by 't Hooft about the structure of planar gauge theories \cite{tHooft:1973alw}. This 't Hooft limit is achieved by scaling the parameter that specifies the internal geometry together with $N$ while keeping their ratio finite. For the 3d $\mathcal{N}\geq2$ SCFTs we consider in this paper, the parameter in question is the order of the orbifold of the Sasaki-Einstein manifold probed by the stack of M2-branes\footnote{In the $N^{0,1,0}$ theory this order is denoted $k$ in Table \ref{table:abc} but one should ultimately set $k=N_f$, see \cite{Cheon:2011th}.}. In ABJM it corresponds to the CS level $k$, while in the other SCFTs it corresponds to the number of pairs of fundamental chiral multiplets $N_f$. We collectively denote these integer parameters by $K \in \{k,N_f\}$ and consider the 't Hooft limit
\begin{equation}
\label{eq:IIA-limit}
N,K \;\rightarrow\; \infty \, , \qquad N/K = \lambda \;\;\; \text{fixed} \, .
\end{equation}
By inspection of Table~\ref{table:abc}, the $\mathcal{C}$ and $\mathcal{B}$ parameters can be expressed in terms of $K$ as
\begin{equation}
\label{eq:rpq-def}
\mathcal{C} = \frac{r}{K}  \, , \qquad \mathcal{B} = \frac{p}{K} + q\,K \, ,
\end{equation}
where the values of $(r,p,q)$ are given in Table~\ref{table:rpq}.
	\begin{table}[t]
		\centering
		\footnotesize
		\renewcommand{\arraystretch}{1.4}
		\begin{tabular}{ |c||c|c|c| } 
			\hline
			& $r$ & $p$ & $q$ \\
			\hline\hline
			\text{ABJM} & $\fft{2}{\pi^2 Q^4\Delta_1\Delta_2\Delta_3\Delta_4}$ & $-\fft{1}{12}\sum_{a=1}^4\fft{1}{\Delta_a}+\fft{4-\sum_{a=1}^4\Delta_a^2}{12Q^2\Delta_1\Delta_2\Delta_3\Delta_4}$ & $\fft{1}{24}$ \\
			\hline
			\text{ADHM} & $\fft{2}{\pi^2 Q^4\tDelta_1\tDelta_2\tDelta_3\tDelta_4}$ & $-\fft{1}{12}(\fft{1}{\tDelta_3}+\fft{1}{\tDelta_4})-\fft{\tDelta_1^2+\tDelta_2^2-2(\tDelta_1+\tDelta_2)+\tDelta_1\tDelta_2}{6Q^2\tDelta_1\tDelta_2\tDelta_3\tDelta_4}$ & $\fft{1}{24}-\fft{1}{12}(\fft{1}{\tDelta_1}+\fft{1}{\tDelta_2})+\fft{1}{6Q^2\tDelta_1\tDelta_2}$ \\
			\hline
			$N^{0,1,0}$ & $\fft{12}{\pi^2 Q^4}$ & $-\fft{1}{3}+\fft{5}{3Q^2}$ & $-\fft{5}{48}+\fft{1}{3Q^2}$ \\ 
			\hline
			$V^{5,2}$ & $\fft{81}{4\pi^2 Q^4}$ & $-\fft{1}{4}+\fft{9}{4Q^2}$ & $-\fft{1}{6}+\fft{3}{4Q^2}$ \\ 
			\hline
			$Q^{1,1,1}$ & $\fft{12}{\pi^2 Q^4}$ & $\fft{1}{Q^2}$ & $-\fft{1}{6}+\fft{1}{3Q^2}$ \\ 
			\hline
		\end{tabular}\caption{The relabelling of the Airy parameters in terms of $(r,p,q)$ defined in \eqref{eq:rpq-def}.}\label{table:rpq}%
	\end{table}
In terms of these parameters, we can use the results of \cite{Bobev:2022eus} to arrange the expansion of the partition in the limit \eqref{eq:IIA-limit} as a genus expansion\footnote{The $\log N$ term, which is omitted here to highlight the structure of the genus expansion, is reinstated in (\ref{genus-exp}). See also \cite{Bobev:2022jte,Bobev:2022eus} for expressions that include the logarithmic contribution.}
\begin{equation} 
-\log Z(N,\lambda) = \sum_{\mathsf{g}\geq0} F_{\mathsf{g}}(\lambda)\,N^{2-2\mathsf{g}} \, ,
\end{equation}
where the coefficients $F_{\mathsf{g}}$ are given by
\begin{align}
	F_{\mathsf{g}}(\lambda) =&\; \sum_{n\geq0}\,\Bigl\{\frac{2}{3\sqrt{r}}\,\mathcal{F}_{\mathsf{g},n}(r,p,q)\,\frac{(-q)^n}{n!}\,\prod_{\ell=0}^{n-1}\Bigl(\frac{3}{2} - \ell\Bigr) + \mathcal{G}_{\mathsf{g},n}(r,p,q)\,\lambda^{-3/2}\Bigr\}\,\lambda^{2\mathsf{g}-1/2 - n} - A_{\mathsf{g}}\,\lambda^{2\mathsf{g}-2} \nn\\[1mm] 
	&\; + \delta_{\mathsf{g},1}\Bigl[\,\frac14\log(r) + \Bigl(\frac14 + A_{\log}\Bigr)\log(\lambda) + \frac12\log(4\pi)\Bigr] \, .\label{eq:Fg}
\end{align}
The functions $\mathcal{F}$ and $\mathcal{G}$ are defined as, see \cite{Bobev:2022eus} for further details,
\begin{equation}
	\label{eq:cal-F}
	\mathcal{F}_{\mathsf{g},n}(r,p,q) = \begin{cases} 1 &\text{for} \;\; \mathsf{g}=0 \\[1mm] n\,p\,q^{-1} &\text{for} \;\; \mathsf{g}=1 \\[1mm] \sum_{m=0}^{\lfloor\frac{\mathsf{g}}{2}\rfloor}\,F^{(m)}_{\mathsf{g},n}(r,p,q) &\text{for} \;\; \mathsf{g} > 1 \, , \end{cases}
\end{equation}
where
\begin{equation}
	F^{(m)}_{\mathsf{g},n}(r,p,q) = \mathcal{P}^{(m)}\,\frac{n(n-1)\ldots(n-\mathsf{g}+1-m)}{(\mathsf{g}-2m)!}\,r^m\,p^{\mathsf{g} - 2m}\,q^{-\mathsf{g}-m} \, ,
\end{equation}
and as
\begin{equation}
	\label{eq:cal-G}
	\mathcal{G}_{\mathsf{g},n\neq0}(r,p,q) = \begin{cases} 0 &\text{for} \;\; \mathsf{g}=0 \\[1mm] -\frac{q^n}{4n} &\text{for} \;\; \mathsf{g}=1 \\[1mm] \sum_{m=1}^{\lceil\frac{\mathsf{g}}{2}\rceil}\,G^{(m)}_{\mathsf{g},n}(r,p,q) &\text{for} \;\; \mathsf{g} > 1  \end{cases} \quad \text{together with} \quad \mathcal{G}_{\mathsf{g},0} = 0 \, ,
\end{equation}
where
\begin{equation}
	G^{(m)}_{\mathsf{g},n}(r,p,q) = \mathcal{Q}^{(m)}\,\frac{(n-1)\ldots(n-\mathsf{g}+3-m)}{(\mathsf{g}-2m+1)!}\,r^{m-1}\,p^{\mathsf{g} - 2m+1}\,q^{n-\mathsf{g}+2-m} \, ,
\end{equation}
respectively. The quantities $\mathcal{P}^{(m)}$ and $\mathcal{Q}^{(m)}$ denote a set of rational numbers whose values can be inferred from the asymptotic expansion of the Airy function (\ref{Airy:asymp}). Some values at low $m$ can be found in Appendix B of \cite{Bobev:2022eus}. Lastly, in writing \eqref{eq:Fg}, we have assumed that the unknown function $\mathcal{A}$ in our Airy conjecture admits a large $K$ expansion of the form
\begin{equation}
\label{eq:A-IIA-expand}
\mathcal{A} = \sum_{\mathsf{g}\geq0} A_\mathsf{g}\,K^{2-2\mathsf{g}} + A_{\log} \log(K) \, .
\end{equation}
This is known to be true for ABJM on the round sphere and with specific mass deformations, see \cite{Hatsuda:2014vsa,Nosaka:2015iiw}. Together with the data in Table~\ref{table:rpq}, \eqref{eq:Fg} determines the coefficients in the genus expansion of $\log Z$ for all SCFTs considered in this paper, modulo the unknown terms in \eqref{eq:A-IIA-expand}. It is worth noting that the resulting expressions can be written compactly for a given genus in terms of a shifted `t Hooft parameter. For instance,
\begin{equation}
\begin{split}
F_0(\lambda) =&\; \lambda^{-2}\Bigl[\frac{2}{3\sqrt{r}}\,\hat{\lambda}^{3/2} - A_0\Bigr] + \mathcal{O}(e^{-\sqrt{\lambda}}) \, , \\
F_1(\lambda) =&\; -\frac{p}{\sqrt{r}}\,\hat{\lambda}^{1/2} - A_1 + A_{\log}\log\lambda + \frac14\log(16\pi^2r) + \frac14\log\hat{\lambda} + \mathcal{O}(e^{-\sqrt{\lambda}}) \, , \\
F_2(\lambda) =&\; \lambda^{2}\Bigl[\frac{p^2}{4\sqrt{r}}\,\hat{\lambda}^{-1/2} - \frac{p}{4}\,\hat{\lambda}^{-1} + \frac{5\sqrt{r}}{48}\,\hat{\lambda}^{-3/2} - A_2\Bigr] + \mathcal{O}(e^{-\sqrt{\lambda}}) \, ,
\end{split}
\end{equation}
where $\hat{\lambda} = \lambda - q$. It is natural to conjecture that the field theory quantities $F_{\mathsf{g}}$ should correspond to genus-${\mathsf{g}}$ free energies of the type IIA string on an appropriate 10d limit of the 11d backgrounds dual to the five holographic SCFTs discussed above. Needless to say, it will be most interesting to explicitly confirm this realization of 't Hooft's ideas.

\subsection{Previous evidence}\label{sec:AiryTale:ex}
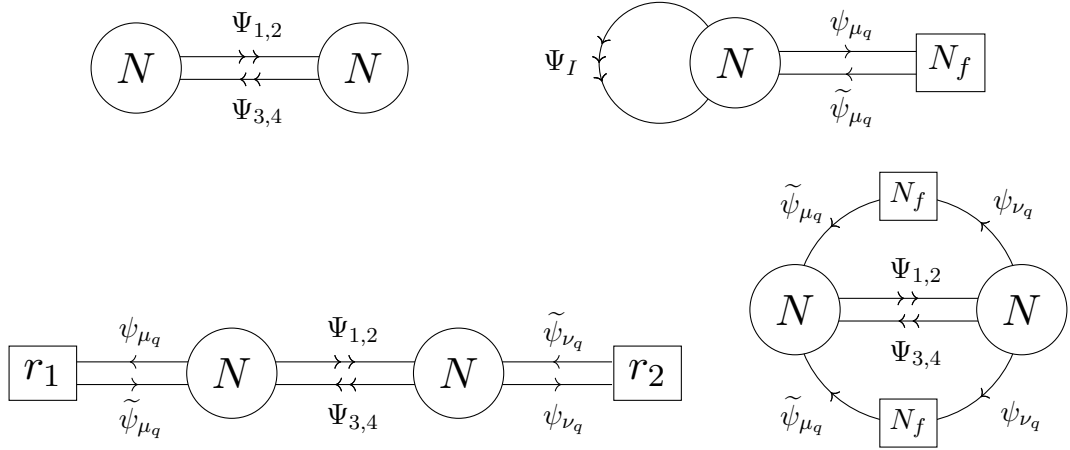
\begin{figure}[t]
	\centering
	\begin{tikzpicture}
		\draw[->-=0.47,->-=0.57] (0.57,0.15) -- (2.43,0.15);
		\draw[->-=0.47,->-=0.57] (2.43,-0.15) -- (0.57,-0.15);
		\node at (0,0) [circle,draw,scale=1.5,fill=white] {$N$};
		\node at (3,0) [circle,draw,scale=1.5,fill=white] {$N$};
		\node at (1.6,0.55) {$\Psi_{1,2}$};
		\node at (1.6,-0.6) {$\Psi_{3,4}$};
	\end{tikzpicture}
	\kern4em
	\begin{tikzpicture}
		\draw[->-=0.52] (0.57,0.15) -- (2.43,0.15);
		\draw[->-=0.52] (2.43,-0.15) -- (0.57,-0.15);
		\draw[->-=0.45,->-=0.5,->-=0.55] (-0.2,0) arc (0:360:0.8) ;
		\node at (0,0) [circle,draw,scale=1.5,fill=white] {$N$};
		\node at (2.85,0) [rectangle,draw,scale=1.2,fill=white] {$N_f$};
		\node at (-2.3,0) {$\Psi_I$};
		\node at (1.55,0.5) {$\psi_{\mu_q}$};
		\node at (1.55,-0.6) {$\tpsi_{\mu_q}$};
	\end{tikzpicture}
	
	\bigskip
	
	\begin{tikzpicture}
		\draw[->-=0.47,->-=0.57] (0.57,0.15) -- (2.43,0.15);
		\draw[->-=0.47,->-=0.57] (2.43,-0.15) -- (0.57,-0.15);
		\draw[->-=0.5] (-0.57,0.15) -- (-2.13,0.15);
		\draw[->-=0.57] (-2.13,-0.15) -- (-0.57,-0.15);
		\draw[->-=0.5] (5.03,0.15) -- (3.47,0.15);
		\draw[->-=0.57] (3.47,-0.15) -- (5.03,-0.15);
		\node at (-2.5,0) [rectangle,draw,scale=1.5,fill=white] {$r_1$};
		\node at (0,0) [circle,draw,scale=1.5,fill=white] {$N$};
		\node at (3,0) [circle,draw,scale=1.5,fill=white] {$N$};
		\node at (5.5,0) [rectangle,draw,scale=1.5,fill=white] {$r_2$};
		\node at (1.6,0.55) {$\Psi_{1,2}$};
		\node at (1.6,-0.6) {$\Psi_{3,4}$};
		\node at (-1.2,0.55) {$\psi_{\mu_q}$};
		\node at (-1.2,-0.6) {$\tpsi_{\mu_q}$};
		\node at (4.4,0.55) {$\tpsi_{\nu_q}$};
		\node at (4.4,-0.6) {$\psi_{\nu_q}$};
	\end{tikzpicture}
	\qquad
	\begin{tikzpicture}
		\draw[->-=0.47,->-=0.57] (0.57,0.15) -- (2.43,0.15);
		\draw[->-=0.47,->-=0.57] (2.43,-0.15) -- (0.57,-0.15);
		\draw[->-=0.28,->-=0.74] (3,0) arc (0:180:1.5);
		\draw[->-=0.28,->-=0.74] (3,0) arc (0:-180:1.5);
		\node at (0,0) [circle,draw,scale=1.5,fill=white] {$N$};
		\node at (3,0) [circle,draw,scale=1.5,fill=white] {$N$};
		\node at (1.5,1.5) [rectangle,draw,scale=1,fill=white] {$N_f$};
		\node at (1.5,-1.5) [rectangle,draw,scale=1,fill=white] {$N_f$};
		\node at (1.6,0.5) {$\Psi_{1,2}$};
		\node at (1.6,-0.6) {$\Psi_{3,4}$};
		\node at (2.9,1.4) {$\psi_{\nu_q}$};
		\node at (3,-1.4) {$\psi_{\nu_q}$};
		\node at (0.1,1.4) {$\tpsi_{\mu_q}$};
		\node at (0.1,-1.4) {$\tpsi_{\mu_q}$};
	\end{tikzpicture}
	\caption{Quiver diagrams for the ABJM (upper-left), ADHM/$V^{5,2}$ (upper-right), $N^{0,1,0}$ (lower-left), and $Q^{1,1,1}$ (lower-right) gauge theories.}\label{quiver}
\end{figure}

We now proceed to summarize the known evidence for the Airy conjecture (\ref{Airy}) with (\ref{mBmC:abc}) for the $\mN\geq2$ holographic SCFTs listed in Table~\ref{table:abc}. We refer the reader to \cite{Bobev:2023lkx,Geukens:2024zmt} and references therein for a detailed description of these SCFTs, and here limit ourselves to only providing their quiver diagrams in Figure~\ref{quiver}.

\medskip

We would like to emphasize that most of the evidence in the previous literature that supports the Airy conjecture is restricted to the round $S^3$ with $b=1$ (or with the specific value $b^2=3$) and often assumes specific values for the real mass parameters. For a general squashing parameter $b$, previous evidence in support of the Airy conjecture is available only at the first two leading terms of order $N^{3/2}$ and $N^{1/2}$ in the M-theory limit where $N$ is taken large with fixed $k_r$ and $N_f$. The  main focus of Sections \ref{sec:evi1}, \ref{sec:evi2}, and \ref{sec:evi3}, is to provide new evidence in favor of the Airy conjecture applicable for generic squashing parameter and real masses.

\subsubsection{ABJM theory}\label{sec:AiryTale:ex:ABJM}
For the $\text{U}(N)_k\times\text{U}(N)_{-k}$ ABJM theory the matter content consists of two pairs of $\mN=2$ bi-fundamental chiral multiplets \cite{Aharony:2008ug}. The general $S^3$ matrix model (\ref{Z}) reduces to
\begin{align}
		Z^\text{ABJM}(N,b,k,\bDelta)&=\fft{1}{(N!)^2}\int\left(\prod_{i=1}^N\fft{d\mu_i}{2\pi}\fft{d\nu_i}{2\pi}\right)\,e^{\fft{\ri k}{4\pi}\sum_{i=1}^N(\mu_i^2-\nu_i^2)}\label{ABJM:Z}\\
		&\quad\times\prod_{i>j}2\sinh\fft{b\mu_{ij}}{2}\,2\sinh\fft{\mu_{ij}}{2b}\,2\sinh\fft{b\nu_{ij}}{2}\,2\sinh\fft{\nu_{ij}}{2b}\nn\\
		&\quad\times\prod_{i,j=1}^N\prod_{a=1}^2s_b\bigg(\fft{\ri Q}{2}(1-\Delta_a)-\fft{\mu_i-\nu_j}{2\pi}\bigg)\prod_{a=3}^4s_b\bigg(\fft{\ri Q}{2}(1-\Delta_a)+\fft{\mu_i-\nu_j}{2\pi}\bigg)\,.\nn
\end{align}
The parameters collectively denoted by $\bxi=\{k,\bDelta\}$ include the CS level $k$ and the $R$ charges of chiral multiplets $\bDelta=\{\Delta_a|\,a\in\{1,2,3,4\}\}$ satisfying the superpotential marginality constraint
\begin{equation}
	\sum_{a=1}^4\Delta_a=2\,.\label{ABJM:constraint}
\end{equation}
The real mass parameters are related to the $R$ charges as
	\begin{equation}
		\begin{alignedat}{2}
			\Delta_1 &= \fft12 - \ri\,\fft{m_1 + m_2 + m_3}{Q} \, ,& \qquad \Delta_2 &= \fft12 - \ri\,\fft{m_1 - m_2 - m_3}{Q} \, , \\
			\Delta_3 &= \fft12 + \ri\,\fft{m_1 + m_2 - m_3}{Q} \, ,& \qquad \Delta_4 &= \frac12 + \ri\,\fft{m_1 - m_2 + m_3}{Q} \, .
		\end{alignedat}\label{Delta:m}
	\end{equation}
The superconformal configuration is determined by imposing vanishing real masses and thus corresponds to $\bDelta_\text{sc}=(\fft12,\fft12,\fft12,\fft12)$. From the localization formula (\ref{ABJM:Z}), it is straightforward to derive that the $S^3_b$ partition function of the ABJM theory is invariant under the three independent permutations\footnote{It is worth noting that the $N$-dependent part of the Airy conjecture (\ref{Airy}) for the ABJM theory, captured by the Airy function, is in fact invariant under the full permutation group of the $\Delta_a$'s. In contrast, the $N$-independent constant term $e^{\mA}$ does not exhibit full permutation symmetry; rather, it is invariant only under the subgroup specified in (\ref{ABJM:sym}). This restricted symmetry is evident in the planar limit expression (\ref{ABJM:mfc:b=1}).}
\begin{equation}
	\Delta_1\leftrightarrow\Delta_2\,,\qquad \Delta_3\leftrightarrow\Delta_4\,,\qquad(\Delta_1,\Delta_2)\leftrightarrow(\Delta_3,\Delta_4)\,.\label{ABJM:sym}
\end{equation}

\medskip

The Airy conjecture for the ABJM theory, given by (\ref{Airy}) with (\ref{mBmC:abc}) and the 1st line of Table~\ref{table:abc}, has been derived or supported in the previous literature in the following special cases. 
\begin{itemize}
	\item On the round $S^3$ with $b=1$, the Airy conjecture was analytically proven for the superconformal configuration $\bDelta=\bDelta_\text{sc}$. This was initially achieved through a series of dualities involving a topological string partition function on local $\mathbb{P}\times\mathbb{P}$ \cite{Fuji:2011km} and then independently derived using the free Fermi-gas formalism \cite{Marino:2011eh}. Subsequently, the free Fermi-gas formalism was applied for more general real mass parameters, leading to the derivation of the Airy conjecture by Nosaka \cite{Nosaka:2015iiw} for the following values of $\Delta_a$ 
	\begin{equation}
		\bDelta=\bDelta_\text{Nosaka}\equiv(\Delta_1,\Delta_2,1-\Delta_2,1-\Delta_1)\,.\label{ABJM:Nosaka}
	\end{equation}
	For the special $R$ charge configuration given by (\ref{ABJM:Nosaka}), the $\mA$ factor in the Airy conjecture (\ref{Airy}) was determined explicitly in \cite{Nosaka:2015iiw} and reads
	\begin{equation}
		\mA^\text{ABJM}(1,k,\bDelta_\text{Nosaka})=\fft14\sum_{a=1}^4A(2\Delta_ak)\,,\label{ABJM:mA:Nosaka}
	\end{equation}
	where the $A$-function is defined in \cite{Hatsuda:2014vsa} as
	\begin{align}
		A(k)&\equiv\fft{2\zeta(3)}{\pi^2 k}\left(1-\fft{k^3}{16}\right)+\fft{k^2}{\pi^2}\int_0^\infty dx\,\fft{x}{e^{kx}-1}\log(1-e^{-2x})\nn\\
		&=-\fft{\zeta(3)}{8\pi^2}k^2+2\zeta'(-1)+\fft16\log\fft{4\pi}{k}+\sum_{g\geq2}\left(\fft{2\pi}{k}\right)^{2g-2}\fft{(-1)^g4^{g-1}|B_{2g}B_{2g-2}|}{g(2g-2)(2g-2)!}\,,\label{def:A}
	\end{align}
where the expression on the second line is valid only for $k>0$. Note that the function $A(k)$ is odd with respect to $k$, which can be shown as follows:
	\begin{align}
		A(-k)&=-\fft{2\zeta(3)}{\pi^2 k}\left(1+\fft{k^3}{16}\right)-\fft{k^2}{\pi^2}\int_0^\infty dx\,\bigg[x+\fft{x}{e^{kx}-1}\bigg]\log(1-e^{-2x})\label{A:odd}\\
		&=-\fft{2\zeta(3)}{\pi^2 k}\left(1+\fft{k^3}{16}\right)-\fft{k^2}{\pi^2}\bigg[-\fft14\zeta(3)+\int_0^\infty dx\,\fft{x}{e^{kx}-1}\log(1-e^{-2x})\bigg]\nn\\
		&=-A(k)\,.\nn
	\end{align}
	%
	
	\item On the squashed $S^3$ with the specific value of the squashing parameter $b=\sqrt{3}$ (or $b=1/\sqrt{3}$) and unit CS level $k=1$, the Airy conjecture was analytically proven for the superconformal configuration $\bDelta=\bDelta_\text{sc}$. This proof employed special properties of the double sine function summarized in Appendix~\ref{app:double-sine} and the mirror symmetry between the ABJM theory and the ADHM theory for $k=N_f=1$ \cite{Hatsuda:2016uqa}, within the framework of the free Fermi-gas formalism \cite{Marino:2011eh}. In this case the $\mA$ factor in the Airy conjecture (\ref{Airy}) was derived in \cite{Hatsuda:2016uqa,Grassi:2014zfa} and reads
	\begin{equation}
		\mA^\text{ABJM}(\sqrt{3},1,\bDelta_\text{sc})=\fft34A(2)-\fft14A(6)=-\fft{\zeta(3)}{3\pi^2}+\fft16\log 3\,.\label{ABJM:mA:sqrt3}
	\end{equation}
	Using similar methods, the Airy conjecture was also derived very recently for more general cases with $k=1$, $b^2\in2\mathbb{N}-1$, and the following $R$ charge configuration \cite{Kubo:2024qhq}\footnote{This configuration is equivalent to the choice of \cite{Kubo:2024qhq} up to the permutation symmetries listed in (\ref{ABJM:sym}).}
	\begin{equation}
		\bDelta_\text{Fermi}^{(N_f=1)}\equiv\bigg(\fft12+\ri\fft{2m}{Q},\fft12-\ri\fft{2\zeta}{Q},\fft12-\ri\fft{2m}{Q},\fft12+\ri\fft{2\zeta}{Q}\bigg)\Bigg|_{m=\fft{(b^2-3)\ri}{4b}}\,,\label{ABJM:bDelta:Fermi}
	\end{equation}
where $\zeta$ is an arbitrary parameter. The closed-form expression for the $\mA$ factor in this case reads
	\begin{align}
		\mA^\text{ABJM}(b,1,\bDelta_\text{Fermi}^{(N_f=1)})&=\fft14\bigg[A\Big(\fft{b^2+1-4\ri b\zeta}{2}\Big)+A\Big(\fft{b^2+1+4\ri b\zeta}{2}\Big)\nn\\
		&\qquad~+A(b^2-1)-A(2b^2)\bigg]\,.\label{ABJM:mA:k=1}
	\end{align}
Note that the expression in (\ref{ABJM:mA:k=1}) is indeed symmetric under the transformation $b\leftrightarrow b^{-1}$, provided the real mass parameter $\zeta$ is chosen appropriately so that $(m,\zeta)$ are paired under $b\leftrightarrow b^{-1}$. Specifically, the $\mA$ factor (\ref{ABJM:mA:k=1}) vanishes for $\zeta=\fft{(b^{-2}-3)\ri}{4b^{-1}}$
	\begin{equation}
		\mA^\text{ABJM}(b,1,\bDelta_\text{Fermi}^{(N_f=1)})\Big|_{\zeta=\fft{(b^{-2}-3)\ri}{4b^{-1}}}=\fft14\big[A(1-b^2)+A(b^2-1)\big]=0\,,
	\end{equation}
	which is trivially symmetric under $b\leftrightarrow b^{-1}$. To obtain this identity we have employed the fact that the $A$-function is odd, see (\ref{A:odd}).
	
	\item On the round $S^3$ with $b=1$, the Airy conjecture is also supported by a numerical evaluation of the matrix model integral in (\ref{ABJM:Z}) via saddle point approximation in the 't~Hooft limit, i.e. $N\gg1$ with fixed $\lambda=N/k$, for generic $\bDelta$ configurations \cite{Geukens:2024zmt}. Specifically, the Airy conjecture in the 't~Hooft limit is consistent with the expression
	\begin{subequations}
	\begin{align}
			-\log Z^\text{ABJM}(N,1,k,\bDelta)&=\fft{4\pi\sqrt{2\Delta_1\Delta_2\Delta_3\Delta_4}}{3}\fft{(\lambda-\fft{1}{24})^\fft32}{\lambda^2}N^2+\fft{\mathfrak{c}^\text{ABJM}(1,\bDelta)}{\lambda^2}N^2\nn\\
			&\quad+\mO(e^{-\#\sqrt{\lambda}})+o(N^2)\,,\label{ABJM:F0:b=1}\\[0.5em]
			\mfc^\text{ABJM}(1,\bDelta)&=\fft{\zeta(3)}{8\pi^2}\Bigg[4-\sum_{a=1}^4\Delta_a^2+\fft{4}{\Delta_{13}\Delta_{14}\Delta_{23}\Delta_{24}}\sum_{a=1}^4\fft{\prod_{b=1}^4\Delta_b}{\Delta_a}\nn\\
			&\kern3em~~-\bigg(\fft{4}{\Delta_{13}\Delta_{24}}+\fft{4}{\Delta_{14}\Delta_{23}}\bigg)\sum_{a=1}^4\fft{\prod_{b=1}^4\Delta_b}{\Delta_a}\Bigg]\,,\label{ABJM:mfc:b=1}
	\end{align}%
	\end{subequations}
	derived in \cite{Geukens:2024zmt} using the asymptotic expansion of the Airy function (\ref{Airy:asymp}). In the expressions above we used the compact notation $\Delta_{ab}\equiv\Delta_a+\Delta_b$. The result above allows us to determine the large $k$ limit of the $\mA$ factor in the Airy conjecture (\ref{Airy}) for generic $R$ charge configurations to be
	\begin{equation}
		\mA^\text{ABJM}(1,k,\bDelta)=-\mfc^\text{ABJM}(1,\bDelta)k^2+o(k^2)\,.\label{ABJM:mA:general}
	\end{equation}
This expression generalizes the large $k$ limit of the $\mA$ factor (\ref{ABJM:mA:Nosaka}) for the special Nosaka configuration (\ref{ABJM:Nosaka}).
	
	\item The Airy conjecture is compatible with the following relation 
	\begin{align}
		&F^\text{ABJM}(N,b,k,m_1,m_2,\ri\fft{b-b^{-1}}{2})\\
		&=F^\text{ABJM}(N,1,k,\fft{b^{-1}(m_1+m_2)+b(m_1-m_2)}{2},\fft{b^{-1}(m_1+m_2)-b(m_1-m_2)}{2},0)\,,\nn
	\end{align}
which is indeed	imposed by the symmetry of the matrix model, see \cite{Chester:2021gdw,Minahan:2021pfv}. Note that here we have used the real mass parameters $m_{1,2,3}$ for the argument of the $S_b^3$ free energy $F=-\log Z$ instead of the $R$ charges. 
	
	\item On the squashed $S^3$ with a generic value of $b$, the Airy conjecture precisely yields the first two leading terms of order $N^{3/2},N^{1/2}$ in the M-theory expansion of the $S_b^3$ free energy at the superconformal configuration $\bDelta=\bDelta_\text{sc}$. These results were obtained using holography and four-derivative supergravity in \cite{Bobev:2020egg,Bobev:2021oku}. To be specific, the Airy conjecture reproduces the following expression for the free energy
	\begin{align}
		-\log Z^\text{ABJM}(N,b,k,\bDelta_\text{sc})&=\fft{\pi\sqrt{2k}}{3}\bigg[\fft{Q^2}{4}\bigg(N^{3/2}+\fft{16-k^2}{16k}N^{1/2}\bigg)-\fft{2}{k}N^{1/2}\bigg]\nn\\
		&\quad+\frac{1}{4}\log N+\mO(N^0)
	\end{align}
	in the M-theory limit with $N\gg1$ and fixed $k$. The coefficient of the $\log N$ term on the second line is independent of $b$ and can be derived using the dual supergravity description, see \cite{Bhattacharyya:2012ye,Bobev:2023dwx}.
	
Yet another check of the Airy conjecture can be performed by using the AdS/CFT correspondence. For $b=1$ and $k=1$ the leading term in the large $N$ ABJM sphere free energy according to the Airy conjecture takes the form\footnote{This result was first computed in \cite{Jafferis:2011zi} by using the large $N$ saddle point approximation to the supersymmetric localization matrix model.} 
\begin{equation}
F^{\rm ABJM} = \frac{4\pi\sqrt{2\Delta_1\Delta_2\Delta_3\Delta_4}}{3}N^{3/2}\,.
\end{equation}
This is indeed in agreement with the regularized on-shell action of a family of dual four-dimensional asymptotically AdS$_4$ Euclidean supergravity solutions constructed in \cite{Freedman:2013oja} for which the parameters $\Delta_a$ are determined by the asymptotic behavior of scalar fields in the 4d maximally supersymmetric gauged supergravity.
	
\end{itemize}

The ABJM results summarized above can be extended to yet another holographic SCFT which we refer to as the mABJM SCFT. This theory can be obtained from the ABJM model with $k=1,2$ by adding a superpotential mass term to the Lagrangian which breaks supersymmetry to $\mathcal{N}=2$ and induces an RG flow that terminates in a strongly interacting SCFT with ${\rm SU}(3)$ flavor symmetry \cite{Benna:2008zy}.\footnote{This relevant deformation of the ABJM theory is available only for the values $k=1,2$ due to the presence of light monopole operators.} The superpotential deformation fixes one of the mass parameters of the ABJM theory $\Delta_1=1$. The remaining two mass parameters correspond to the Cartan generators of the ${\rm SU}(3)$ flavor group and are parametrized by $\Delta_{2,3,4}$ with the constraint 
\begin{equation}\label{eq:mABJMDeltas}
\Delta_2+\Delta_3+\Delta_4=1\,.
\end{equation}
As discussed in \cite{Jafferis:2011zi}, the $S^3$ partition function of the mABJM theory can be computed by supersymmetric localization and the resulting matrix model takes the form \eqref{Z}. We propose that to all orders in the $1/N$ expansion, the partition function of the mABJM theory in the presence of squashing and real mass deformations is given by an Airy function and takes the form \eqref{Airy}. The quantities $\mathcal{A}$, $\mathcal{B}$, and $\mathcal{C}$ are the same as for the ABJM theory but with the fixed values $k=1,2$ and $\Delta_1=1$ and the constraint \eqref{eq:mABJMDeltas}. This conjecture is supported by some of the evidence described above that also supports the ABJM Airy conjecture for fixed $k=1,2$. In addition, the leading $N^{3/2}$ term in the large $N$ expansion of the mABJM free energy can be derived also from the holographically dual asymptotically AdS$_4$ background, see \cite{Warner:1983vz,Corrado:2001nv,Bobev:2018uxk,Bobev:2018wbt}. Moreover, for vanishing real masses and general values of $b$ the subleading $N^{1/2}$ and $\log N$ corrections to the free energy are also in agreement with the Airy conjecture as shown in \cite{Bobev:2020egg,Bobev:2021oku} and \cite{Bobev:2023dwx}, respectively.  


Yet another generalization of the Airy conjecture applies to the ABJ theory \cite{Aharony:2008gk} which has the same matter content as the ABJM theory but the more general gauge group ${\rm U}(N)_{k} \times {\rm U}(N+M)_{-k}$. The integer $M$ determines the number of fractional M2-branes localized at the $\mathbb{C}^4/\mathbb{Z}_k$ singularity in the dual M-theory description. It is possible to compute the path integral of this SCFT on $S^3$ by supersymmetric localization and the resulting matrix model is very similar to the one in \eqref{ABJM:Z} for the ABJM theory. It was shown in \cite{Honda:2014npa,Codesido:2014oua} that the large $N$ $S^3$ partition function of the ABJ theory for $b=1$ and vanishing real masses is determined by an Airy function and takes the form in \eqref{Airy}. This was further extended in \cite{Agmon:2017xes} and \cite{Binder:2020ckj} to include the dependence on two real mass parameters, see also \cite{Chester:2021gdw}. Based on these results we conjecture that the general large $N$ $S^3$ partition function of the ABJ theory with all three real masses and general squashing parameter $b$ also has the Airy function form in \eqref{Airy}. To determine the parameters $(\mathcal{A},\mathcal{B},\mathcal{C})$ we note that it was shown in \cite{Agmon:2017xes} how to calculate the stress-energy two-point function coefficient $C_T$ for the ABJ theory to all orders in the $1/N$ expansion for fixed $M$. This was done by exploiting a superconformal Ward identity that relates the coefficient $C_T$ to the second derivative of the sphere partition function with respect to real mass parameters. This, together with \eqref{eq:CT5holoex} above and the known expressions for $(\mathcal{A},\mathcal{B},\mathcal{C})$ for the ABJ theory with two real masses \cite{Agmon:2017xes,Binder:2020ckj,Chester:2021gdw}, suggests that $\mathcal{C}$ for the ABJ theory is the same for the ABJM one, i.e. it does not depend on $M$ and is explicitly given by \eqref{mBmC:abc} and the first row of Table~\ref{table:abc}. To conjecture the form of $\mathcal{B}$ we in addition use Equation (2.5) of \cite{Chester:2021gdw} to propose that $\gamma(\bxi)$ of the ABJ theory is also the same as that of the ABJM one.\footnote{Notice that we use a convention in which the ABJ theory has an ${\rm U}(N)_{k}\times {\rm U}(N+M)_{-k}$ gauge group which agrees with the one in \cite{Chester:2021gdw} and is different from \cite{Agmon:2017xes}.} If our proposal is correct then the only difference in the form of $\mathcal{B}$ between the ABJ and ABJM models comes from the expression for $\beta(\bxi)$ which we conjecture takes the form
\begin{equation}\label{eq:ABJbetaconj}
\beta(\bxi)=\frac{k}{8}\left(1-\frac{2M}{k}\right)^2-\fft{k}{12}-\fft{1}{12k}\sum_{a=1}^4\fft{1}{\Delta_a}\,.
\end{equation}
Clearly for $M=0$ this reduces back to the result for the ABJM theory and also this form of $\beta$ ensures that the large $N$ expansion for $C_T$ in the ABJ theory agrees with the results in~\cite{Agmon:2017xes}. Since in the ABJM theory we do not know the explicit form of the function $\mathcal{A}(b,\bxi)$ for general values of the arguments we will refrain from conjecturing the corresponding function in the ABJ theory. We stress that this form of the ABJ Airy conjecture and the proposed expression for $\beta(\bxi)$ in \eqref{eq:ABJbetaconj} are on less solid ground than the rest of the Airy conjectures discussed in this work. This is due to the fact that for the ABJ theory with general values of $M$ not much is known about the holographically dual description for general values of the real masses and the squashing parameter. We hope that the Airy conjecture we formulated above will inspire further holographic studies of the ABJ theory.

\subsubsection{ADHM theory}\label{sec:AiryTale:ex:ADHM}
The $\text{U}(N)$ ADHM theory has a single U$(N)$ gauge group with vanishing CS term, together with three $\mN=2$ adjoint chiral multiplets and $N_f$ pairs of $\mN=2$ fundamental \& anti-fundamental chiral multiplets. The theory preserves $\mathcal{N}=4$ supersymmetry and the general matrix model (\ref{Z}) reduces to 
\begin{equation}
	\begin{split}
		Z^\text{ADHM}(N,b,N_f,\btDelta)&=\fft{1}{N!}\int\prod_{i=1}^N\Big(\fft{d\mu_i}{2\pi}\,e^{-\Delta_m\mu_i}\Big)\,\prod_{i>j}^N2\sinh\fft{b\mu_{ij}}{2}2\sinh\fft{\mu_{ij}}{2b}\\
		&\quad\times\prod_{a=1}^3\prod_{i,j=1}^Ns_b\bigg(\fft{\ri Q}{2}(1-\Delta_a)-\fft{\mu_{ij}}{2\pi}\bigg)\\
		&\quad\times\prod_{i=1}^N\prod_{q=1}^{N_f}\Bigg[s_b\bigg(\fft{\ri Q}{2}(1-\Delta_{\mu_q})-\fft{\mu_i}{2\pi}\bigg)s_b\bigg(\fft{\ri Q}{2}(1-\tDelta_{\mu_q})+\fft{\mu_i}{2\pi}\bigg)\Bigg]\,.
	\end{split}\label{ADHM:Z}
\end{equation}
In the ADHM theory we introduce the collective parameter $\bxi=\{N_f,\btDelta\}$ where $\btDelta$ is expressed in terms of the $R$ charges as
\begin{equation}
	\btDelta=\bigg(\Delta_1,\Delta_2,\fft{2-\Delta-\tDelta}{2}-\fft{2}{Q}\fft{\Delta_m}{N_f},\fft{2-\Delta-\tDelta}{2}+\fft{2}{Q}\fft{\Delta_m}{N_f}\bigg)\,.\label{ADHM:btDelta}
\end{equation}
For simplicity we will from now on take all $R$ charges of the fundamental \& anti-fundamental multiplets to be equal, i.e. $(\Delta,\tDelta)=(\Delta_{\mu_q},\tDelta_{\mu_q})$. The superpotential marginality constraints are given by
\begin{equation}
	\sum_{I=1}^3\Delta_I=\Delta_3+\Delta+\tDelta=2\quad\to\quad\sum_{a=1}^4\tDelta_a=2\,.\label{ADHM:constraint}
\end{equation}
The superconformal configuration reads $\btDelta_\text{sc}=(\fft12,\fft12,\fft12,\fft12)$ in terms of the $R$ charges in (\ref{ADHM:btDelta}). Furthermore, we will identify $\Delta=\tDelta$, since the difference $\Delta-\tDelta$ only affects the phase of the partition function (\ref{ADHM:Z}) in a trivial manner. See \cite{Geukens:2024zmt} for further details regarding this choice. 

\medskip

The Airy conjecture for the ADHM theory, given by (\ref{Airy}) with (\ref{mBmC:abc}) and the second line of Table~\ref{table:abc}, has been established in various special cases which we now summarize. 
\begin{itemize}
	\item On the round $S^3$ with $b=1$, the free Fermi-gas formalism \cite{Marino:2011eh} was employed to demonstrate the Airy conjecture for the superconformal configuration $\btDelta=\btDelta_\text{sc}$ \cite{Mezei:2013gqa,Hatsuda:2014vsa}. Recently, this analysis has been extended to include more general $R$ charge configurations \cite{Hatsuda:2021oxa}, specifically\footnote{\label{footnote6} We refer to this configuration as ``Nosaka-like'' because a natural parameter identification for the mirror duality between the ABJM theory and the ADHM theory \cite{Geukens:2024zmt},
	\begin{equation}
		\bDelta=(\Delta_1,\Delta_2,\Delta_3,\Delta_4)~\leftrightarrow~(\tDelta_1,\tDelta_3,\tDelta_2,\tDelta_4)=\btDelta\,,\label{Map:Nosaka}
	\end{equation}
	maps (\ref{ADHM:Nosaka-like}) to the Nosaka configuration (\ref{ABJM:Nosaka}).
	}
	\begin{equation}
		\btDelta_\text{Nosaka-like}\equiv(\tDelta_1,1-\tDelta_1,\tDelta_3,1-\tDelta_3)\,,\label{ADHM:Nosaka-like}
	\end{equation}
	where \cite{Chester:2023qwo} independently addressed the case with the additional constraint $\tDelta_3=\fft12$.  The $\mA$ factor in the Airy conjecture (\ref{Airy}) was derived in \cite{Hatsuda:2014vsa,Chester:2023qwo,Geukens:2024zmt} and reads
	\begin{align}
		\mA^\text{ADHM}(1,N_f,\btDelta_\text{Nosaka-like})&=\fft{N_f^2}{2}\Big(A(2\tDelta_1)+A(2\tDelta_2)\Big)\nn\\
		&\quad+\fft14\Big(A(2\tDelta_3N_f)+A(2\tDelta_3N_f)\Big)\label{ADHM:mA:special}
	\end{align}
	where the $A$-function is defined in (\ref{def:A}). See \cite{Geukens:2024zmt} for further details on the derivation presented in \cite{Hatsuda:2021oxa,Chester:2023qwo}.	
	
	\item On the squashed $S^3$ with $b=\sqrt3$ (or $b=1/\sqrt3$), the Airy conjecture was derived analytically for the superconformal configuration $\btDelta=\btDelta_\text{sc}$ \cite{Hatsuda:2016uqa}. The $\mA$ factor in the Airy conjecture (\ref{Airy}) for this special case was found to be \cite{Hatsuda:2016uqa,Grassi:2014zfa}
	\begin{equation}
		\mA^\text{ADHM}(\sqrt{3},1,\btDelta_\text{sc})=\fft34A(2)-\fft14A(6)=-\fft{\zeta(3)}{3\pi^2}+\fft16\log 3\,.\label{ADHM:mA:sqrt3}
	\end{equation}
	As in the ABJM case, this result was recently extended in \cite{Kubo:2024qhq} to more general cases with $N_f=1$, $b^2\in2\mathbb{N}-1$, and the $R$ charge configuration 
	\begin{equation}
		\btDelta_\text{Fermi}^{(N_f=1)}\equiv\bigg(\fft12+\ri\fft{2m}{Q},\fft12-\ri\fft{2m}{Q},\fft12-\ri\fft{2\zeta}{Q},\fft12+\ri\fft{2\zeta}{Q}\bigg)\Bigg|_{m=\fft{(b^2-3)\ri}{4b}}\,,\label{ADHM:btDelta:Fermi:Nf=1}
	\end{equation}
	which corresponds to the ABJM configuration (\ref{ABJM:bDelta:Fermi}) under the mirror duality map~(\ref{Map:Nosaka}). The corresponding $\mA$ factor in the Airy conjecture is identical to (\ref{ABJM:mA:k=1}) as 
	\begin{align}
		&\mA^\text{ADHM}(b,1,\btDelta_\text{Fermi}^{(N_f=1)})=\mA^\text{ABJM}(b,1,\bDelta_\text{Fermi}^{(N_f=1)})\nn\\
		&=\fft14\bigg[A\Big(\fft{b^2+1-4\ri b\zeta}{2}\Big)+A\Big(\fft{b^2+1+4\ri b\zeta}{2}\Big)+A(b^2-1)-A(2b^2)\bigg]\,.\label{ADHM:mA:Nf=1}
	\end{align}

	\item On the round $S^3$ with $b=1$, it was shown in \cite{Geukens:2024zmt} that the Airy conjecture is consistent with the numerical evaluation of the matrix model (\ref{ADHM:Z}) via the saddle point approximation in the 't~Hooft limit, i.e. $N\gg1$ with fixed $\lambda=N/N_f$, for generic $\btDelta$ configurations. More explicitly, the Airy conjecture in the 't~Hooft limit reads
	\begin{subequations}
		\begin{align}
			-\log Z^\text{ADHM}(N,1,N_f,\btDelta)&=\fft{4\pi\sqrt{2\tDelta_1\tDelta_2\tDelta_3\tDelta_4}}{3}\fft{\Big(\lambda-\fft{1-2(\tDelta_1+\tDelta_2)+\tDelta_1\tDelta_2}{24\tDelta_1\tDelta_2}\Big)^\fft32}{\lambda^2}N^2\nn\\
			&\quad+\fft{\mfc^\text{ADHM}(1,\btDelta)}{\lambda^2}N^2+\mO(e^{-\#\sqrt{\lambda}})+o(N^2)\,,\label{F:ADHM:num}\\
			\mathfrak{c}^\text{ADHM}(1,\btDelta_\text{Nosaka-like})&=\fft{\mA(2\tDelta_1)+\mA(2\tDelta_2)}{4}-\fft{\zeta(3)}{8\pi^2}\big(\tDelta_3^2+\tDelta_4^2\big)\,,\label{ADHM:mfc:special}
		\end{align}
	\end{subequations}
	which is identical to the result from the saddle point approximation in \cite{Geukens:2024zmt}. Note that the functional structure (\ref{F:ADHM:num}) was confirmed for generic $R$ charges, which determines the large $N_f$ behavior of the $\mA$ factor in the Airy conjecture to be
	\begin{equation}
		\mA^\text{ADHM}(1,N_f,\btDelta)=-\mfc^\text{ADHM}(1,\btDelta)N_f^2+o(N_f^2)\,.
	\end{equation}
	The analytic expression of the $\mfc^\text{ADHM}(1,\btDelta)$ coefficient, however, has been deduced only for the Nosaka-like configuration as presented in (\ref{ADHM:mfc:special}). 
	
	\item The Airy conjecture at the superconformal configuration $\btDelta=\btDelta_\text{sc}$ can be expanded in the M-theory limit, i.e. $N\gg1$ with fixed $N_f$, and leads to the following expression for the $S^3$ free energy
	\begin{align}
		-\log Z^\text{ADHM}(N,b,N_f,\btDelta_\text{sc})&=\fft{\pi\sqrt{2N_f}}{3}\bigg[\fft{Q^2}{4}\bigg(N^{3/2}+\fft{8+7N_f^2}{16N_f}N^{1/2}\bigg)-\fft{N_f^2+5}{4N_f}N^{1/2}\bigg]\nn\\
		&\quad+\mO(\log N)\,.
	\end{align}
	Similarly to the ABJM case, this result is consistent with the expression obtained using holographic methods and higher-derivative supergravity in \cite{Bobev:2020egg,Bobev:2021oku}. 
\end{itemize}
Based on the results summarized above, the Airy formula for the round $S^3$ partition function of the ADHM theory with generic $\btDelta$ configurations was proposed in \cite{Geukens:2024zmt}. The Airy conjecture in (\ref{Airy}) above, in conjunction with (\ref{mBmC:abc}), extends the proposal of \cite{Geukens:2024zmt} to the squashed $S^3$. Finally, we note that the ADHM theory with $N_f=1$ is dual under 3d mirror symmetry to the ABJM theory with $k=1$. This duality is also manifest in the Airy conjecture since the first two lines of Table~\ref{table:abc} are the same for $N_f=k=1$ after taking into account the parameter identification in Footnote~\ref{footnote6}. 

\subsubsection{$N^{0,1,0}$ theory}\label{sec:AiryTale:ex:N010}
The $N^{0,1,0}$ theory can be constructed by adding $r_{1,2}$ pairs of $\mN=2$ fundamental \& anti-fundamental chiral multiplets to each gauge group in the ABJM theory. The $S_b^3$ partition function of the $N^{0,1,0}$ theory can then be obtained from the general matrix model (\ref{Z}) and reads
\begin{align}
	Z^{N^{0,1,0}}(N,b,k,r_1,r_2,\bDelta)&=\fft{1}{(N!)^2}\int(\prod_{i=1}^N\fft{d\mu_i}{2\pi}\fft{d\nu_i}{2\pi})\,e^{\fft{\ri k}{4\pi}\sum_{i=1}^N(\mu_i^2-\nu_i^2)}\label{N010:Z}\\
	&\quad\times\prod_{i>j}2\sinh\fft{b\mu_{ij}}{2}\,2\sinh\fft{\mu_{ij}}{2b}\,2\sinh\fft{b\nu_{ij}}{2}\,2\sinh\fft{\nu_{ij}}{2b}\nn\\
	&\quad\times\prod_{i,j=1}^N\prod_{a=1}^2s_b\bigg(\fft{\ri Q}{2}(1-\Delta_a)-\fft{\mu_i-\nu_j}{2\pi}\bigg)\prod_{a=3}^4s_b\bigg(\fft{\ri Q}{2}(1-\Delta_a)+\fft{\mu_i-\nu_j}{2\pi}\bigg)\nn\\
	&\quad\times\prod_{q=1}^{r_1}\prod_{i=1}^Ns_b\bigg(\fft{\ri Q}{2}(1-\Delta_{\mu_q})-\fft{\mu_i}{2\pi}\bigg)s_b\bigg(\fft{\ri Q}{2}(1-\tDelta_{\mu_q})+\fft{\mu_i}{2\pi}\bigg)\nn\\
	&\quad\times\prod_{q=1}^{r_2}\prod_{i=1}^Ns_b\bigg(\fft{\ri Q}{2}(1-\Delta_{\nu_{q}})-\fft{\nu_i}{2\pi}\bigg)s_b\bigg(\fft{\ri Q}{2}(1-\tDelta_{\nu_{q}})+\fft{\nu_i}{2\pi}\bigg)\,.\nn
\end{align}
Since the matrix model (\ref{N010:Z}) is similar to the one for the ABJM theory examined extensively for generic $R$ charge configurations, here we restrict ourselves to the superconformal configuration
\begin{equation}
	\Delta_a=\Delta_{\mu_q}=\tDelta_{\mu_q}=\Delta_{\nu_q}=\tDelta_{\nu_q}=\fft12\,,\label{N010:sc}
\end{equation}
for the $N^{0,1,0}$ theory. Moreover, we focus on the symmetric case with
\begin{equation}
	r_1=r_2=\fft{N_f}{2}\,.\label{N010:sym}
\end{equation}
The collective parameter for these choices reduces to $\bxi=\{k,r_1,r_2,\bDelta\}\to\{k,N_f\}$ and will be the one we will use for the $N^{0,1,0}$ theory in the rest of the paper. A complete formulation of the Airy conjecture for the $N^{0,1,0}$ theory with generic real mass deformations and an asymmetric configuration $r_1\neq r_2$ is left for future work. In the following subsections, we likewise restrict our analysis of the $V^{5,2}$/$Q^{1,1,1}$ theories to the superconformal configuration.

\medskip

The Airy conjecture for the $N^{0,1,0}$ theory for the choice of parameters in (\ref{N010:sc}) and (\ref{N010:sym}), given by (\ref{Airy}) with (\ref{mBmC:abc}) and the third line of Table~\ref{table:abc}, is supported by the following evidence. 
\begin{itemize}
	\item On the round sphere with $b=1$, the Airy conjecture was proven analytically based on the free Fermi-gas formalism \cite{Marino:2011eh} recently reviewed in \cite{Bobev:2023lkx}.
	
	\item For a generic squashing parameter $b$, the M-theory expansion ($N\gg1$ with fixed $N_f$) of the Airy conjecture leads to the following free energy
	\begin{align}
		-\log Z^{N^{0,1,0}}(N,b,k,N_f)&=\fft{2\pi(k+N_f)}{3\sqrt{2k+N_f}}\Bigg[\fft{Q^2}{4}\bigg(N^{3/2}+\bigg(\fft{7N_f-2k}{32}+\fft{1}{k+N_f}\bigg)N^{1/2}\bigg)\nn\\
		&\kern7em-\bigg(\fft{N_f}{8}+\fft{3k+2N_f}{2(k+N_f)^2}\bigg)N^{1/2}\Bigg]+\mO(\log N)\,.
	\end{align}
	This is consistent with the expression obtained in \cite{Bobev:2023lkx} using holography and the four-derivative supergravity analysis \cite{Bobev:2020egg,Bobev:2021oku}.
\end{itemize}
%

\subsubsection{$V^{5,2}$ theory}\label{sec:AiryTale:ex:V52}
For the $V^{5,2}$ theory, the $S_b^3$ partition function can be written as (\ref{ADHM:Z}) but with different constraints on the $R$ charges
\begin{equation}
	\Delta_1+\Delta_2=\fft43\,,\qquad \Delta_3=\fft23\,,\qquad\Delta_{\mu_q}+\tDelta_{\mu_q}=\fft23\,.\label{V52:constraints}
\end{equation}
This difference is to due to the different superpotentials for the ADHM theory and the $V^{5,2}$ theory. Since the complicated dependence of the sphere partition function on $R$ charges has already been studied thoroughly for the ADHM theory, we focus on the superconformal configuration for the $V^{5,2}$ theory, namely
\begin{equation}
	(\Delta_1,\Delta_2,\Delta_3)=(\fft23,\fft23,\fft23)\,,\qquad (\Delta_{\mu_q},\tDelta_{\mu_q})=(\fft13,\fft13)\,.\label{V52:sc}
\end{equation}
Put differently, the collective parameter simplifies to $\bxi=\{N_f,\btDelta\}\to\{N_f\}$ for the $V^{5,2}$ theory we study here. 

\medskip

The Airy conjecture for the $V^{5,2}$ theory with the superconformal configuration (\ref{V52:sc}), given in (\ref{Airy}) with the parameters from (\ref{mBmC:abc}) and the fourth line of Table~\ref{table:abc}, has been substantiated in the following two cases.
\begin{itemize}
	\item On the round $S^3$ with $b=1$, the Airy conjecture is consistent with the 't~Hooft limit ($N\to\infty$ with fixed $\lambda=N/N_f$) of the matrix model (\ref{ADHM:Z}) evaluated at the superconformal configuration of the $V^{5,2}$ theory (\ref{V52:sc}) through the saddle point approximation, see~\cite{Geukens:2024zmt}, namely
	\begin{align}
		-\log Z^{V^{5,2}}(1,N_f)=\fft{16\sqrt{2}\pi}{27}\fft{\big(\lambda-\fft{1}{48}\big)^\fft32}{\lambda^2}N^2+\fft{\mfc^{V^{5,2}}(1)}{\lambda^2}N^2+\mO(e^{-\#\sqrt{\lambda}})+o(N^2)\,.\label{F:V52:num}
	\end{align}
	Even though the analytic expression of the $\mfc^{V_{5,2}}(b)$ coefficient is not presently known, the second term in (\ref{F:V52:num}) indicates that the $\mA$ factor in the Airy conjecture for the round $S^3$ should be expanded in the large $N_f$ limit as
	\begin{equation}
		\mA^{V^{5,2}}(1,N_f)=-\mfc^{V^{5,2}}(1)N_f^2+o(N_f^2)\,.
	\end{equation}

	\item The Airy conjecture expanded in the M-theory limit ($N\gg1$ with fixed $N_f$) leads to the following sphere free energy
	\begin{align}
		-\log Z^{V^{5,2}}(N,b,N_f)&=\fft{16\pi\sqrt{2N_f}}{27}\Bigg[\fft{Q^2}{4}\bigg(N^{3/2}+\bigg(\fft{N_f}{4}+\fft{3}{8N_f}\bigg)N^{1/2}\bigg)\nn\\
		&\kern6em~-\fft{9(N_f^2+3)}{32N_f}N^{1/2}\Bigg]+\mO(\log N)\,.
	\end{align}
	This is in agreement with the expression obtained by applying the relation between the topologically twisted index and the superconformal index \cite{Bobev:2024mqw} to the holographic and four-derivative supergravity analysis in \cite{Bobev:2020egg,Bobev:2021oku}.
\end{itemize}
These two observations have prompted the authors of \cite{Geukens:2024zmt} to propose the Airy formula for the $S^3$ partition function of the $V^{5,2}$ theory. The Airy conjecture (\ref{Airy}) in conjunction with (\ref{mBmC:abc}) generalizes the proposal in \cite{Geukens:2024zmt} to an arbitrary squashing parameter $b$. We also note that in~\cite{Marino:2012az}  it was pointed out that for single-node 3d CS-matter quiver gauge theories, such as the $V^{5,2}$ theory, one can use the formalism of interacting Fermi gases to deduce that the large $N$ sphere partition function is determined by an Airy function and takes the form in \eqref{Airy}. The results in~\cite{Marino:2012az}   however do not lead to a direct calculation of the quantities $\mathcal{A}$ and $\mathcal{B}$ for the $V^{5,2}$ theory.

\subsubsection{$Q^{1,1,1}$ theory}\label{sec:AiryTale:ex:Q111}
For the $Q^{1,1,1}$ theory reviewed in \cite{Geukens:2024zmt}, the $S^3_b$ partition function can be obtained from the general localization formula (\ref{Z}) and reads 
\begin{equation}
	\begin{split}
		Z^{Q^{1,1,1}}(N,b,N_f,\bDelta)&=\fft{1}{(N!)^2}\int(\prod_{i=1}^N\fft{d\mu_i}{2\pi}\fft{d\nu_i}{2\pi})\,\prod_{i>j}2\sinh\fft{b\mu_{ij}}{2}\,2\sinh\fft{\mu_{ij}}{2b}\,2\sinh\fft{b\nu_{ij}}{2}\,2\sinh\fft{\nu_{ij}}{2b}\\
		&\quad\times\prod_{i,j=1}^N\prod_{a=1}^2s_b\bigg(\fft{\ri Q}{2}(1-\Delta_a)-\fft{\mu_i-\nu_j}{2\pi}\bigg)\prod_{a=3}^4s_b\bigg(\fft{\ri Q}{2}(1-\Delta_a)+\fft{\mu_i-\nu_j}{2\pi}\bigg)\\
		&\quad\times\prod_{q=1}^{N_f}\prod_{i=1}^Ns_b\bigg(\fft{\ri Q}{2}(1-\tDelta_{\mu_q})+\fft{\mu_i}{2\pi}\bigg)s_b\bigg(\fft{\ri Q}{2}(1-\Delta_{\nu_q})-\fft{\nu_i}{2\pi}\bigg)\\
		&\quad\times\prod_{q=N_f+1}^{2N_f}\prod_{i=1}^Ns_b\bigg(\fft{\ri Q}{2}(1-\tDelta_{\mu_q})+\fft{\mu_i}{2\pi}\bigg)s_b\bigg(\fft{\ri Q}{2}(1-\Delta_{\nu_q})-\fft{\nu_i}{2\pi}\bigg)\,.
	\end{split}\label{Q111:Z}
\end{equation}
Here we focus on the superconformal configuration 

\begin{equation}
	\Delta_a=\fft12\,,\quad \tDelta_{\mu_q}=\Delta_{\nu_q}=\fft34\,,\label{Q111:sc}
\end{equation}
as in the $N^{0,1,0}$/$V^{5,2}$ theories, and thereby the collective parameter $\bxi=\{N_f,\bDelta\}$ reduces to $\bxi=\{N_f\}$.

\medskip

The Airy conjecture for the $Q^{1,1,1}$ theory at the superconformal configuration (\ref{Q111:sc}), given by (\ref{Airy}) with (\ref{mBmC:abc}) and the fifth line of Table~\ref{table:abc}, has been supported in a manner similar to that for the $V^{5,2}$ theory as summarized below. 
\begin{itemize}
	\item On the round $S^3$ with $b=1$, the Airy conjecture reproduces the matrix model (\ref{Q111:Z}) evaluated at the superconformal configuration of the $Q^{1,1,1}$ theory (\ref{V52:sc}) via the saddle point approximation in the 't~Hooft limit ($N\to\infty$ with fixed $\lambda=N/N_f$), see~\cite{Geukens:2024zmt}. The result is given by
	\begin{align}
		-\log Z^{Q^{1,1,1}}(1,N_f)=\fft{4\pi}{3\sqrt{3}}\fft{\big(\lambda+\fft{1}{12}\big)^\fft32}{\lambda^2}N^2+\fft{\mfc^{Q^{1,1,1}}(1)}{\lambda^2}N^2+\mO(e^{-\#\sqrt{\lambda}})+o(N^2)\,.\label{F:Q111:num}
	\end{align}
	There is presently no known analytic expression for the coefficient $\mfc^{Q_{1,1,1}}(b)$. It is however clear that this coefficient governs the large $N_f$ limit of the $\mA$ coefficient in the Airy conjecture and on the round $S^3$ it reads
	\begin{equation}
		\mA^{Q^{1,1,1}}(1,N_f)=-\mfc^{Q^{1,1,1}}(1)N_f^2+o(N_f^2)\,.
	\end{equation}

	\item The M-theory expansion ($N\gg1$ with fixed $N_f$) of the Airy conjecture results in the following two leading terms for the sphere free energy
	\begin{align}
		-\log Z^{Q^{1,1,1}}(N,b,N_f)&=\fft{4\pi\sqrt{N_f}}{3\sqrt{3}}\Bigg[\fft{Q^2}{4}\bigg(N^{3/2}+\fft{N_f}{4}N^{1/2}\bigg)-\fft{N_f^2+3}{8N_f}N^{1/2}\Bigg]\nn\\
		&\quad+\mO(\log N)\,.
	\end{align}
	This results agrees with the expression derived from the relationship between the topologically twisted index and the superconformal index in \cite{Bobev:2024mqw} together with the holographic and four-derivative supergravity analysis of \cite{Bobev:2020egg,Bobev:2021oku}.
\end{itemize}
The two observations above motivate the formulation of the complete Airy formula for the round $S^3$ partition function of the $Q^{1,1,1}$ theory \cite{Geukens:2024zmt}. The Airy conjecture (\ref{Airy}) with (\ref{mBmC:abc}) extends this previous conjecture to  an arbitrary squashing parameter $b$.

\section{New evidence I: direct numerical integration}\label{sec:evi1}

In this section, we discuss new evidence in favor of the Airy conjecture for the specific values of the squashing parameter $b^2\in2\mathbb{N}-1$, by directly implementing a numerical integration of the matrix model for the $S^3_b$ partition function. We focus on the ADHM theory with certain configurations of $R$ charges, where the $S^3_b$ partition function can be reformulated in terms of a free Fermi-gas model. This allows the numerical integration to be performed efficiently using the Bornemann method outlined in Appendix~\ref{app:num:Bornemann}. The same model was recently explored analytically in \cite{Kubo:2024qhq} focusing on the case with a single pair of $\mN=2$ fundamental \& anti-fundamental chiral multiplets $N_f=1$. Below we extend their results to arbitrary values of $N_f$ using numerical methods.

\medskip

The starting point is to take the following configuration of $R$ charges
\begin{equation}
	\btDelta_\text{Fermi}\equiv\bigg(\fft12+\ri\fft{2m}{Q},\fft12-\ri\fft{2m}{Q},\fft12-\ri\fft{2\zeta}{QN_f},\fft12+\ri\fft{2\zeta}{QN_f}\bigg)\Bigg|_{m=\fft{(b^2-3)\ri}{4b}}\,,\label{ADHM:btDelta:Fermi}
\end{equation}
which generalizes the choice of \cite{Kubo:2024qhq} presented in (\ref{ADHM:btDelta:Fermi:Nf=1}) by restoring the $N_f$ dependence. For the $R$ charge configuration (\ref{ADHM:btDelta:Fermi}), the $S^3_b$ partition function of the ADHM theory (\ref{ADHM:Z}) with $b^2\in2\mathbb{N}-1$ can be rewritten in terms of a free Fermi-gas model as \cite{Kubo:2024qhq}
\begin{align}
	Z^\text{ADHM}(N,b,N_f,\btDelta_\text{Fermi})=\fft{1}{N!}\int \prod_{i=1}^Nd\mu_i\,\det_{i,j=1}^N\langle \mu_i|\hat{\rho}|\mu_j\rangle\,,\label{ADHM:Z:Fermi}
\end{align}
where the density operator reads 
\begin{equation}
	\hat{\rho}=e^{\fft{\ri b\zeta}{2}\hat{q}}s_b(\fft{b\hat{q}}{2\pi}+\fft{\ri Q}{4})^{N_f}\fft{e^{-\fft{\hat{p}}{2b^2}}}{2\cosh\fft{\hat{p}}{2}}\fft{1}{s_b(\fft{b\hat{q}}{2\pi}-\fft{\ri Q}{4})^{N_f}}e^{\fft{\ri b\zeta}{2}\hat{q}}\,.\label{rho}
\end{equation}
Note that this expression is not manifestly invariant under $b \to 1/b$ since we have already assumed that $b^2$ is an odd integer. This free Fermi-gas formulation allows for the analytic calculation of the $S^3_b$ partition function by implementing the following steps, see \cite{Marino:2011eh,Hatsuda:2015oaa}.
\begin{enumerate}
	\item Evaluate the spectral trace as
	\begin{align}
		\mZ(s)=\Tr[\hat{\rho}^s]=\int\fft{dpdq}{2\pi\hbar}(\rho^s)_W(p,q)=\sum_{n=0}^\infty\hbar^{2n}\underbrace{\int\fft{dpdq}{2\pi\hbar}(\rho^s)^{(n)}_W(p,q)}_{=\mZ^{(n)}(s)}\,,\label{spectral:trace}
	\end{align}
	where the Wigner transform introduced in Appendix \ref{app:Wigner} is employed, and the expression is expanded in the small $\hbar$ limit. 
	
	\item Determine the grand potential $\mJ^\text{ADHM}$ defined through
	\begin{align}
		e^{\mJ^\text{ADHM}(\mu;b,N_f,\btDelta_\text{Fermi})}=1+\sum_{N=1}^\infty Z^\text{ADHM}(N;b,N_f,\btDelta_\text{Fermi})e^{\mu N}
	\end{align}
	by substituting the spectral trace (\ref{spectral:trace}) into the following Mellin-Barnes type contour integral
	\begin{equation}
		\mJ^\text{ADHM}(\mu;b,N_f,\btDelta_\text{Fermi})=-\int_{c-\ri\infty}^{c+\ri\infty}\fft{ds}{2\pi\ri}\Gamma(s)\Gamma(-s)\mZ(s)e^{s\mu}\qquad(c\in(0,1))\,.
	\end{equation}

	\item Apply the Laplace transform to determine the canonical partition function $Z^\text{ADHM}$ from the grand potential $\mJ^\text{ADHM}$ as follows
	\begin{equation}
		Z^\text{ADHM}(N;b,N_f,\btDelta_\text{Fermi})=\int_{-\pi\ri}^{\pi\ri}\fft{d\mu}{2\pi\ri}\,\exp[\mJ^\text{ADHM}(\mu;b,N_f,\btDelta_\text{Fermi})-\mu N]\,.
	\end{equation}
\end{enumerate}

\medskip

The authors of \cite{Kubo:2024qhq} implemented this procedure for the special case of $N_f=1$, where the spectral trace (\ref{spectral:trace}) can be evaluated order by order in the small $\hbar$ expansion. As a result, the ADHM $S^3_b$ partition function (\ref{ADHM:Z:Fermi}) is shown to take the Airy form (\ref{Airy}) with the coefficients presented in the second row of Table~\ref{table:abc}. Moreover, an analytic expression for the $\mA$ factor in the Airy conjecture (\ref{Airy}) was derived explicitly, see (\ref{ADHM:mA:Nf=1}), which passed several non-trivial consistency checks.

\medskip

For generic values of $N_f$, however, the evaluation of the spectral trace (\ref{spectral:trace}) becomes much more involved. To illustrate this technical difficulty, let us focus on the semi-classical contribution to the spectral trace in the small $\hbar$ expansion (\ref{spectral:trace}) for the density operator (\ref{rho}), which can be written explicitly as
\begin{align}
	\mZ^{(0)}(s)=\int\fft{dpdq}{2\pi\hbar}\big(e^{-\ri b\zeta q+(\fft{1}{2b^2}+\fft12)p}+\sum_{n=0}^{N_f}\binom{N_f}{n}e^{((2n-N_f)\fft{b^2}{2}-\ri b\zeta)q+(\fft{1}{2b^2}-\fft12)p}\big)^{-s}\,.\label{mZ:0}
\end{align}
As noted earlier, for $N_f=1$, this phase space integral can be evaluated explicitly, yielding a closed-form expression in terms of a multivariate Beta function \cite{Hatsuda:2015oaa,Kubo:2024qhq}. For generic $N_f$, however, the phase space integral (\ref{mZ:0}) becomes significantly more intricate and consequently determining the semi-classical contribution to the spectral trace becomes a non-trivial task. An exception is the special case with $b^2=3$ discussed in \cite{Hatsuda:2016uqa} which can be treated analytically. Calculating the phase space integral (\ref{mZ:0}) for a generic $N_f$, and subsequently investigating quantum corrections beyond the semi-classical contribution, will constitute an analytic proof of the Airy conjecture (\ref{Airy}) for the ADHM $S^3_b$ partition function for the $R$ charge configuration (\ref{ADHM:btDelta:Fermi}). In the absence of such an analytic calculation we will approach the problem numerically.

\medskip

The numerical method delineated in Appendix \ref{app:num:Bornemann} can be applied to the evaluation of the $S^3_b$ partition function (\ref{ADHM:Z:Fermi}) without any restriction on the values $b^2\in2\mathbb{N}-1$ and $N_f$. We have implemented this approach and  have numerically confirmed that the $S^3_b$ partition function (\ref{ADHM:Z:Fermi}) of the ADHM theory for the $R$ charge configuration (\ref{ADHM:btDelta:Fermi}) with $b^2\in2\mathbb{N}-1$ agrees with good numerical accuracy with the Airy formula (\ref{Airy}) with the coefficients in Table~\ref{table:abc}. Furthermore, we can numerically confirm that the $\mA$ factor for $N_f=1$ in (\ref{ADHM:mA:Nf=1}) generalizes to
\begin{align}
	\mA^\text{ADHM}(b,N_f,\btDelta_\text{Fermi})&=\fft14\bigg[A\Big(\fft{(b^2+1)N_f-4\ri b\zeta}{2}\Big)+A\Big(\fft{(b^2+1)N_f+4\ri b\zeta}{2}\Big)\nn\\
	&\qquad~+N_f^2\big(A(b^2-1)-A(2b^2)\big)\bigg]\,.\label{ADHM:mA:Nf}
\end{align}

Below we outline the numerical analysis that supports the Airy conjecture and leads to the proposal (\ref{ADHM:mA:Nf}). 
\begin{enumerate}
	\item Evaluate the $S^3_b$ free energy $F=-\log Z$ of the ADHM theory for values of $N$ in the range $N \in (1,20)$ and a given configuration of $(b,N_f,\btDelta_\text{Fermi})$ using the numerical algorithms introduced in Appendix \ref{app:num:Bornemann}. This gives the numerical function
	\begin{equation}
		F^\text{ADHM}_\text{num}(N,b,N_f,\btDelta_\text{Fermi})\,.\label{F:ADHM:num:Fermi}
	\end{equation}

	\item Subtract the Airy formula, except for the undetermined $\mA$ factor, from the numerical value of the free energy. This defines another numerical function 
	\begin{align}
		D^\text{ADHM}(N,b,N_f,\btDelta_\text{Fermi})&\equiv F^\text{ADHM}_\text{num}(N,b,N_f,\btDelta_\text{Fermi})\nn\\
		&\quad+\log \Big[\mC^{-1/3}\text{Ai}[\mC^{-1/3}(N-\mB)]\Big]\,.
	\end{align}
	Here the coefficients $(\mB,\mC)$ are determined explicitly by substituting the given configuration $(b,N_f,\btDelta_\text{Fermi})$ into Table~\ref{table:abc} along with (\ref{mBmC:abc}). 
	
	\item To support the Airy conjecture (\ref{Airy}) combined with the proposed $\mA$ factor (\ref{ADHM:mA:Nf}), one must show
	\begin{equation}
		D^\text{ADHM}(N,b,N_f,\btDelta_\text{Fermi})\overset{!}{=}-\mA^\text{ADHM}(b,N_f,\btDelta_\text{Fermi})+\mO(e^{-\#\sqrt{N}})\,.\label{D:mA}
	\end{equation}
	This was verified by observing that the following ratio approaches zero
	\begin{equation}
		R^\text{ADHM}(N,b,N_f,\btDelta_\text{Fermi})\equiv \fft{D^\text{ADHM}(N,b,N_f,\btDelta_\text{Fermi})}{-\mA^\text{ADHM}(b,N_f,\btDelta_\text{Fermi})}\,-\,1~\to~ 0\,,\label{Rto0}
	\end{equation}
	as $N$ increases for various configurations of $(b,N_f,\zeta)$. Figure~\ref{fig:ADHM:R} illustrates this behavior for two representative examples, with the numerical values of the ratio (\ref{Rto0}) for $N=20$ indeed approaching zero as
	\begin{align}
		R^\text{ADHM}(20,\sqrt{5},2,\btDelta_\text{Fermi})\Big|_{\zeta=\fft{\ri}{10}}&=-1.168\times 10^{-9}\,,\nn\\
		R^\text{ADHM}(20,\sqrt{7},3,\btDelta_\text{Fermi})\Big|_{\zeta=\fft15}&=-1.595\times 10^{-6}\,,\nn
	\end{align}
	in these cases. See Appendix \ref{app:num:data} for a more complete list of data that supports the behavior in (\ref{Rto0}). 
	\begin{figure}[t]
		\centering
		\includegraphics[width=0.45\textwidth]{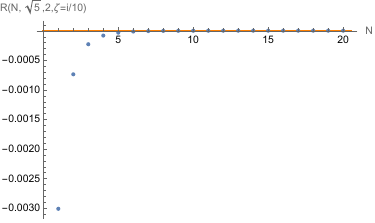}
		\quad
		\includegraphics[width=0.45\textwidth]{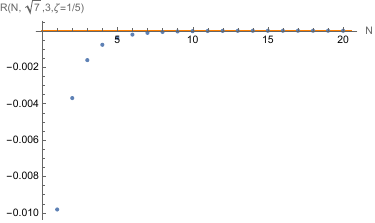}
		\caption{Blue dots represent numerical values of $R^\text{ADHM}(N,b,N_f,\btDelta_\text{Fermi})$ for $(b,N_f,\zeta)=(\sqrt{5},2,\fft{\ri}{10})$ and $(\sqrt{7},3,\fft15)$ respectively. They asymptotically approach zero (orange lines) as $N$ increases.}\label{fig:ADHM:R}%
	\end{figure}
\end{enumerate}

\medskip

To rigorously support the Airy conjecture (\ref{Airy}) along with the $\mA$ factor (\ref{ADHM:mA:Nf}), it is important to demonstrate that the ratio (\ref{Rto0}) is exponentially suppressed as $\sim e^{-\#\sqrt{N}}$, rather than exhibiting power-law suppression of the form $\sim N^{-\#}$, in the large $N$ limit. To accomplish this, we use Mathematica and perform a \texttt{LinearModelFit} on the logarithm of the difference between the left-hand side and right-hand side of (\ref{D:mA}) for the range $N=1\ldots 20$. This results in an expression of the following form
\begin{align}
	&\log\Big|D^\text{ADHM}(N,b,N_f,\btDelta_\text{Fermi})+\mA^\text{ADHM}(b,N_f,\btDelta_\text{Fermi})\Big|\nn\\
	&=\mfe_{1/2}^\text{(lmf)}N^{1/2}+\mfe_\text{log}^\text{(lmf)}\log N+\sum_{g=0}^{3}\mfe_g^\text{(lmf)}N^{-g/2}\,.\label{np:fitting}
\end{align}
If the coefficient $\mfe_{1/2}^\text{(lmf)}$ in the \texttt{LinearModelFit} (\ref{np:fitting}) is non-zero, this implies the exponential suppression of the ratio (\ref{Rto0}) and therefore confirms the validity of the Airy conjecture along with the $\mA$ factor (\ref{ADHM:mA:Nf}). Alternatively, if the Airy conjecture does not fully capture the all order $1/N$-perturbative expansion, and thus the difference between the numerical free energy (\ref{F:ADHM:num:Fermi}) and the corresponding expression from the Airy conjecture is power-law suppressed, the leading behavior in the \texttt{LinearModelFit} (\ref{np:fitting}) would be logarithmic. We find that in all examples we have studied, the leading order coefficients $\mfe_{1/2}^\text{(lmf)}$ take non-zero values, thereby confirming the Airy conjecture (\ref{Airy}) along with the $\mA$ factor (\ref{ADHM:mA:Nf}). Several examples of the leading order coefficients $\mfe_{1/2}^\text{(lmf)}$, together with the associated standard errors, are presented in Table~\ref{table:non-pert} for cases with $b^2\in\{3,7\}$, $N_f\in\{2,5\}$, and $\zeta\in\{0,\fft{\ri}{5}\}$.
\begin{table}[t]
	\centering
	\footnotesize
	\renewcommand{\arraystretch}{1.3}
	\begin{tabular}{|c|c|c||c|c|c|}
		\hline
		$(b^2,N_f,\zeta)$ & $\mfe_{1/2}^\text{(lmf)}$ & Standard error & $(b^2,N_f,\zeta)$ & $\mfe_{1/2}^\text{(lmf)}$ & Standard error \\
		\hline\hline
		$(3,2,0)$ & $-6.27478965$ &  $1.698\times10^{-4}$ & $(5,2,0)$ & $-3.61888083$ &  $2.913\times10^{-4}$ \\
		\hline
		$(3,5,0)$ &  $-3.95851786$ &  $3.338\times10^{-4}$ & $(5,5,0)$ & $-2.29316899$ &  $4.984\times10^{-4}$ \\
		\hline
		$(3,2,\fft{\ri}{5})$ &  $-5.28818832$ &  $8.471\times10^{-4}$ & $(5,2,\fft{\ri}{5})$ & $-3.18353969$ &  $2.519\times10^{-3}$ \\
		\hline
		$(3,5,\fft{\ri}{5})$ &  $-3.59847102$ &  $2.322\times10^{-3}$ & $(5,5,\fft{\ri}{5})$ & $-2.15336338$ &  $1.505\times10^{-3}$ \\
		\hline
	\end{tabular}
	\caption{Leading order coefficients $\mfe_{1/2}^\text{(lmf)}$ in the \texttt{LinearModelFit} (\ref{np:fitting}), along with their associated standard errors}\label{table:non-pert}
\end{table}

\medskip

The analytic expression for the leading order coefficients $\mfe_{1/2}^\text{(lmf)}$ in the \texttt{LinearModelFit} (\ref{np:fitting}) (if it is non-vanishing) is not needed to support the Airy conjecture alongside the $\mA$ factor (\ref{ADHM:mA:Nf}). Nevertheless, this coefficient captures interesting physics from the perspective of the holographically dual M-theory. More specifically, this leading exponential correction should be captured by  probe M2-branes wrapping specific 3-cycles in the internal manifold in the M-theory limit, or worldsheet instantons in the IIA string theory regime, see \cite{Hatsuda:2014vsa} as well as \cite{Drukker:2010nc,Nosaka:2015iiw,Cagnazzo:2009zh,Hatsuda:2012dt,Gautason:2023igo,Beccaria:2023ujc,Nosaka:2024gle} for a discussion on this in the ABJM theory case. It would therefore be very interesting to derive an analytic expression for the leading order coefficients $\mfe_{1/2}^\text{(lmf)}$ presented in Table~\ref{table:non-pert} either by field theory methods or from a dual string/M-theory perspective. We leave this analysis for future work.

\section{New evidence II: the 't~Hooft limit}\label{sec:evi2}

In this section, we provide new evidence for the Airy conjecture in the 't~Hooft limit following the strategy of \cite{Geukens:2024zmt} but importantly keeping a generic value of the squashing parameter $b$. Specifically, we determine the $S_b^3$ planar free energy $F_0$ defined through the genus expansion\footnote{Usually the genus expansion does not include the logarithmic contribution, which we include in (\ref{genus-exp}). The coefficient $\#$ depends in general on the theory studied.}
\begin{equation}
	F(N,b,\bxi)=-\log Z(N,b,\bxi)=\sum_{\mathsf{g}=0}^\infty N^{2-2\mathsf{g}}F_{\mathsf{g}}(b,\bxi)+\#\log N\,,\label{genus-exp}
\end{equation}
by employing the saddle point approximation, and show that the result is consistent with the Airy conjecture (\ref{Airy}) expanded via the asymptotic formula (\ref{Airy:asymp}). For this analysis we focus on the ABJM and ADHM theories.

\subsection{ABJM theory}\label{sec:evi2:ABJM}

Before we discuss some details of our analysis we present the result for the $S_b^3$ planar free energy of the ABJM theory which reads
\begin{subequations}
\begin{align}
	F_0^\text{ABJM}(b,\lambda,\bDelta)&=\fft{4\pi\sqrt{2\Delta_1\Delta_2\Delta_3\Delta_4}}{3}\fft{Q^2}{4}\fft{(\lambda-\fft{1}{24})^{3/2}}{\lambda^2}+\fft{\mfc^\text{ABJM}(b,\bDelta)}{\lambda^2}+\mO(e^{-\#\sqrt{\lambda}})\,,\label{ABJM:F0}\\
	\mfc^\text{ABJM}(b,\bDelta)&=\fft{\zeta(3)}{8\pi^2}\Bigg[\fft{Q^2}{4}\bigg(4-\sum_{a=1}^4\Delta_a^2\bigg)\nn\\
	&\quad+\bigg[\fft{4}{\Delta_{13}\Delta_{14}\Delta_{23}\Delta_{24}}-Q^2\bigg(\fft{1}{\Delta_{13}\Delta_{24}}+\fft{1}{\Delta_{14}\Delta_{23}}\bigg)\bigg]\sum_{a=1}^4\fft{\prod_{b=1}^4\Delta_b}{\Delta_a}\Bigg]\label{ABJM:mfc}
\end{align}\label{ABJM:F0:mfc}%
\end{subequations}
This expression is obtained by using a saddle point approximation at large $N$ and large  $\lambda=N/k$. The result (\ref{ABJM:F0:mfc}) provides non-trivial evidence for the Airy conjecture (\ref{Airy}) for the squashed $S^3$ partition function of ABJM theory with an arbitrary squashing parameter $b$ in the 't~Hooft limit. Moreover, the large $k$ expansion of the $\mA$ factor in the Airy conjecture is determined by the expression in (\ref{ABJM:F0:mfc}) as
\begin{equation}
	\mA^\text{ABJM}(b,k,\bDelta)=-\mfc^\text{ABJM}(b,\bDelta)k^2+o(k^2)\,.\label{ABJM:mA:general+b}
\end{equation}
This generalizes the result in (\ref{ABJM:mA:general}) on the round $S^3$. We now proceed to explain how the $S_b^3$ planar free energy (\ref{ABJM:F0:mfc}) is derived via the saddle point approximation. 

\subsubsection{Direct calculation for Nosaka configuration $\bDelta_\text{Nosaka}$}\label{sec:evi2:ABJM:Nosaka}
We first focus on the Nosaka configuration (\ref{ABJM:Nosaka}) for which the matrix model (\ref{ABJM:Z}) can be recast into  
\begin{equation}
	\begin{split}
		Z^\text{ABJM}_{S^3_b}(N,b,k,\bDelta_\text{Nosaka})&=\fft{1}{(N!)^2}\int(\prod_{i=1}^N\fft{d\mu_i}{2\pi}\fft{d\nu_i}{2\pi})\,e^{\fft{\ri k}{4\pi}\sum_{i=1}^N(\mu_i^2-\nu_i^2)}\\
		&\quad\times\prod_{i>j}2\sinh\fft{b\mu_{ij}}{2}\,2\sinh\fft{\mu_{ij}}{2b}\,2\sinh\fft{b\nu_{ij}}{2}\,2\sinh\fft{\nu_{ij}}{2b}\\
		&\quad\times\prod_{i,j=1}^N\mD_b(\fft{\mu_i-\nu_j}{2\pi}+\fft{m_1+m_2}{2})\mD_b(\fft{\mu_i-\nu_j}{2\pi}+\fft{m_1-m_2}{2})\\
		&=\fft{1}{(N!)^2}\int(\prod_{i=1}^Nd\mu_id\nu_i)\,\exp[-S_\text{eff}^\text{ABJM}[\mu,\nu;N,b,\lambda,\bDelta_\text{Nosaka}]]
	\end{split}\label{ABJM:Z:Nosaka}
\end{equation}
in terms of the effective action $S_\text{eff}^\text{ABJM}$. Here $\bDelta$ is reparametrized in terms of real masses as in (\ref{Delta:m}) and we have also utilized the $\mD_b$-function introduced in Appendix~\ref{app:double-sine}. Taking the Nosaka configuration enables us to implement the numerical analysis described below, using the closed-form expressions of the $\mD_b$-function presented in~(\ref{mD:closed}) for odd integer values of $b^2$.

\medskip

With this at hand we then implement the following procedure
\begin{enumerate}
	\item Construct numerical solutions $\{\mu^\star,\nu^\star\}$ to the saddle point equations associated with the matrix model (\ref{ABJM:Z:Nosaka}), namely
	\begin{subequations}
		\begin{align}
			0&=\fft{\ri k}{2\pi}\mu_i+\sum_{j=1\,(\neq i)}^N\bigg[\fft{b}{2}\coth\fft{b\mu_{ij}}{2}+\fft{1}{2b}\coth\fft{\mu_{ij}}{2b}\bigg]\nn\\
			&\quad+\fft{1}{2\pi}\sum_{j=1}^N\bigg[\fft{\mD_b'(\fft{\mu_i-\nu_j}{2\pi}+\fft{m_1+m_2}{2})}{\mD_b(\fft{\mu_i-\nu_j}{2\pi}+\fft{m_1+m_2}{2})}+\fft{\mD_b'(\fft{\mu_i-\nu_j}{2\pi}+\fft{m_1-m_2}{2})}{\mD_b(\fft{\mu_i-\nu_j}{2\pi}+\fft{m_1-m_2}{2})}\bigg]\,,\\
			0&=-\fft{\ri k}{2\pi}\nu_j+\sum_{i=1\,(\neq j)}^N\bigg[\fft{b}{2}\coth\fft{b\nu_{ji}}{2}+\fft{1}{2b}\coth\fft{\nu_{ji}}{2b}\bigg]\nn\\
			&\quad-\fft{1}{2\pi}\sum_{i=1}^N\bigg[\fft{\mD_b'(\fft{\mu_i-\nu_j}{2\pi}+\fft{m_1+m_2}{2})}{\mD_b(\fft{\mu_i-\nu_j}{2\pi}+\fft{m_1+m_2}{2})}+\fft{\mD_b'(\fft{\mu_i-\nu_j}{2\pi}+\fft{m_1-m_2}{2})}{\mD_b(\fft{\mu_i-\nu_j}{2\pi}+\fft{m_1-m_2}{2})}\bigg]\,,
		\end{align}\label{ABJM:Nosaka:saddle-eq}%
	\end{subequations}
	for various $(N,b,\lambda,\bDelta)$ configurations. In this step, we used \texttt{FindRoot} in \emph{Mathematica} at \texttt{WorkingPrecision} 200.
	
	\item Evaluate the on shell effective action on the solutions of the saddle point equations in order to find the classical action
	\begin{equation}
		S_\text{cl}^\text{ABJM}(N,b,\lambda,\bDelta_\text{Nosaka})\equiv S_\text{eff}^\text{ABJM}[\mu^\star,\nu^\star;N,b,\lambda,\bDelta_\text{Nosaka}]\,.
	\end{equation}
	This classical action captures the leading $N^2$ planar free energy $F_0^\text{ABJM}(b,\lambda,\bDelta_\text{Nosaka})$ in the genus expansion (\ref{genus-exp})
	\begin{equation}
		\begin{split}
			S_\text{cl}^\text{ABJM}(N,b,\lambda,\bDelta_\text{Nosaka})=F_0^\text{ABJM}(b,\lambda,\bDelta_\text{Nosaka})N^2+o(N^2)\,.
		\end{split}\label{ABJM:Scl:F0}
	\end{equation}

	\item Implement the \texttt{LinearModelFit} function in \emph{Mathematica} for the classical action $S_\text{cl}^\text{ABJM}$ by varying $N$ in the range $N=100\sim350$ (in step of 10) for given configurations of $(b,\lambda,\bDelta_\text{Nosaka})$ in order to find the large $N$ expansion
	\begin{align}
		&S_\text{cl}^\text{ABJM}(N,b,\lambda,\bDelta_\text{Nosaka})\nn\\
		&=S_{2}^\text{ABJM,(lmf)}(b,\lambda,\bDelta_\text{Nosaka})N^2+S_\text{1-log}^\text{ABJM,(lmf)}(b,\lambda,\bDelta_\text{Nosaka})N\log N\nn\\
		&\quad+S_{1}^\text{ABJM,(lmf)}(b,\lambda,\bDelta_\text{Nosaka})N+S_\text{log}^\text{ABJM,(lmf)}(b,\lambda,\bDelta_\text{Nosaka})\log N\nn\\
		&\quad+\sum_{L=0}^{20}S_{-L}^\text{ABJM,(lmf)}(b,\lambda,\bDelta_\text{Nosaka})N^{-L}\,.\label{ABJM:lmf}
	\end{align}
	Note that we have included a $N\log N$ term above. This term is not expected to arise in the genus expansion (\ref{genus-exp}), and therefore must cancel out upon incorporating loop corrections to the matrix integral (\ref{ABJM:Z:Nosaka}) beyond the classical on-shell action contribution. For related discussions, see \cite{Liu:2019tuk,Hong:2021bsb,Geukens:2024zmt}. In the \texttt{LinearModelFit} (\ref{ABJM:lmf}), we find the following logarithmic coefficients
	\begin{equation}
		S_\text{1-log}^\text{ABJM,(lmf)}(b,\lambda,\bDelta_\text{Nosaka})\approx-2\,,\qquad S_\text{log}^\text{ABJM,(lmf)}(b,\lambda,\bDelta_\text{Nosaka})\approx-\fft16\,.
	\end{equation}
	As in many similar large $N$ calculations these coefficients are universal, i.e. they do not depend on any continuous parameters (see \cite{Bobev:2023dwx} for a discussion on this). Using this information we improve the numerical accuracy of the original \texttt{LinearModelFit} (\ref{ABJM:lmf}) by taking
	\begin{align}
		&S_\text{cl}^\text{ABJM}(N,b,\lambda,\bDelta_\text{Nosaka})+2N\log N+\fft16\log N\nn\\
		&=S_{2}^\text{ABJM,(lmf)}(b,\lambda,\bDelta_\text{Nosaka})N^2+S_{1}^\text{ABJM,(lmf)}(b,\lambda,\bDelta_\text{Nosaka})N\nn\\
		&\quad+\sum_{L=0}^{22}S_{-L}^\text{ABJM,(lmf)}(b,\lambda,\bDelta_\text{Nosaka})N^{-L}\,,\label{ABJM:lmf:improve}
	\end{align}
This analytic implementation of the coefficients of the $\log$ terms on the left hand side increases the precision in the fitting procedure for the numerical coefficients on the right hand side. 
	
	\item Deduce the analytic expression for the leading coefficient $S_{2}^\text{ABJM,(lmf)}(b,\lambda,\bDelta_\text{Nosaka})$ in the \texttt{LinearModelFit} (\ref{ABJM:lmf:improve}) by exploring their numerical values for various $(b,\lambda,\bDelta_\text{Nosaka})$ configurations. Consequently, we determine the ABJM planar free energy for the Nosaka configuration based on the relation (\ref{ABJM:Scl:F0}) as 
	\begin{align}
		F_0^\text{ABJM}(b,\lambda,\bDelta_\text{Nosaka})&=\fft{4\pi\sqrt{2\Delta_1\Delta_2\Delta_3\Delta_4}}{3}\fft{Q^2}{4}\fft{(\lambda-\fft{1}{24})^{3/2}}{\lambda^2}+\fft{\mfc^\text{ABJM}(b,\bDelta_\text{Nosaka})}{\lambda^2}\nn\\
		&\quad+\mO(e^{-\#\sqrt{\lambda}})\,.\label{ABJM:F0:Nosaka}
	\end{align}
	Based on the numerical data we were also able to find the following analytic expression for the coefficient $\mfc^\text{ABJM}(b,\bDelta_\text{Nosaka})$
	\begin{align}
		\mfc^\text{ABJM}(b,\bDelta_\text{Nosaka})&=\fft{\zeta(3)}{8\pi^2}\bigg[-\fft18(b+b^{-1})^2(b^2-4+b^{-2})-\fft{b^2+b^{-2}}{2}(m_1^2+m_2^2)\label{ABJM:mfc:Nosaka}\\
		&\quad+\fft{(b-b^{-1})^2}{4}\bigg(m_2^2+\fft{(b-b^{-1})^2}{4}\bigg)\fft{2\big(1+\fft{4}{(b+b^{-1})^2}(m_1^2+m_2^2)\big)}{1+\fft{4}{(b+b^{-1})^2}m_2^2}\bigg]\,.\nn
	\end{align}
	The result (\ref{ABJM:F0:Nosaka}) with (\ref{ABJM:mfc:Nosaka}) yields the $S^3_b$ ABJM planar free energy presented in (\ref{ABJM:F0:mfc}) for the Nosaka configuration (\ref{ABJM:Nosaka}).
\end{enumerate}
We refer the reader to Appendix~\ref{app:data:ABJM} for a summary of the numerical data and further details of the numerical analysis that led us to the analytic expressions (\ref{ABJM:F0:Nosaka}) and (\ref{ABJM:mfc:Nosaka}).

\subsubsection{Indirect analysis for general $\bDelta$ configurations}\label{sec:evi2:ABJM:general}
For general $\bDelta$ configurations that go beyond the Nosaka configuration (\ref{ABJM:Nosaka}), we were unable to perform the numerical saddle point analysis presented in Section~\ref{sec:evi2:ABJM:Nosaka} due to the complicated closed-form expression for the double sine function $s_b(x)$ introduced in~(\ref{sbclosedform}). For the Nosaka configuration, we were able to rewrite the matrix model integrand in terms of the $\mD_b(x)$ function, as shown in~(\ref{ABJM:Z}), which was crucial for enabling the numerical analysis. Nevertheless, as outlined below, we still have some indirect evidence for the result in \eqref{ABJM:F0:mfc} for general values of the parameters $\bDelta$.

\medskip

The first term in the expression (\ref{ABJM:F0}) in the strict large $\lambda$ limit of the $S^3_b$ ABJM planar free energy has been analytically evaluated in \cite{Hosseini:2019and} 
\begin{equation}
	F_0^\text{ABJM}(b,\lambda,\bDelta)=\fft{4\pi\sqrt{2\Delta_1\Delta_2\Delta_3\Delta_4}}{3}\fft{Q^2}{4}\lambda^{-1/2}+o(\lambda^{-1/2})\,,
\end{equation}
for generic $(b,\bDelta)$ configurations. On the other hand, the round-sphere case with $b=1$ has been analyzed in \cite{Geukens:2024zmt} for a finite 't~Hooft parameter $\lambda$, yielding
\begin{equation}
	F_0^\text{ABJM}(1,\lambda,\bDelta)=\fft{4\pi\sqrt{2\Delta_1\Delta_2\Delta_3\Delta_4}}{3}\fft{(\lambda-\fft{1}{24})^{3/2}}{\lambda^2}+\fft{\mfc^\text{ABJM}(1,\bDelta)}{\lambda^2}+\mO(e^{-\#\sqrt{\lambda}})\,,
\end{equation}
for generic $\bDelta$ configurations as already introduced in (\ref{ABJM:F0:b=1}). These two observations, together with our direct calculation (\ref{ABJM:F0:Nosaka}) for the Nosaka configuration, strongly suggest that the first term in the expression (\ref{ABJM:F0})
\begin{equation}
	\fft{4\pi\sqrt{2\Delta_1\Delta_2\Delta_3\Delta_4}}{3}\fft{Q^2}{4}\fft{(\lambda-\fft{1}{24})^{3/2}}{\lambda^2}\,,
\end{equation}
is valid for finite $\lambda$ and generic $(b,\bDelta)$ configurations, not just for $b=1$ or $\bDelta=\bDelta_\text{Nosaka}$. A complete proof of this claim is left for future work.

\medskip

We deduce the analytic form of $\mfc^\text{ABJM}(b,\bDelta)$ presented in (\ref{ABJM:mfc}) based on the following ingredients. 
\begin{itemize}
	\item $\mfc^\text{ABJM}(1,\bDelta)$ in (\ref{ABJM:mfc:b=1});
		
	\item $\mfc^\text{ABJM}(b,\bDelta_\text{Nosaka})$ in (\ref{ABJM:mfc:Nosaka});
		
	\item The symmetry under three independent permutations of the $\bDelta$ parameters described in (\ref{ABJM:sym});
		
	\item The constraint derived in \cite{Chester:2021gdw} based on the structure of the matrix model 
	\begin{align}
		\mfc^\text{ABJM}(b,\bDelta)\Big|_{m_3\to\ri\fft{b-b^{-1}}{2}}=\mfc^\text{ABJM}(1,\bDelta)\Big|_{(m_1,m_2,m_3)\to(\fft{b^{-1}m_+-bm_-}{2},\fft{b^{-1}m_++bm_-}{2},0)}\,,\label{Shai:constraint}
	\end{align}
	where $m_\pm\equiv m_2\pm m_1$ and we used the reparametrization (\ref{Delta:m}).
\end{itemize}
In Section~\ref{sec:evi3}, we present additional, highly non-trivial, evidence for the analytic expression~(\ref{ABJM:mfc}). We leave a direct proof of (\ref{ABJM:mfc}) for future work.

\subsection{ADHM theory}\label{sec:evi2:ADHM}
For the ADHM theory, we focus on the superconformal configuration $\btDelta_\text{sc}=(\fft12,\fft12,\fft12,\fft12)$. In this case the matrix model for the $S^3_b$ partition function (\ref{ADHM:Z}) reduces to
\begin{align}
	Z^\text{ADHM}_{S^3_b}(N,b,N_f)&=\fft{1}{N!}\int\prod_{i=1}^N\fft{d\mu_i}{2\pi}\,\prod_{i>j}^N2\sinh\fft{b\mu_{ij}}{2}2\sinh\fft{\mu_{ij}}{2b}\prod_{i,j=1}^N\mD_b(\fft{\mu_{ij}}{2\pi})\times\prod_{i=1}^N\mD_b(\fft{\mu_i}{2\pi})^{N_f}\nn\\
	&=\fft{1}{N!}\int(\prod_{i=1}^Nd\mu_i)\,\exp[-S_\text{eff}[\mu;N,b,\lambda]]\,.\label{ADHM:Z:sc}
\end{align}
Applying the saddle point approximation at large $N$ and large $\lambda$ we find the following $S^3_b$ ADHM planar free energy
\begin{align}
	F_0^\text{ADHM}(b,\lambda)=\fft{\pi\sqrt{2}}{3}\fft{Q^2}{4}\fft{(\lambda-\fft{1}{24}+\fft{b^2+b^{-2}}{3Q^2})^{3/2}}{\lambda^2}+\fft{\mfc^\text{ADHM}(b)}{\lambda^2}+\mO(e^{-\#\sqrt{\lambda}})\,,\label{ADHM:F0}
\end{align}
in terms of the 't~Hooft parameter $\lambda=N/N_f$. Note that we omit $\btDelta=\btDelta_\text{sc}$ in the argument of $F_0$ for notational convenience. The expression (\ref{ADHM:F0}) supports the Airy conjecture (\ref{Airy}) for the $S^3_b$ partition function of the ADHM theory at the superconformal configuration in the 't~Hooft limit. Furthermore, the large $N_f$ expansion of the $\mA$ factor in the Airy conjecture is determined by (\ref{ADHM:F0})
\begin{equation}
	\mA^\text{ADHM}(b,N_f)=-\mfc^\text{ADHM}(b)N_f^2+o(N_f^2)\,,
\end{equation}
where the analytic expression for $\mfc^\text{ADHM}(b)$ is known only for $b\in\{1,\sqrt{3}\}$ and is given by
\begin{equation}
	\mfc^\text{ADHM}(b)=\begin{cases}
		\fft{\zeta(3)}{16\pi^2}-\fft12 A(1)=\fft{\zeta(3)}{8\pi^2}-\fft18\log2 & (b=1) \\
		\fft{\zeta(3)}{4\pi^2}-\fft14( A(2)- A(6))=\fft{\zeta(3)}{3\pi^2}-\fft16\log 3 & (b=\sqrt{3})
	\end{cases}\,.\label{ADHM:mfc}
\end{equation}
The above coefficients for $b=1$ and $b=\sqrt{3}$ are obtained by applying the large $k$ expansion of the $A$-function (\ref{def:A}) and its closed-form expression at integer arguments presented in \cite{Hatsuda:2014vsa} to the corresponding $\mA$ factors (\ref{ADHM:mA:special}) and (\ref{ADHM:mA:Nf}) respectively. Below we explain how we deduce the analytic expression (\ref{ADHM:F0}) for the $S^3_b$ planar free energy of ADHM theory.

\medskip

\begin{enumerate}
	\item We first construct numerical solutions $\{\mu^\star\}$ to the saddle point equations for the matrix model (\ref{ADHM:Z:sc}), namely
	\begin{equation}
		0=\sum_{j=1\,(\neq i)}^N\bigg[\fft{b}{2}\coth\fft{b\lambda_{ij}}{2}+\fft{1}{2b}\coth\fft{\lambda_{ij}}{2b}\bigg]+\fft{1}{\pi}\sum_{j=1}^N\fft{\mD_b'(\fft{\lambda_{ij}}{2\pi})}{\mD_b(\fft{\lambda_{ij}}{2\pi})}+\fft{N_f}{2\pi}\fft{\mD_b'(\fft{\lambda_i}{2\pi})}{\mD_b(\fft{\lambda_i}{2\pi})}\,,\label{ADHM:sc:saddle-eq}
	\end{equation}
	for various $(N,b,\lambda)$ configurations as we did for the ABJM theory.
	
	\item We then evaluate the on shell effective action capturing the planar free energy $F_0^\text{ADHM}(b,\lambda)$ in the genus expansion (\ref{genus-exp}) as
	\begin{align}
		S_\text{cl}^\text{ADHM}(N,b,\lambda)&\equiv S_\text{eff}^\text{ADHM}[\mu^\star;N,b,\lambda]=F_0^\text{ADHM}(b,\lambda)N^2+o(N^2)\,.\label{ADHM:Scl:F0}
	\end{align}

	\item We then employ the \texttt{LinearModelFit} function in \textit{Mathematica} for the classical action $S_\text{cl}^\text{ADHM}$ in the large $N$ approximation by taking $N$ in the range $N=100\sim350$ (in step of 10) for any given $(b,\lambda)$ configurations
	\begin{align}
		&S_\text{cl}^\text{ADHM}(N,b,\lambda)+N\log N+\fft{1}{12}\log N\nn\\
		&=S_{2}^\text{ADHM,(lmf)}(b,\lambda)N^2+S_{1}^\text{ADHM,(lmf)}(b,\lambda)N+\sum_{L=0}^{22}S_{-L}^\text{ADHM,(lmf)}(b,\lambda)N^{-L}\,,\label{ADHM:lmf:improve}
	\end{align}
	where we have already employed the universal logarithmic coefficients
	\begin{equation}
		S_\text{1-log}^\text{ADHM,(lmf)}(b,\lambda)\approx-1\,,\qquad S_\text{log}^\text{ADHM,(lmf)}(b,\lambda)\approx-\fft{1}{12}\,,
	\end{equation}
	to improve the fitting as in the ABJM case (\ref{ABJM:lmf:improve}).
	
	\item Finally, we deduce the analytic expression for the leading coefficient $S_{2}^\text{ADHM,(lmf)}(b,\lambda)$ in the \texttt{LinearModelFit} (\ref{ADHM:lmf:improve}) by exploring their numerical values for various $(b,\lambda)$ configurations. The resulting expression determines the ADHM planar free energy (\ref{ADHM:F0}) immediately through (\ref{ADHM:Scl:F0}).
\end{enumerate}
We refer the reader to Appendix~\ref{app:data:ADHM} for more details on the data and subsequent numerical analysis that lead to the expression (\ref{ADHM:F0}).

\section{New evidence III: the Cardy-like expansion}\label{sec:evi3}
In this section, we present additional evidence for the Airy conjecture in the limit of small squashing parameter to all orders in the $1/N$-perturbative expansion. The analysis is valid for all five holographic SCFTs listed in Table~\ref{table:abc} and is based on the following two key building blocks. 
\begin{itemize}
	\item Holomorphic factorization of 3d partition functions that relates the $S^3_b$ partition function of interest here and the $S^1\times_\omega S^2$ superconformal index (SCI) in the small $b$ limit as \cite{Choi:2019dfu}\footnote{The relation between the $S^3_b$ partition function and the SCI presented in \cite{Choi:2019dfu} applies to the chemical potentials associated with flavor symmetries  --- related to the $R$ charges of chiral multiplets for the sphere partition function --- in the `region I' defined in \cite{Choi:2019zpz}. This becomes manifest when comparing (4.1,4.5) of \cite{Choi:2019zpz} with (4.3) of \cite{Choi:2019dfu}. In this work, we instead adopt the convention for `region II', where the index relation is as in (\ref{SCI:S3b}) with the parameter identification $\beta^\text{\cite{Choi:2019dfu}}=-\ri\pi b^2=-\ri\pi\omega$.}
	\begin{align}
		F_{S^1\times_\omega S^2}(N,\omega,\bxi)\Big|_{\omega=b^2}=2F_{S^3_b}(N,b,\bxi)+\mO(b^2)\,,\label{SCI:S3b}
	\end{align}
	where the free energy is defined by $F=-\log Z$. 
	
	\item Closed-form expressions for the first two leading terms of the $S^1\times_\omega S^2$ SCI in the Cardy-like expansion ($\omega\to\ri0^+$), as derived in \cite{Bobev:2022wem,Bobev:2024mqw}\footnote{For comparison with the $S^3_b$ partition function through the factorization formula (\ref{SCI:S3b}), we identify the `flux' parameters $\mn_a$ with the chemical potentials $\Delta_a$ in the results of \cite{Bobev:2022wem,Bobev:2024mqw}. Note that $\mn_a$ do not represent actual flavor magnetic fluxes on $S^2$, but are merely auxiliary parameters introduced in a reparametrization step, see \cite{Bobev:2024mqw} for further details.}
	\begin{align}
		F_{S^1\times_\omega S^2}(N,\omega,\bxi)&=\fft{2}{\omega}\bigg[\fft{\pi\alpha(\bxi)}{4}(N-\beta(\bxi))^\fft32+\hat{g}_0(\bxi)+\mO(e^{-\#\sqrt{N}})\bigg]\nn\\
		&\quad+\bigg[\pi\alpha(\bxi)\Big((N-\beta(\bxi))^\fft32+\gamma(\bxi)(N-\beta(\bxi))^\fft12\Big)+\fft12\log (N-\beta(\bxi))\nn\\
		&\kern3em-\hat{f}_0(\bxi)+\mO(e^{-\#\sqrt{N}})\bigg]+\mO(\omega)\,,\label{SCI}
	\end{align}
	where the $(\alpha,\beta,\gamma)$ coefficients are given explicitly for the holographic SCFTs of interest here in Table~\ref{table:abc}. The above expression for the SCI is valid to all orders in the $1/N$-perturbative expansion up to exponentially suppressed terms. From here on, we often omit the $\bxi$ parameters in the arguments of $(\alpha,\beta,\gamma)$ coefficients for notational convenience.
\end{itemize}
New evidence for the Airy conjecture (\ref{Airy}) can be obtained by employing the above two pieces of information as follows. First, combining the factorization formula (\ref{SCI:S3b}) and the closed-form expression for the SCI (\ref{SCI}), we derive the following small $b$ expansion of the $S^3_b$ free energy
\begin{align}
	F_{S^3_b}(N,b,\bxi)\Big|_\text{(\ref{SCI:S3b})+(\ref{SCI})}&=\fft{1}{b^2}\bigg[\fft{\pi\alpha}{4}(N-\beta)^\fft32+\hg_0(\bxi)\bigg]+\bigg[\fft{\pi\alpha}{2}\Big(\alpha(N-\beta)^{\fft32}+\gamma(N-\beta)^{\fft12}\Big)\nn\\
	&\quad+\fft14\log(N-\beta)-\fft12\hf_0(\bxi)\bigg]+\mO(b^2,e^{-\#\sqrt{N}})\label{F:S3b:evi}
\end{align}
On the other hand, applying the asymptotic expansion of the Airy function (\ref{Airy:asymp}) to the Airy conjecture (\ref{Airy}) yields the following alternative expression for the $S^3_b$ free energy
\begin{align}
	F_{S^3_b}(N,b,\bxi)\Big|_\text{(\ref{Airy})}&=\fft{1}{b^2}\bigg[\fft{\pi\alpha}{4}(N-\beta)^\fft32\bigg]+\log b+\bigg[\fft{\pi\alpha}{2}\Big(\alpha(N-\beta)^{\fft32}+\gamma(N-\beta)^{\fft12}\Big)\nn\\
	&\quad+\fft14\log(N-\beta)+\fft12\log\fft{32}{3\alpha}\bigg]-\mA(b,\bxi)+\mO(b^2,e^{-\#\sqrt{N}})\,.\label{F:S3b:Airy}
\end{align}
Comparing the $S^3_b$ free energy (\ref{F:S3b:evi}) derived from the factorization formula and the all order SCI results, with the alternative expression (\ref{F:S3b:Airy}) based on the Airy conjecture, we obtain the following two pieces of evidence supporting the Airy conjecture (\ref{Airy}).
\begin{enumerate}
	\item The agreement of the $N$-dependent terms at order $b^{-2}$ and $b^0$ in the small $b$ expansion of \eqref{F:S3b:evi} and \eqref{F:S3b:Airy} provides a highly non-trivial consistency check for the Airy conjecture (\ref{Airy}) to all orders in the $1/N$-perturbative expansion.
	
	\item A comparison of the $N$-independent terms in \eqref{F:S3b:evi} and \eqref{F:S3b:Airy} leads to the following small $b$ expansion of $\mA(b,\bxi)$
	\begin{equation}
		\mA(b,\bxi)\overset{!}{=}-\fft{1}{b^2}\hat{g}_0(\bxi)+\log b+\fft12\bigg[\hat{f}_0(\bxi)+\log\fft{32}{3\alpha(\bxi)}\bigg]+\mO(b^2)\,,\label{mA:smallb}
	\end{equation}
	which was first conjectured in \cite{Bobev:2022wem}. To support the relation in (\ref{mA:smallb}) by a direct calculation we evaluate the left-hand side numerically in the small $b$ limit based on the Airy conjecture (\ref{Airy}) as follows
	\begin{align}
		\mA^\text{(num)}(N,b,\bxi)&=\log Z^\text{(num)}_{S^3_b}(N,b,\bxi)-\log\bigg[\mC^{-1/3}\text{Ai}\Big[\mC^{-1/3}(N-\mB)\Big]\bigg]\nn\\
		&=-\fft{1}{b^2}\hat{g}_0^\text{num}(\bxi)+\mO(\log b)\,,\label{mA:num:smallb}
	\end{align}
	where we numerically calculate the $b^{-2}$ leading order coefficient $\hat{g}_0^\text{num}(\bxi)$. In (\ref{mA:num:smallb}), the numerical values of $Z^\text{(num)}_{S^3_b}(N,b,\bxi)$ are computed via a saddle point approximation in the small $b$ limit outlined in Appendix \ref{app:smallb:saddle}, focusing on the ADHM and $V^{5,2}$ theories. Our results demonstrate that the numerically determined leading order coefficients $\hat{g}_0^\text{num}(\bxi)$, which is based on the Airy conjecture, match the known values of $\hat{g}_0(\bxi)$ obtained by analyzing the superconformal index in \cite{Bobev:2022wem,Bobev:2024mqw} without relying on the Airy conjecture. This confirms the relation (\ref{mA:smallb}) at leading order in the small $b$ expansion. Examples of this numerical comparison are presented in Table~\ref{table:evi3:ADHM:sc}. 
	\begin{table}[t]
		\centering
		\footnotesize
		\renewcommand*{\arraystretch}{1.4}
		\begin{tabular}{ |c||c||c| } 
			\hline $\btDelta=\btDelta_\text{sc}$ & $\hat g^\text{ADHM,(num)}_0(N_f,\btDelta)$
			& $\hat g^\text{ADHM}_0(N_f,\btDelta)$ \\
			\hline\hline
			$N_f=1$ & $-0.1838410233$ & $-0.183841023338$ \\
			\hline
			$N_f=2$ & $-0.33723359$  & $-0.337233589618$ \\
			\hline
			$N_f=3$ &$-0.62879449$   & $-0.628794493649$ \\ 
			\hline
			$N_f=4$ & $-1.043299$  & $-1.043299506902$ \\ 
			\hline
			$N_f=5$ & $-1.57819$  & $-1.578200494091$ \\ 
			\hline
			$N_f=6$ & $-2.2327$  & $-2.232772433361$ \\ 
			\hline
		\end{tabular}
		
		\smallskip
		
		\begin{tabular}{ |c||c||c| } 
			\hline $\btDelta=(\fft37,\fft47,\fft12,\fft12)$ & $\hat g^\text{ADHM,(num)}_0(N_f,\btDelta)$
			& $\hat g^\text{ADHM}_0(N_f,\btDelta)$ \\
			\hline\hline
			$N_f=1$ & $-0.184873819$ & $-0.184873819329$  \\
			\hline
			$N_f=2$ & $-0.3430694$ & $-0.343069446632$ \\
			\hline
			$N_f=3$ & $-0.6425169$ & $-0.642516986634$ \\ 
			\hline
		\end{tabular}
		
		\smallskip
		
		\begin{tabular}{ |c||c||c| } 
			\hline $\btDelta=(\fft38,\fft58,\fft25,\fft35)$ & $\hat g^\text{ADHM,(num)}_0(N_f,\btDelta)$
			& $\hat g^\text{ADHM}_0(N_f,\btDelta)$ \\
			\hline\hline
			$N_f=1$ & $-0.18922588$  & $-0.189225885117$  \\
			\hline
			$N_f=2$ & $-0.3541838$  & $-0.354183848923$ \\
			\hline
			$N_f=3$ & $-0.66685$ & $-0.666855367656$ \\ 
			\hline
		\end{tabular}
		
		\smallskip
		
		\begin{tabular}{ |c||c||c| } 
			\hline  & $\hg^{V^{5,2},\text{(num)}}_0(N_f)$
			& $\hg^{V^{5,2}}_0(N_f)$ \\
			\hline\hline
			$N_f=1$ & $-0.1394556$ & $-0.139455670622$  \\
			\hline
			$N_f=2$ & $-0.188884$ & $-0.188884534562$ \\
			\hline
			$N_f=3$ & $-0.2965$  & $-0.296573183635$ \\ 
			\hline
		\end{tabular}
		\caption{Comparison between $\hg^\text{(num)}_0(N_f,\btDelta)$ numerically estimated from (\ref{mA:num:smallb}) and the known values of $\hg_0(N_f,\btDelta)$ from \cite{Bobev:2022wem,Bobev:2024mqw}. The values of $\hat g^\text{ADHM}_0(N_f,\btDelta)$ for $N_f=5,6$ in the first table are newly obtained in this work by following the approach described in \cite{Bobev:2022wem}.}\label{table:evi3:ADHM:sc}
	\end{table}
\end{enumerate}
It is worth highlighting that the evidence presented above holds to all orders in the $1/N$ expansion and at small $b$ but with generic values of the parameters collectively denoted by $\bxi$. Most of the previous evidence for the Airy conjecture has been established in the large $N$ limit or for specific values of the squashing parameter $b$. Thus, the new evidence presented here provides a qualitative new confirmation of the Airy conjecture (\ref{Airy}).

\bigskip
\noindent\textbf{Intermezzo: supporting (\ref{ABJM:mfc})}
\medskip

Recall that we deduced the coefficient  $\mfc^\text{ABJM}(b,\bDelta)$ in (\ref{ABJM:mfc}) for generic configurations of $(b,\bDelta)$, employing a numerical analysis alongside various constraints outlined in Section~\ref{sec:evi2:ABJM}. However, we did not provide direct evidence for the analytic expression (\ref{ABJM:mfc}) beyond the Nosaka configuration (\ref{ABJM:Nosaka}). We now present highly non-trivial evidence for the validity of (\ref{ABJM:mfc}) for general deformation parameters by employing the small $b$ expansion of the $\mA$ factor~(\ref{mA:smallb}).

The first step is to take the small $b$ expansion of the coefficient in (\ref{ABJM:mfc}) which reads
\begin{align}
	&\mfc^\text{ABJM}(b,\bDelta)\nn\\
	&=\fft{1}{b^2}\fft{\zeta(3)}{8\pi^2}\bigg[\fft{4-\sum_{a=1}^4\Delta_a^2}{4}-\bigg(\fft{1}{\Delta_{13}\Delta_{24}}+\fft{1}{\Delta_{14}\Delta_{23}}\bigg)\sum_{a=1}^4\fft{\prod_{b=1}^4\Delta_b}{\Delta_a}\bigg]\nn\\
	&\quad+\fft{\zeta(3)}{8\pi^2}\bigg[\fft{4-\sum_{a=1}^4\Delta_a^2}{2}+\bigg(\fft{4}{\Delta_{13}\Delta_{14}\Delta_{23}\Delta_{24}}-\fft12\bigg(\fft{1}{\Delta_{13}\Delta_{24}}+\fft{1}{\Delta_{14}\Delta_{23}}\bigg)\bigg)\sum_{a=1}^4\fft{\prod_{b=1}^4\Delta_b}{\Delta_a}\bigg]\nn\\
	&\quad+\mO(b^2)\,.\label{ABJM:mf:smallb}
\end{align}
Substituting the large $k$ expansion of the $\mA$ factor (\ref{ABJM:mA:general}) together with the expression (\ref{ABJM:mf:smallb}) into the small $b$ expansion of the $\mA$ factor (\ref{mA:smallb}), derived by applying the factorization formula and the all order SCI expression to the Airy conjecture (\ref{Airy}), then determines the large $k$ limit of the constants $\hat{g}_0$ and $\hat{f}_0$ for the ABJM theory. The results are given by
\begin{subequations}
	\begin{align}
		\hat{g}_0^\text{ABJM}(k,\bDelta)&=\hat{g}_{0,2}^\text{ABJM}(\bDelta)\fft{\zeta(3)}{8\pi^2}k^2+\mO(\log k)\,,\label{ABJM:g0}\\
		\hat{g}_{0,2}^\text{ABJM}(\bDelta)&=\fft{4-\sum_{a=1}^4\Delta_a^2}{4}-\bigg(\fft{1}{\Delta_{13}\Delta_{24}}+\fft{1}{\Delta_{14}\Delta_{23}}\bigg)\sum_{a=1}^4\fft{\prod_{b=1}^4\Delta_b}{\Delta_a}\,,\nn\\[0.5em]
		\hat{f}_0^\text{ABJM}(k,\bDelta)&=\hat{f}_{0,2}^\text{ABJM}(\bDelta)\fft{\zeta(3)}{8\pi^2}k^2+\mO(\log k)\,,\label{ABJM:f0}\\
		\hat{f}_{0,2}^\text{ABJM}(\bDelta)&=-\bigg(4-\sum_{a=1}^4\Delta_a^2\bigg)+\bigg(\fft{1}{\Delta_{13}\Delta_{24}}+\fft{1}{\Delta_{14}\Delta_{23}}-\fft{8}{\Delta_{13}\Delta_{14}\Delta_{23}\Delta_{24}}\bigg)\sum_{a=1}^4\fft{\prod_{b=1}^4\Delta_b}{\Delta_a}\,.\nn
	\end{align}\label{ABJM:g0:f0}%
\end{subequations}
By independently checking the above expressions for $\hg_{0,2}^\text{ABJM}(\bDelta)$ and $\hf_{0,2}^\text{ABJM}(\bDelta)$ beyond the Nosaka configurations without resorting to the expression (\ref{mA:smallb}), we find that the small $b$ expansion (\ref{ABJM:mf:smallb}) is indeed valid. This provides compelling evidence for the closed-form expression of the coefficient $\mfc^\text{ABJM}(b,\bDelta)$ presented in (\ref{ABJM:mfc}). For further details on the numerical data and methodology employed in this analysis, we refer the reader to Appendix~\ref{app:smallb:mfc}.

\section{A topological string interpretation?}
\label{sec:TS-ST}

In this section, we will attempt to frame the Airy conjecture in the context of the TS/ST correspondence \cite{Grassi:2014zfa}. As we will review, there are a few instances where this correspondence has proved invaluable to resum the perturbative part of the large $N$ expansion, and it is natural to ask whether these isolated examples point to a more general structure.\\

We begin with a general discussion. Consider a partition function $Z(N,b,\bxi)$ of the type \eqref{Z} and introduce the grand canonical partition function as the generating series
\begin{equation}
\label{eq:Xi}
\Xi(\mu,b,\bxi) = \sum_{N\geq0} Z(N,b,\bxi) e^{N \mu} \quad \text{with} \quad Z(0,b,\bxi) = 1 \, .
\end{equation}
We define the grand potential as
\begin{equation}
\mathcal{J}(\mu,b,\bxi) = \log \Xi(\mu,b,\bxi) \, .
\end{equation}
The generating series \eqref{eq:Xi} is invariant under shifts of $\mu$ by $2\mathrm{i}\pi\mathbb{Z}$, which motivates the introduction of a modified grand potential $J$ through the relation
\begin{equation}
e^{\mathcal{J}(\mu,b,\bxi)} = \sum_{n\in\mathbb{Z}} e^{J(\mu +2\pi\ri n,b,\bxi)} \, .
\end{equation}
This gives an alternative representation of the partition function as a contour integral
\begin{equation}
\label{eq:Z-int}
Z(N,b,\bxi) = \frac{1}{2\mathrm{i}\pi} \int_{-\mathrm{i}\infty}^{\mathrm{i}\infty} \mathrm{d}\mu\, \exp\Bigl[J(\mu,b,\bxi) - N\mu\Bigr] \, .
\end{equation}
Recalling the integral representation of the Airy function
\begin{equation}\label{eq:Airyintrep}
{\rm Ai}(z) = \frac{1}{2\pi}\int_{-\infty}^{+\infty}dt\, {\rm exp}\left({\ri \frac{t^3}{3}+\ri tz}\right)\,,
\end{equation}
this formulation of $Z(N,b,\bxi)$ shows that the Airy conjecture presented in Section \ref{sec:AiryTale:statement} amounts to the following polynomial behavior of the modified grand potential in the large $\mu$ limit:
\begin{equation}
\label{eq:J}
J(\mu,b,\bxi) = \frac{\mathcal{C}(b,\bxi)}{3}\,\mu^3 + \mathcal{B}(b,\bxi)\,\mu + \mathcal{A}(b,\bxi) + \mathcal{O}(e^{-\mu}) \, .
\end{equation}
This structure can also be explained by holography and the perturbative expansion of M-theory on asymptotically AdS$_4$ backgrounds of 11d supergravity. Indeed, it is well known that in the large $N$ limit, the SCFTs describing the world-volume of coincident M2-branes have a free energy that scales as $N^{3/2}$. It is straightforward to check using the saddle-point approximation that the integral \eqref{eq:Z-int} behaves as
\begin{equation}
\log Z \sim N^{\frac{p}{p-1}} \quad \text{as} \quad N \rightarrow +\infty \, ,
\end{equation}
whenever $J$ is a polynomial of degree $p$. Thus, holography and M-theory predict that the modified grand potential of 3d $\mathcal{N} \geq 2$ SCFTs describing stacks of M2-branes has degree $p=3$. Furthermore, the leading correction to the $N^{3/2}$ scaling of the free energy has been shown to be of order $N^{1/2}$ using causality arguments \cite{Camanho:2014apa} or the structure of perturbative M-theory corrections to 11d supergravity, see for example \cite{Chester:2018aca}. Via the saddle-point evaluation of \eqref{eq:Z-int}, this implies that the quadratic term in the large $\mu$ limit of the modified grand potential should vanish and the leading correction to the leading $\mu^3$ behavior is by a linear term. The coefficients of the cubic and linear terms can then be derived using higher-derivative supergravity and precision holography by applying the idea outlined in \cite{Bobev:2023lkx}. These results can be derived for all 3d $\mathcal{N}=2$ SCFTs described in Section~\ref{sec:AiryTale} and are in agreement with the first two terms in \eqref{eq:J}, i.e. with the Airy conjecture formulated in Section~\ref{sec:AiryTale}. We stress that there is as of yet no general explanation from M-theory or holography for the fact that the subleading term in the large $\mu$ limit of $J$ is determined solely by the $\mu$-independent function $\mathcal{A}$ up to exponentially suppressed terms. This is in essence the non-trivial prediction of our Airy conjecture when rephrased in the language of this grand potential. Only in some special cases related to specific values of the $(b,\bxi)$ parameters, and described in detail in Section~\ref{sec:AiryTale}, has \eqref{eq:J} been proven to hold using the TS/ST correspondence, as we now review.

\subsection{A lightning review of the TS/ST correspondence}

There exists, in some cases, a remarkable connection between the modified grand potential encoding the partition function of 3d $\mathcal{N}\geq2$ SCFTs and topological string theory on a toric CY manifold. This relation is known as the TS/ST correspondence, see \cite{Grassi:2014zfa,Marino:2015nla,Rella:2022bwn} and references therein. To introduce the correspondence, let $X$ be a toric CY three-fold with $\vec{t} = (t_1,\ldots, t_{b_2(X)})$ complexified Kähler moduli. By local mirror symmetry, we associate a mirror CY $\hat{X}$ to $X$ in such a way that the complex structure moduli of $\hat{X}$ are encoded in the mirror curve to $X$, which is a Riemann surface $\Sigma$ embedded in $\mathbb{C}^* \times \mathbb{C}^*$. The complex structure deformations of $\hat{X}$ are split into $\mathfrak{g}_\Sigma$ moduli of the geometry, which we collect in a vector $\vec{\kappa}$, and $b_2(X) - \mathfrak{g}_\Sigma$ ``mass parameters'' collected in the vector $\vec{\lambda}$. Introducing the chemical potentials via $\kappa_j = e^{\mu_j}$, a set of coordinates $\vec{z} = (z_1,\ldots,z_{b_2(X)})$ called Batyrev coordinates for the mirror $\hat{X}$ is given by the relation
\begin{equation}
\label{eq:Batyrev}
-\log z_i = \sum_{j=1}^{\mathfrak{g}_\Sigma} r_{ij} \mu_j + \sum_{k=1}^{b_2(X) - \mathfrak{g}_\Sigma} m_{ik}\log\lambda_k \, .
\end{equation}
The mirror map now relates the Kähler moduli of $X$ to the complex structure moduli of $\hat{X}$ via
\begin{equation}
\label{eq:mirror-map-class}
t_i(\vec{z}) = -\log z_i - \Pi_i(\vec{z}) \, ,
\end{equation}
where $\Pi_i(\vec{z})$ is a power series in $\vec{z}$ with finite radius of convergence. Together with \eqref{eq:Batyrev}, this implies 
\begin{equation}
\label{eq:t-mu-xi}
t_i(\vec{\mu},\vec{\lambda}) = \sum_{j=1}^{\mathfrak{g}_\Sigma} r_{ij} \mu_j + \sum_{k=1}^{b_2(X) - \mathfrak{g}_\Sigma} m_{ik}\log\lambda_k + \mathcal{O}(e^{-\mu_j}) \, .
\end{equation}

The total free energy of the A-model topological string with target $X$ has the following expansion in the large radius limit $\text{Re}(t_i) \rightarrow \infty$, see \cite{Gopakumar:1998jq}: 
\begin{equation}
\label{eq:F-A-large-volume}
F(\vec{t},g_{\text{top}}) = \frac{1}{6(g_{\text{top}})^2}\sum_{i,j,k=1}^{b_2(X)}c_{ijk}t_i t_j t_k + \sum_{i=1}^{b_2(X)} d_i t_i + \sum_{\mathfrak{g}\geq 2} c_\mathfrak{g} (g_{\text{top}})^{2\mathfrak{g} - 2} + F^{\text{GV}}(\vec{t},g_{\text{top}}) \, , 
\end{equation}
where $g_{\text{top}}$ is the topological string coupling, and the Gopakumar-Vafa contribution is exponentially suppressed and given by the formal power series
\begin{equation}
\label{eq:F-GV}
F^{\text{GV}}(\vec{t},g_{\text{top}}) = \sum_{\mathfrak{g} \geq 0} \sum_{\vec{d} \in H_2(X,\mathbb{Z})} \sum_{w\geq 1} \frac{n_{\mathfrak{g}}^{\vec{d}}}{w}\Bigl(2\sin\frac{w g_{\text{top}}}{2}\Bigr)^{2\mathfrak{g}-2} e^{-w \vec{d}\cdot\vec{t}} \, ,
\end{equation}
with $n_{\mathfrak{g}}^{\vec{d}} \in \mathbb{Z}$ the Gopakumar-Vafa invariants. When $X$ is a toric CY manifold, the above A-model free energy can be obtained as a limit of the refined topological string theory on $X$. This refinement splits the coupling constant as $g_{\text{top}}^2 = -\epsilon_1 \epsilon_2$ and introduces a new parameter $\hbar = \epsilon_1 + \epsilon_2$ that controls the quantization of the mirror curve $\Sigma$ in the B-model. The refined A-model free energy at large radius has a double perturbative expansion~\cite{Iqbal:2007ii,Krefl:2010fm,Huang:2010kf}
\begin{equation}
F(\vec{t},\epsilon_1,\epsilon_2) = \sum_{\mathfrak{g},n\geq0} F_{\mathfrak{g},n}(\vec{t})(g_{\text{top}})^{2\mathfrak{g} - 2} \hbar^{2n} \, . 
\end{equation}
The unrefined theory is obtained by setting $\epsilon_1 = -\epsilon_2 = g_{\text{top}}$, while the Nekrasov-Shatashvili (NS) free energy is defined as the power series~\cite{Nekrasov:2009rc}
\begin{equation}
F^{\text{NS}}(\vec{t},\hbar) = - \lim_{\epsilon_2 \rightarrow 0} \epsilon_2 F(\vec{t},\epsilon_1,\epsilon_2) = \sum_{n\geq0} F^{\text{NS}}_n(\vec{t}) \hbar^{2n-1} \, .
\end{equation}
At large radius, there is a perturbative expansion
\begin{equation}
F^{\text{NS}}(\vec{t},\hbar) = \frac{1}{6\hbar}\sum_{i,j,k=1}^{b_2(X)}c_{ijk}t_i t_j t_k + \hbar\sum_{i=1}^{b_2(X)} d^{\text{NS}}_i t_i + F^{\text{GV,ref}}(\vec{t},\hbar) \, ,
\end{equation}
where the contribution from the Gopakumar-Vafa (GV) invariants $N_{j_L,j_R}^{\vec{d}}$ refined by the spin of the corresponding BPS state~\cite{Choi:2012jz,Nekrasov:2014nea} is again exponentially suppressed as $\text{Re}(t_i) \rightarrow \infty$ and reads
\begin{equation}
F^{\text{GV,ref}}(\vec{t},\hbar) = \sum_{2j_L,2j_R \geq 0}\sum_{\vec{d}}\sum_{w\geq1} (-1)^{2(j_L+j_R)} \frac{N_{j_L,j_R}^{\vec{d}}}{w^2}\frac{\sin\bigl(\frac{\hbar w}{2}(2j_L+1)\bigr)\sin\bigl(\frac{\hbar w}{2}(2j_R+1)\bigr)}{2\sin^3\bigl(\frac{\hbar w}{2}\bigr)}\,e^{-w\vec{d}\cdot\vec{t}} \, .
\end{equation}

We emphasize again that the relevance of the NS limit is that it takes into account the quantization of the mirror curve $\Sigma$. This quantization also promotes the mirror map \eqref{eq:mirror-map-class} to its quantum version
\begin{equation}
\label{eq:quantum-mirror}
t_i(\vec{z},\hbar) = -\log z_i - \Pi_i(\vec{z},\hbar) \, .
\end{equation}
We now define a ``grand potential'' for the NS topological string theory on $X$ as~\cite{Grassi:2014zfa,Hatsuda:2013oxa,Rella:2022bwn} 
\begin{equation}
\label{eq:JX}
\begin{split}
J_X(\vec{\mu},\vec{\lambda},\hbar) =&\; \frac{1}{2\pi}\Bigl[\sum_{i=1}^{b_2(X)} t_i(\hbar)\partial_{t_i}F^{\text{NS}}(\vec{t}(\hbar),\hbar) + \hbar^2 \partial_{\hbar}\Bigl(\frac{1}{\hbar}F^{\text{NS}}(\vec{t}(\hbar),\hbar)\Bigr) + \frac{4\pi^2}{\hbar} \sum_{i=1}^{b_2(X)} d_i t_i(\hbar)\Bigr] \\
&\; + \mathcal{A}(\vec{\lambda},\hbar) + F^{\text{GV}}\Bigl(\frac{2\pi}{\hbar}\vec{t}(\hbar) + \mathrm{i}\pi\vec{B},\frac{4\pi^2}{\hbar}\Bigr) \, ,
\end{split}
\end{equation}
where the derivative w.r.t. $\hbar$ in the second term does not act on the $\hbar$-dependence of the Kähler moduli $\vec{t}(\hbar)$. Furthermore, $\mathcal{A}$ is the same quantity as in the Airy conjecture \eqref{Airy} and $\vec{B}$ is the Kalb-Ramond field through the two-cycles on the CY. Note that upon using \eqref{eq:t-mu-xi}, the grand potential $J_X$ is seen as a function of the fugacities $\vec{\mu}$ and the mass parameters $\vec{\lambda}$. Assembling the previous results, we obtain the behavior of $J_X$ at the large radius point:
\begin{equation}
\label{eq:JX-large-radius}
J_X(\vec{\mu},\vec{\lambda},\hbar) = \frac{1}{12\pi\hbar}\sum_{i,j,k=1}^{b_2(X)} c_{ijk} t_i t_j t_k + \sum_{i=1}^{b_2(X)} \Bigl(\frac{2\pi}{\hbar}d_i + \frac{\hbar}{2\pi}d_i^{\text{NS}}\Bigr) t_i + \mathcal{A}(\vec{\lambda},\hbar) + \mathcal{O}(e^{-t_i},e^{-2\pi t_i/\hbar}) \, .
\end{equation}

The claim of the TS/ST correspondence is that the grand potential $J_X$ computes the spectral determinant of a quantum-mechanical operator that arises naturally from quantizing $\Sigma$. More precisely, there is a family of parametrizations of the mirror curve as
\begin{equation}
O_j(x,y) + \kappa_j = 0 \, , \quad j = 1,\ldots,\mathfrak{g}_\Sigma \, ,
\end{equation}
where $O_j(x,y)$ is a polynomial in the variables $e^x$ and $e^y$. This can be quantized by promoting $x,y\in\mathbb{C}$ to operators $\mathsf{x},\mathsf{y}$ satisfying the commutation relation $[\mathsf{x},\mathsf{y}] = \mathrm{i}\hbar$. This promotes $O_j$ to quantum-mechanical operators $\mathsf{O}_j$ acting on the Hilbert space $L^2(\mathbb{R})$. The inverse operators
\begin{equation}
\label{eq:rho-hat}
\hat{\rho}_j = \mathsf{O}_j^{-1} \, , \quad j = 1,\ldots,\mathfrak{g}_\Sigma \, ,
\end{equation}
are conjectured to be positive-definite and trace class~\cite{Grassi:2014zfa,Kashaev:2015kha}, which allows one to define their spectral determinant\footnote{This is slightly imprecise when $\mathfrak{g}_\Sigma > 1$, see \cite{Codesido:2015dia,Codesido:2016ixn} for details.}
\begin{equation}
\label{eq:Xi-JX}
\Xi_X(\vec{\mu},\vec{\lambda},\hbar) = \text{det}\Bigl(1 + \sum_{j=1}^{\mathfrak{g}_\Sigma}\kappa_j \hat{\rho}_j\Bigr) \, .
\end{equation}
The TS/ST correspondence now states
\begin{equation}
\Xi_X(\vec{\mu},\vec{\lambda},\hbar) = \sum_{\vec{n}\in\mathbb{Z}^{\mathfrak{g}_\Sigma}} \text{exp}\Bigl[J_X(\vec{\mu} + 2\mathrm{i}\pi\vec{n},\vec{\lambda},\hbar)\Bigr] \, ,
\end{equation}
with $J_X$ given by \eqref{eq:JX}. Moreover, the spectral determinant has a well-defined expansion around $\vec{\kappa} = \vec{0}$ given by
\begin{equation}
\label{eq:Xi-ZX}
\Xi_X(\vec{\mu},\vec{\lambda},\hbar) = \sum_{N_1 \geq 0} \ldots \sum_{N_{\mathfrak{g}_\Sigma}\geq0} Z_X(\vec{N},\vec{\lambda},\hbar)\,\kappa_1^{N_1}\ldots \kappa_{\mathfrak{g}_\Sigma}^{N_{\mathfrak{g}_\Sigma}} \, ,
\end{equation}
which can be inverted to obtain the coefficients
\begin{equation}
\label{eq:ZX}
Z_X(\vec{N},\vec{\lambda},\hbar) = \frac{1}{(2\mathrm{i}\pi)^{\mathfrak{g}_\Sigma}}\int_{-\mathrm{i}\infty}^{\mathrm{i}\infty} d\mu_1 \ldots \int_{-\mathrm{i}\infty}^{\mathrm{i}\infty} d\mu_{\mathfrak{g}_\Sigma} \,\text{exp}\Bigl[J_X(\vec{\mu},\vec{\lambda},\hbar) - \sum_{j=1}^{\mathfrak{g}_\Sigma} N_j \mu_j\Bigr] \, .
\end{equation}

In the case where the mirror curve $\Sigma$ has genus one, the above formula gives a simple integral representation of $Z_X$ in terms of the grand potential $J_X$ associated to $X$. From the large radius behavior \eqref{eq:JX-large-radius}, it is clear that the grand potential is a cubic polynomial in $\mu$ up to exponentially small terms. In these cases the coefficients of the cubic and linear terms can be computed using topological string theory on $X$ and $Z_X$ takes the form of an Airy function. In the context of the TS/ST correspondence, our Airy conjecture can thus be potentially proved in two steps:
\begin{enumerate}
\item Identify the toric geometry $X$ that allows one to interpret the SCFT partition function $Z(N,b,\bxi)$ as the coefficient $Z_X$ in the expansion of the spectral determinant \eqref{eq:Xi-JX}.
\item Compute the geometric data of $X$, and in particular the numbers $c_{ijk}$, $d_i$ and $d_i^{\text{NS}}$ that control the Airy parameters $(\mathcal{C},\mathcal{B},\mathcal{A})$.
\end{enumerate}
When available, the ideal Fermi gas formulation of $Z(N,b,\bxi)$ naturally associates a curve (the Fermi surface) to the theory being considered. When that curve can be identified with the mirror of a toric geometry, this achieves step 1. The density matrix of the gas is identified with the operator $\hat{\rho}$ in \eqref{eq:rho-hat}, and quantization of the Fermi surface amounts to quantizing $\Sigma$ by turning on the refinement in the topological string free energy. This gives a concrete and rigorous way to establish the Airy conjecture in the cases reviewed in Section~\ref{sec:AiryTale:ex}, see \cite{Marino:2009jd,Marino:2011eh,Hatsuda:2016uqa}.\\

In the absence of a Fermi gas description, one can instead simply assume that step 1. above can be implemented and proceed to compute the data required for step 2. in a variety of toric CYs. Examples of such computations have been provided in the topological string literature over the years, see e.g. \cite{Chiang:1999tz,Haghighat:2008gw,Huang:2013yta}. Unfortunately, we were not able to find a new example of a manifold $X$ that yields the $(\mathcal{C},\mathcal{B},\mathcal{A})$ coefficients of the Airy conjecture for the holographic SCFTs discussed in Section~\ref{sec:AiryTale} for general values of the parameters $(b,\bxi)$. This search was conducted by computing the geometric data associated to regular local CYs with $\mathfrak{g}_\Sigma = 1$ and $b_2(X) < 3$. While the first restriction is natural as our conjecture pertains to SCFTs with a single large $N$ parameter, the second restriction is merely technical. It would be interesting to extend the search to higher values of $b_2$. The condition of regularity should also be relaxed in some controllable way. Indeed, it was recently shown that the geometry comprising the canonical bundle over the weighted projective space $\mathbb{P}(1,m,n)$, while singular for $(m,n) \neq (1,1)$, is relevant to the partition function of mass-deformed ADHM on a squashed sphere \cite{Kubo:2024qhq}.

While the negative result of our search may point to the absence of a suitable topological string theory that computes the $S^3$ partition functions of 3d $\mathcal{N}\geq2$ SCFTs with general deformation parameters, we choose to be more optimistic and view the implications of our Airy conjecture regarding the cubic behavior of the modified grand potential \eqref{eq:J} as worthy of further investigation. The fact that this cubic structure naturally arises from three independent vantage points, namely the large $N$ expansion of the  $S^3$ matrix model,  the geometry of Calabi-Yau manifolds, and the semiclassical expansion of M-theory on asymptotically AdS$_4$ backgrounds, hints at a larger framework connecting quantum geometry, holography and SCFT partition functions that has yet to be fleshed out in full generality.

\section{Discussion}
\label{sec:discussion}

In this work we formulated a conjecture giving a compact expression for the large $N$ $S^3$ partition function of holographic SCFTs in terms of an Airy function. We summarized a plethora of previously available and newly derived evidence in support of this conjecture. While we are reasonably confident in the validity of our conjectured results, it will clearly be beneficial to prove the Airy conjecture through a rigorous derivation. Perhaps our attempts to this end, described in Section~\ref{sec:TS-ST}, can be improved and will lead to a derivation of the conjecture. Or perhaps an entirely different approach is needed. 

In view of the strong evidence in favor of the Airy conjecture presented above it is perhaps prudent to discuss some open questions and possible generalizations.

\begin{itemize}

\item While we are confident to propose the Airy conjecture \eqref{Airy}, we have not been able to determine the explicit form of the function $\mathcal{A}(b,\bxi)$ for general values of the real mass and squashing parameters in the holographic SCFTs we have considered. This remains an important open problem and it will be very interesting to find the general form of $\mathcal{A}(b,\bxi)$. Relatedly, it is important to understand how to derive the $N^0$ term in the large $N$ expansion of the sphere free energy, which is controlled by $\mathcal{A}(b,\bxi)$, using holography and the dual M-theory description.

\item It is natural to wonder whether there are similar closed-form conjectures for the $S^3$ partition functions of other large $N$ SCFTs, including 3d $\mathcal{N}=2$ holographic theories arising on the worldvolume of D2-, D3-, D4- or M5-branes. To the best of our knowledge there is no current evidence that would suggest the existence of such a structure, but perhaps there is something new to be uncovered?

\item The Airy conjecture captures all the perturbative terms in the $1/N$ expansion of the $S^3$ partition function in a compact resummed expression. It is important to understand whether there is a similarly elegant structure that controls the exponentially suppressed contributions denoted by $\mathcal{O}(e^{-\#\sqrt{N}})$ in \eqref{Airy}. In some of the cases controlled by the TS/ST correspondence discussed in Section~\ref{sec:TS-ST} these exponentially suppressed corrections can be systematically computed. Perhaps it is too ambitious to aim for such explicit analytic control for general values of the parameters $(b,\bxi)$. Nevertheless, it is important to study this question since these exponential corrections of the SCFT partition function capture non-perturbative effects like M2-brane or worldsheet instantons in the dual M-theory and string theory description. Indeed, studying the quantum dynamics of probe branes and strings in the holographically dual 10d or 11d AdS$_4$ supergravity solutions could be helpful in uncovering the structure of these exponential corrections. Promising work along these lines has recently appeared in \cite{Gautason:2023igo} and \cite{Beccaria:2023ujc}.

\item The Airy conjecture discussed in this work, together with the recent results for the compact resummation of the large $N$ expansion of the topologically twisted index \cite{Bobev:2022jte,Bobev:2022eus,Bobev:2023lkx}, the superconformal index \cite{Bobev:2022wem,Bobev:2024mqw}, and the partition function on more general Seifert manifolds \cite{Hong:2024uns}, strongly suggest that the large $N$ path integrals of many holographic SCFTs on compact Euclidean manifolds can be resummed into simple compact expressions. It will be most interesting to understand the reason behind this remarkable structure and to attempt to generalize it to other examples. Perhaps the holomorphic blocks perspective of 3d $\mathcal{N}=2$ Euclidean partition functions \cite{Beem:2012mb} will prove useful in this regard?

\item From the perspective of the AdS/CFT correspondence it is important to understand how to reproduce the Airy form of the SCFT partition function from a dual string or M-theory analysis. This appears to be a tall order since one would naively need to understand all perturbative corrections to the leading supergravity approximation. A way around this impasse may be provided by supersymmetric localization of the gravitational path integral. An early attempt in this direction was indeed made in \cite{Dabholkar:2014wpa}. While this analysis appears incomplete since it does not include details of the higher-derivative corrections to supergravity, it still remains a very interesting avenue for future studies. We should also mention that it will be interesting to understand whether there is a relation with \cite{Ooguri:2005vr,Caputa:2018asc} where the Airy function makes an appearance in semiclassical approximation calculations of 4d quantum gravity path integrals.

\item The general form of the Airy conjecture formulated here should have interesting implications for the study of integrated correlators in 3d SCFTs, see \cite{Agmon:2017xes,Chester:2018aca,Binder:2018yvd,Binder:2019mpb,Agmon:2019imm,Chester:2024bij}. The Airy form of the $S^3$ partition function can be readily expanded for small values of the real mass parameters and the squashing deformation. This leads to, in principle, infinitely many integrated correlators on $S^3$ of the holographic SCFT at hand. Unpacking this wealth of information and utilizing it for bootstrap studies of these SCFTs or to understand the structure of the quantum effective action or scattering amplitudes in the dual string and M-theory will clearly be very interesting to explore.

\item The Airy conjecture should also teach us important lessons about the structure of the perturbative corrections to the type IIA and 11d supergravity effective action. Given our limited understanding of these corrections it is clearly important to uncover this structure through the lens of the AdS/CFT correspondence, see \cite{Beccaria:2023hhi} for a recent discussion on this. For instance, it will be very interesting to understand why in the Airy conjecture the integer $N$ that indicates the number of M2-branes in M-theory appears shifted by $\mathcal{B}$. As pointed out in \cite{Bergman:2009zh} a part of this shift is due to the modified charge quantization in M-theory in the presence of higher-derivative corrections to the 11d supergravity action. Given that $\mathcal{B}$ is a function of the continuous $(b,\bxi)$ it is clear that a further understanding of this shift is needed. 

\item As shown in the seminal work \cite{Kapustin:2009kz}, supersymmetric localization allows for the explicit calculation of supersymmetric Wilson loop expectation values in 3d $\mathcal{N}=2$ SCFTs on $S^3$. For the ABJM theory with no squashing and vanishing real masses this Wilson loop vev was calculated in \cite{Klemm:2012ii} to all orders in the $1/N$ expansion and takes the form of a ratio of two Airy functions of the form \eqref{Airy} with the same $\mathcal{C}$ and different $\mathcal{B}$'s, see also \cite{Armanini:2024kww} for recent work where this result was generalized to the ABJM theory in the presence of real masses. It is tempting to speculate that a similar compact form for the large $N$ Wilson loop vev in terms of an Airy function also exists for the other holographic SCFTs we discussed in this work. It will be interesting to explore and test this hypothesis further.

\item Understanding the generalization of the OSV conjecture \cite{Ooguri:2004zv} to asymptotically AdS$_4$ black holes is a long-standing open problem in string theory. One of the obstacles in formulating a similar conjecture is that there is no clear method, analogous to the one in \cite{Bershadsky:1993cx}, for systematically calculating the higher-derivative prepotential that controls the action of 4d $\mathcal{N}=2$ \textit{gauged} supergravity. It was proposed in \cite{Bobev:2022jte,Bobev:2022eus}, see also \cite{Bobev:2020egg,Bobev:2021oku,Hristov:2024cgj,Hristov:2022plc,Hristov:2022lcw,Hristov:2021qsw} for related work, that the $S^3$ partition function of the dual 3d SCFT may be the answer to this question. As shown in \cite{Bobev:2022eus} if one assumes the Airy conjecture of the form \eqref{Airy} for the ABJM theory with $b=1$ and three real mass parameters then one finds that the large $N$ expansion of the sphere free energy has intriguing homogeneity properties that directly mimic the expected homogeneity of the supergravity prepotential as a function of the vector multiplet scalar fields. It is unlikely that this remarkable structure is an accident and we are tempted to speculate that $S^3$ free energy obtained from the Airy conjecture \eqref{Airy} for all holographic SCFTs discussed here can similarly be used to obtain a supergravity prepotential that determines the higher-derivative action of the corresponding gauged supergravity model. If this proposal is indeed true it will pave the way for a better understanding of the quantum nature of black holes in AdS$_4$.

\end{itemize}

\section*{Acknowledgments}

We are grateful to Francesco Benini, Shai Chester, Sunjin Choi, Fri\dh rik Freyr Gautason, Seppe Geukens, Yasuyuki Hatsuda, Kiril Hristov, Zohar Komargodski, Marcos Mari\~no, Silviu Pufu, and Jesse van Muiden for valuable discussions. NB, PJDS, and XZ are supported in part by FWO projects G003523N, G094523N, and G0E2723N, as well as  by the Odysseus grant G0F9516N from the FWO. JH is supported by the National Research Foundation of Korea(NRF) grant funded by the Korea government(MSIT) with grant number RS-2024-00449284, the Sogang University Research Grant of 202410008.01, the Basic Science Research Program of the National Research Foundation of Korea (NRF) funded by the Ministry of Education through the Center for Quantum Spacetime (CQUeST) with grant number RS-2020-NR049598, and the Fonds Wetenschappelijk Onderzoek--Vlaanderen (FWO) Junior Postdoctoral Fellowship with grant number 1203024N. VR is partly supported by a Visibilit\'e Scientifique Junior Fellowship from LabEx LMH, and by ANR grant ANR-21-CE31-0021. VR is grateful to the CCPP at New York University and KU Leuven for hospitality during parts of this project. XZ is also supported by the Fundamental Research Funds for the Central Universities NSFC NO. 12175237, and by funds from the University of Chinese Academy of Sciences.

\appendix

\addtocontents{toc}{\protect\setcounter{tocdepth}{1}}

\section{Double sine functions}
\label{app:double-sine}
The double sine function is defined as (see \cite{Hatsuda:2016uqa})
\begin{equation}
	\begin{split}
		s_b(z)&\equiv\fft{\Gamma_2(\fft{Q}{2}+\ri z;b,b^{-1})}{\Gamma_2(\fft{Q}{2}-\ri z;b,b^{-1})}=\prod_{m,n=0}^\infty\fft{mb+nb^{-1}+\fft{Q}{2}-\ri z}{mb+nb^{-1}+\fft{Q}{2}+\ri z}\,,
	\end{split}\label{bsine}
\end{equation}
where $Q=b^{-1}+b$, $b$ is a positive real number, and the multiple Hurwitz zeta and Gamma functions are defined as (see section 2 of \cite{Kurokawa2003MultipleSF})
\begin{equation}
	\begin{split}
		\zeta_r(s,z;\vec{\omega})&\equiv\sum_{n_1,\cdots,n_r=0}^\infty\fft{1}{(\vec{n}\cdot\vec{\omega}+z)^s}\,,\\
		\Gamma_r(z;\vec{\omega})&\equiv\exp[\fft{\partial}{\partial s}\zeta_r(s,z;\vec{\omega})\bigg|_{s=0}]=\prod_{n_1,\cdots,n_r=0}^\infty\fft{1}{(\vec{n}\cdot\vec{\omega}+z)}\,,
	\end{split}\label{HurwitzZetaGamma}
\end{equation}
for a given vector $\vec{\omega}=(\omega_1,\cdots,\omega_r)$. The expressions in terms of the infinite sum or product are valid as long as they converge, and they are defined outside the region of convergence via analytic continuation. The double sine function satisfies the following useful properties \cite{Hatsuda:2016uqa}
\begin{equation}
	\begin{split}
		s_b(z)s_b(-z)&=1\,,\\
		\overline{s_b(z)}&=s_b(-\bar{z})\,,\\
		s_b\left(\fft{\ri}{2}b^{\pm1}+z\right)s_b\left(\fft{\ri}{2}b^{\pm1}-z\right)&=\fft{1}{2\cosh(\pi b^{\pm1}z)}\,.
	\end{split}\label{bsine:property}
\end{equation}

\medskip

There exists a closed-form expression for the double sine function when $b^2$ is rational. If the square of the squashing parameter is written as $b^2 = p / q$ with $\gcd(p,q) = 1$, the closed-form expression reads 
\begin{equation}
\begin{split}
	\log s_b(x) & = \sum_{c=1}^{ p q - 1} \large\left[ \frac{1}{4} \log( pq) B_2(v) + v \log\Gamma(v) - \log G (v+1) \right.\\
	& \kern4em + \large\left. ( N(c) - v )( \log \Gamma (v) - \frac{1}{2} \log(2 \pi) +\frac{1}{2} \log(p q) B_1(v)) \large\right]\\
	&\quad - (x \to -x)\qquad\bigg(v \equiv \frac{ Q/2 + \ri x}{ \sqrt{p q} } + \frac{c}{p q}\bigg)\,.
\end{split}\label{sbclosedform}
\end{equation}
In this expression, $B_n(x)$ are the Bernoulli polynomials, $G$ is the Barnes $G$-function and the function $N(c)$ counts the number of solutions to the linear equation
\begin{equation}
	N(c) = \# \left\{ k , l \in \mathbb{Z}_{\ge 0} \ \middle\vert\ k p + l q = c  \right\} = 0\ \text{or}\ 1\,.
\end{equation} 
The formula~\eqref{sbclosedform} is derived by rewriting the double Hurwitz zeta function in~\eqref{HurwitzZetaGamma} as a finite sum of single Hurwitz zeta functions and simplifying the resulting expression.

\medskip 

For the numerical computation of the double sine function, we also introduce its integral representation
\begin{equation}\label{sbcontourintegral}
	\log s_b(x) = \frac{1}{2 \pi \ri} \int \frac{dt}{t}  (\gamma + \log t) \frac{1} { ( 1 - e^{b t} )( 1 - e^{t/b} )}\left[ e^{ ( Q/2 + \ri x) t} - e^{ ( Q/2 - \ri x) t}\right]\,,
\end{equation}
where the integration is carried out over a contour that circles the negative real axis counterclockwise. This integral representation \eqref{sbcontourintegral} is derived by applying the contour integral formula for the multiple Hurwitz zeta function to equations (\ref{bsine}) and \eqref{HurwitzZetaGamma}. The integral (\ref{sbcontourintegral}) converges for $0 < \Re(Q/2 + \ri x) < Q$, and values of $x$ outside this range can be shifted into this horizontal strip on the complex $x$ plane by employing the last identity in \eqref{bsine:property}.
The efficiency of the numerical integration depends on the shape of the chosen contour. Mathematica code for this calculation is available in the Wolfram Function Repository \cite{WolframFunctionRepository}.

\medskip

Finally, we introduce the $\mD_b$-function as
\begin{equation}
	\mD_b(x)\equiv\fft{s_b(x+\ri Q/4)}{s_b(x-\ri Q/4)}=s_b\left(\fft{\ri Q}{4}+x\right)s_b\left(\fft{\ri Q}{4}-x\right)\,,
\label{mD}
\end{equation}
following equation (A.9) of \cite{Hatsuda:2016uqa}. The $\mD_b$-function could be more useful for numerical analysis compared to the double-sine function thanks to its closed-form expression in terms of simple hyperbolic functions for $b^2\in2\mathbb{N}-1$, see \cite{Kubo:2024qhq} and Footnote 2 of \cite{Hatsuda:2016uqa}:
\begin{align}
	\mD_b(x)=\prod_{\ell=1}^n\fft{1}{2\cosh(\fft{\pi}{b}x+\fft{\pi\ri}{b^2}(\fft{n+1}{2}-\ell))}\qquad(b^2=2n-1,~n\in\mathbb{N})\,.\label{mD:closed}
\end{align}
%

\section{Wigner transform}
\label{app:Wigner}

Here we briefly introduce the Wigner transform. We start with the following conventions in quantum mechanics:
\begin{equation}
\begin{alignedat}{4}
	[\hat q,\hat p]&=\ri\hbar\,,&\quad \hat q|q\rangle&=q|q\rangle\,,&\quad	\hat p|p\rangle&=p|p\rangle\,,&\quad 1&=\int_{-\infty}^{\infty} dq |q\rangle\langle q|=\int_{-\infty}^{\infty} dp |p\rangle\langle p|\,,\\
	&&\langle q|q'\rangle&=\delta(q-q')\,,&\quad\langle p|p'\rangle&=\delta(p-p')\,,&\quad \langle q|p\rangle&=\fft{1}{\sqrt{2\pi\hbar}}e^{\ri pq/\hbar}\,.
\end{alignedat}
\end{equation}
The Wigner transform of an operator is defined by
\begin{equation}
	A_W(p,q)\equiv\int_{-\infty}^{\infty} dq'\langle q-\fft{q'}{2}|\hat A|q+\fft{q'}{2}\rangle e^{ipq'/\hbar}\,,\label{Wigner}
\end{equation}
which can be used to evaluate the trace as 
\begin{equation}
	\Tr\hat A=\int_{-\infty}^{\infty}\fft{dpdq}{2\pi\hbar}A_W(p,q)\,.\label{Wigner:trace}
\end{equation}
For a composite operator, the Wigner transform can be written as (see \cite{Marino:2011eh})
\begin{equation}
	\begin{split}
		(\hat A\hat B)_W(p,q)&=A_W(p,q)\star B_W(p,q)\\
		&=\sum_{n=0}^\infty\fft{1}{n!}\left(\fft{\ri\hbar}{2}\right)^n\sum_{k=0}^\infty(-1)^k\binom{n}{k}(\partial_p^k\partial_q^{n-k}A_W(p,q))(\partial_p^{n-k}\partial_q^kB_W(p,q))\,,\label{Wigner:product}
	\end{split}
\end{equation}
in terms of the ``$\star$'' operation defined by
\begin{equation}
	\star=\exp[\fft{\ri\hbar}{2}(\overleftarrow{\partial}_q\overrightarrow{\partial}_p-\overleftarrow{\partial}_p\overrightarrow{\partial}_q)].\label{star:oper}
\end{equation}
%

\section{Numerical integration}
\label{app:num}
\subsection{Bornemann method}\label{app:num:Bornemann}
This Appendix introduces a numerical method, which we refer to as the Bornemann method, that efficiently computes multi-dimensional integrals of the form
\begin{equation}
	Z_N= \frac{1}{N!}\int_{-\infty}^{\infty}\!\!\! d^N x\, \det_{i,j=1}^N f(x_i,x_j)\qquad(Z_0=1)\label{Bornemann:ZN}
\end{equation}
even for large values of $N$. The approach relies on the fact that the grand canonical partition function associated with the canonical partition function in (\ref{Bornemann:ZN}) is a Fredholm determinant, which can be computed numerically with high efficiency as demonstrated in \cite{Bornemann1}. 

\medskip

To begin, we discretize the real line using a set of points $\{x_1, \ldots, x_m\}$, which are evenly spaced with $\Delta x = x_{i+1} - x_i$ for simplicity. The following Bornemann algorithm is then implemented.
\begin{algorithm}[H]
	\caption{Bornemann}\label{alg:Bornemann}
	\begin{algorithmic}
		\State \textbf{Input:} A positive integer $N$, a function $f(x,y)$, a set of points $\{x_1, \ldots, x_m\}$
		\State \textbf{Step 1:} Construct the $m \times m$ matrix $A$, $A_{ab} = \Delta x f(x_a,x_b)$
		\State \textbf{Step 2:} Calculate its eigenvalues $\lambda_1, \ldots, \lambda_m$
		\State \textbf{Step 3:} Evaluate $Z_N = e_N(\lambda_1, \ldots, \lambda_m)$
	\end{algorithmic}
\end{algorithm}
\noindent Here, $e_N(\lambda_1, \cdots, \lambda_m)$ represents the elementary symmetric polynomial of degree $N$ in $m$ variables $\lambda_1, \ldots, \lambda_m$, defined as
\begin{equation}
	e_N(\lambda_1, \cdots, \lambda_m)=\sum_{1\leq j_1<j_2<\cdots<j_N\leq m}\lambda_{j_1}\cdots\lambda_{j_N}\,.
\end{equation}
Algorithm \ref{alg:Bornemann} is based on the following expansion for the grand canonical partition function 
\begin{align}
	\Xi(z) &= \sum_{N=0}^{\infty} Z_N z^N\nn\\
	& \approx 1 + z \sum_i A_{ii} + \frac{z^2}{2!} \sum_{i,j} \begin{vmatrix}
		A_{ii} & A_{ij}\\
		A_{ji} & A_{jj}
	\end{vmatrix} +\frac{z^3}{3!} \sum_{i,j,k} \begin{vmatrix}
		A_{ii} & A_{ij}&A_{ik}\\
		A_{ji} & A_{jj}&A_{jk}\\
		A_{ki} & A_{kj}&A_{kk}
	\end{vmatrix}+ \cdots  \nn\\
	& = \det(1 + z A) \nn\\
	& =\prod_{a=1}^m (1 + \lambda_a z)= \sum_{N=0}^{\infty} e_N(\lambda_1, \ldots, \lambda_m) z^N \,.\label{Xi}
\end{align}
In the 2nd line, we have assumed that the integral is well approximated by its discretization. 

\medskip

Numerical calculations of the elementary symmetric polynomial in Step 3 of Algorithm~\ref{alg:Bornemann} can be technically problematic, however, for example when $N$ and/or $m$ are large. In such cases, the elementary symmetric polynomial can be computed more efficiently using contour integration
\begin{equation}
	Z_N = \oint \frac{dz}{2 \pi \ri} \frac{1}{z^{N+1}} \Xi(z) =
	r^{-N} \int_0^{2 \pi} \frac{d\theta}{2 \pi} e^{- \ri N \theta} \Xi( r e^{i \theta})\,.
\end{equation}
Although in theory this integral is independent of the radius $r$, in practice, $r$ must be chosen appropriately to avoid excessive oscillations in the integrand, see \cite{Bornemann2}. If all eigenvalues $\lambda_a$ are positive, an optimal radius $r$ is the one that minimizes
\begin{equation}\label{eq:goodradius}
	-N \log r + \log \Xi(r)\,,
\end{equation}
according to Theorem 12.1 in \cite{Bornemann2}. This process is summarized in Algorithm~\ref{alg:ek}.

\begin{algorithm}[H]
	\caption{Numerical calculation of elementary symmetric polynomials}\label{alg:ek}
	\begin{algorithmic}
		\State \textbf{Input:} A positive integer $N$, a set of positive real numbers $\{\lambda_1, \ldots, \lambda_m\}$
		\State \textbf{Step 1:} Define $\Xi(z) = \prod_{a=1}^m (1 + \lambda_a z)$
		\State \textbf{Step 2:} Determine the `good' radius $r$ as $r \gets \underset{r}{\operatorname{argmin}} ( 	-N \log r + \log \Xi(r))$ 
		\State \textbf{Step 3:} Evaluate $I = \int_0^{2 \pi} \frac{d\theta}{2 \pi} e^{- \ri N \theta} \Xi( r e^{\ri \theta})$ with the radius chosen in Step 2
		\State \textbf{Output:} $\log Z_N=\log e_N(\lambda_1, \ldots, \lambda_m) = - N \log r + \log I$
	\end{algorithmic}
\end{algorithm}

\noindent Improving Step 3 of Algorithm \ref{alg:Bornemann} with Algorithm \ref{alg:ek} then gives the numerical value of the free energy $F_N=-\log Z_N$.

\medskip
\noindent\textbf{Example}
\smallskip

We apply the above numerical method to the ADHM model at the superconformal configuration, namely for the partition function \eqref{ADHM:Z} with parameters $\Delta_m = 0$, $b=1$, $\Delta_a=1/2$, $\Delta_{\mu_q} = \tDelta_{\mu_q}=1/2$. Using the Cauchy determinant formula, the partition function $Z_N$ of the ADHM theory at the superconformal configuration can be expressed in the form \eqref{Bornemann:ZN} with~\cite{Marino:2011eh}
\begin{equation}
	f(x,y) = \frac{1}{\sqrt{2 \cosh (\pi x)}} \frac{1}{\sqrt{2 \cosh (\pi y)}} \frac{1}{2 \cosh (\pi (x - y))}\,.
\end{equation}
The numerical values of the corresponding free energy $F_N=-\log Z_N$, computed through Algorithms \ref{alg:Bornemann}\footnote{In Algorithm \ref{alg:Bornemann}, we truncated the real line to the interval $[-L ,L ]$ and took $m$ equidistant points in this interval: for $N=1,2$ we have taken $L=8$ and $m=150$; for $N=10$, $L=8, m = 200$; for $N=20$, $L=10, m = 250$; for $N=40$, $L=12, m = 400$; for $N=100$, $L=12, m = 600$. We have checked that all reported digits are correct with these choices.} \& \ref{alg:ek}, are presented in Table \ref{tableNumericalExampleBornemann1} for different values of $N$. These numerical results agree with the analytic results from \cite{Hatsuda:2012dt}, providing a good consistency check for the numerical method.
\begin{table}[H]
	\centering
	\footnotesize
	\renewcommand{\arraystretch}{1.1}
	$$
	\begin{array}{|c|c|c|}
		\hline
		N & \text{Numerical}~F_N & \text{Analytic}~F_N  \\
		\hline\hline
		1 & 1.386294361 &  1.386294361\\
		\hline
		2 &  3.917318608 &  3.917318608\\
		\hline
		10 &  45.49866122 & 45.49866122 \\
		\hline
		20 &  130.2063341&  130.2063341\\
		\hline
		40 &  371.0289665&  371.0289665\\
		\hline
		100 &  1474.4973& \text{N/A}  \\
		\hline
	\end{array}
	$$
	\caption{The free energy $F_N = -\log Z_N$ of the ADHM theory at the superconformal configuration. Note that the free energy for $N=100$ is not explicitly available via the analytic method of \cite{Hatsuda:2012dt} but can be obtained easily with our numerical method.}\label{tableNumericalExampleBornemann1}
\end{table}
%

\subsection{Numerical data}\label{app:num:data}
Here we present the numerical data that supports the Airy conjecture for the $S^3_b$ partition function of the ADHM theory at the $R$ charge configuration (\ref{ADHM:btDelta:Fermi}) and for the proposed $\mA$ factor (\ref{ADHM:mA:Nf}). 

\medskip

We calculate the $S^3_b$ free energy $F=-\log Z$ for the ADHM theory with the following configurations
\begin{equation}
	N=1\sim20\,,\qquad b^2\in\{3,5,7\}\,,\qquad N_f\in\{1,2,3,4,5\}\,,\qquad \zeta\in\{0,\fft15,\fft{\ri}{5},\fft{\ri}{10}\}\,,
\end{equation}
using the numerical algorithms described in the previous Appendix \ref{app:num:Bornemann}. We then determine the ratio (\ref{Rto0}) for all cases and confirm that these ratios indeed approach zero as expected, thereby supporting the Airy conjecture (\ref{Airy}) with the $\mA$ factor (\ref{ADHM:mA:Nf}) for the ADHM model under consideration. The following tables display these ratios $R^\text{ADHM}(N,b,N_f,\btDelta_\text{Fermi})$ for $N=20$. 

\begin{table}[H]
	\centering
	\footnotesize
	\renewcommand{\arraystretch}{1.2}
	\begin{tabular}{ |c||c|c|c| } 
		\hline
		$\zeta=0$ & $b=\sqrt{3}$ & $b=\sqrt{5}$ & $b=\sqrt{7}$ \\
		\hline\hline 
		$N_f=1$ & $2.301\times10^{-12}$ & $-1.240\times10^{-10}$ & $-3.206\times10^{-9}$ \\
		\hline
		$N_f=2$ & $-2.015\times10^{-9}$ & $-2.201\times10^{-8}$ & $3.743\times10^{-7}$ \\
		\hline
		$N_f=3$ & $5.868\times10^{-9}$ & $-2.268\times10^{-7}$ & $1.433\times10^{-5}$ \\
		\hline
		$N_f=4$ & $3.469\times10^{-8}$ & $2.935\times10^{-7}$ & $2.244\times10^{-5}$ \\
		\hline
		$N_f=5$ & $-1.078\times10^{-6}$ & $-1.333\times10^{-6}$ & $-1.555\times10^{-4}$ \\
		\hline
	\end{tabular}%
\end{table}
\begin{table}[H]
	\centering
	\footnotesize
	\renewcommand{\arraystretch}{1.2}
	\begin{tabular}{ |c||c|c|c| } 
		\hline
		$\zeta=\fft15$ & $b=\sqrt{3}$ & $b=\sqrt{5}$ & $b=\sqrt{7}$ \\
		\hline\hline 
		$N_f=1$ & $-7.197\times10^{-10}$ & $1.988\times10^{-8}$ & $-1.113\times10^{-8}$ \\
		\hline
		$N_f=2$ & $-3.665\times10^{-10}$ & $-1.090\times10^{-8}$ & $-7.673\times10^{-8}$ \\
		\hline
		$N_f=3$ & $-1.949\times10^{-9}$ & $-1.886\times10^{-7}$ & $-1.595\times10^{-6}$ \\
		\hline
		$N_f=4$ & $-4.454\times10^{-9}$ & $-3.156\times10^{-6}$ & $-3.100\times10^{-5}$ \\
		\hline
		$N_f=5$ & $-1.008\times10^{-8}$ & $-1.625\times10^{-5}$ & $-2.761\times10^{-4}$ \\
		\hline
	\end{tabular}%
\end{table}
\begin{table}[H]
	\centering
	\footnotesize
	\renewcommand{\arraystretch}{1.2}
	\begin{tabular}{ |c||c|c|c| } 
		\hline
		$\zeta=\fft{\ri}{5}$ & $b=\sqrt{3}$ & $b=\sqrt{5}$ & $b=\sqrt{7}$ \\
		\hline\hline 
		$N_f=1$ & $-1.345\times10^{-9}$ & $-8.574\times10^{-10}$ & $-1.490\times10^{-8}$ \\
		\hline
		$N_f=2$ & $2.498\times10^{-9}$ & $-2.417\times10^{-8}$ & $-5.890\times10^{-7}$ \\
		\hline
		$N_f=3$ & $1.620\times10^{-8}$ & $-3.621\times10^{-7}$ & $-1.729\times10^{-6}$ \\
		\hline
		$N_f=4$ & $-2.979\times10^{-8}$ & $-3.326\times10^{-6}$ & $-8.288\times10^{-5}$ \\
		\hline
		$N_f=5$ & $1.002\times10^{-6}$ & $-8.411\times10^{-6}$ & $-1.471\times10^{-4}$ \\
		\hline
	\end{tabular}%
\end{table}
\begin{table}[H]
	\centering
	\footnotesize
	\renewcommand{\arraystretch}{1.2}
	\begin{tabular}{ |c||c|c|c| } 
		\hline
		$\zeta=\fft{\ri}{10}$ & $b=\sqrt{3}$ & $b=\sqrt{5}$ & $b=\sqrt{7}$ \\
		\hline\hline 
		$N_f=1$ & $-4.244\times10^{-10}$ & $-1.739\times10^{-9}$ & $-5.772\times10^{-9}$ \\
		\hline
		$N_f=2$ & $-9.711\times10^{-11}$ & $-1.168\times10^{-9}$ & $-7.800\times10^{-8}$ \\
		\hline
		$N_f=3$ & $3.599\times10^{-9}$ & $-4.187\times10^{-8}$ & $-4.511\times10^{-6}$ \\
		\hline
		$N_f=4$ & $-1.380\times10^{-5}$ & $-4.786\times10^{-4}$ & $3.858\times10^{-5}$ \\
		\hline
		$N_f=5$ & $-2.661\times10^{-7}$ & $-8.303\times10^{-6}$ & $-1.340\times10^{-4}$ \\
		\hline
	\end{tabular}%
\end{table}

\noindent In our numerical evaluation of the $S^3_b$ free energy, the precision achieved is approximately $8-N_f~\sim~10-N_f$ digits, which accounts for the error scale presented in the tables above. Further improvements in the precision of the numerical algorithms will be pursued in future work.

\section{The 't~Hooft limit}\label{app:data}
In this appendix, we present the numerical data and the corresponding analysis used to support the Airy conjecture in the 't Hooft limit.
\subsection{ABJM theory}\label{app:data:ABJM}
The following list of $(b,\lambda,\bDelta_\text{Nosaka})$ configurations has been used to derive the analytic expression for the $S^3_b$ ABJM planar free energy given in (\ref{ABJM:F0:Nosaka}) along with (\ref{ABJM:mfc:Nosaka}). 

\medskip

\noindent Case I. $m_1=m_2=m_3=0$
\begin{equation}
	\begin{alignedat}{3}
		&\text{i)}&~~\lambda&\in\{30,32,34,36,38,40\}\,,&\quad b&=\sqrt{7}\,,\label{ABJM:data:i}\\
		&\text{ii)}&~~\lambda&\in\{30,35,40\}\,,&\quad b&\in\{\sqrt{3},\sqrt{5},\sqrt{9}\}\,.
	\end{alignedat}
\end{equation}
\noindent Case II. $m_2=m_3=0$ with $\lambda=30$
\begin{equation}
	\begin{alignedat}{3}
		&\text{iii)}&~~b&=\sqrt{3}\,,&\quad \bDelta_\text{Nosaka}&\in\{(\fft14,\fft14,\fft34,\fft34),(\fft13,\fft13,\fft23,\fft23),(\fft25,\fft25,\fft35,\fft35)\}\,,\\
		&\text{iv)}&~~b&=\sqrt{5}\,,&\quad \bDelta_\text{Nosaka}&\in\{(\fft14,\fft14,\fft34,\fft34),(\fft13,\fft13,\fft23,\fft23)\}\,,\\
		&\text{v)}&~~b&\in\{\sqrt{7},\sqrt{9}\}\,,&\quad\bDelta_\text{Nosaka}&=(\fft14,\fft14,\fft34,\fft34)\,.
	\end{alignedat}
\end{equation}
\noindent Case III. $m_1=m_3=0$ with $\lambda=30$
\begin{equation}
	\begin{alignedat}{3}
		&\text{vi)}&~~b&=\sqrt{3}\,,&\quad \bDelta_\text{Nosaka}&\in\{(\fft58,\fft38,\fft38,\fft58),(\fft35,\fft25,\fft25,\fft35),(\fft{7}{12},\fft{5}{12},\fft{5}{12},\fft{7}{12})\}\,,\\
		&\text{vii)}&~~b&\in\{\sqrt{5},\sqrt{7}\}\,,&\quad \bDelta_\text{Nosaka}&=(\fft58,\fft38,\fft38,\fft58)\,.
	\end{alignedat}
\end{equation}
\noindent Case IV. $m_3=0$
\begin{equation}
	\begin{alignedat}{4}
		&\text{iix)}&~~b&=\sqrt{3}\,,&\quad \lambda&=30\,,&\quad\bDelta_\text{Nosaka}&\in\{(\fft35,\fft{5}{12},\fft25,\fft{7}{12}),(\fft{7}{12},\fft38,\fft{5}{12},\fft58)\}\,,\\
		&\text{ix)}&~~b&=\sqrt{5}\,,&\quad \lambda&\in\{30,32,34,36,38,40\}\,,&\quad \bDelta_\text{Nosaka}&=(\fft58,\fft25,\fft38,\fft35)\,,\label{ABJM:data:ix}\\
		&\text{x)}&~~b&=\sqrt{7}\,,&\quad \lambda&=30\,,&\quad \bDelta_\text{Nosaka}&=(\fft58,\fft25,\fft38,\fft35)\,.
	\end{alignedat}
\end{equation}
Specifically, the numerical values of the leading $N^2$ term in the classical action for the above configurations, namely $S_{2}^\text{ABJM,(lmf)}(b,\lambda,\bDelta_\text{Nosaka})$ in the \texttt{LinearModelFit} (\ref{ABJM:lmf:improve}), were utilized in two different ways as outlined below. 
\begin{itemize}
	\item \texttt{LinearModelFit} the numerical values of $S_2^\text{ABJM,(lmf)}$ for the i) \& ix) configurations with respect to $\lambda\in\{30,32,\cdots,40\}$, employing the two fitting functions as
	\begin{align}
		S_2^\text{ABJM,(lmf)}(b,\lambda,\bDelta_\text{Nosaka})&=	S_\text{2,lead}^\text{ABJM,(lmf)}(b,\bDelta_\text{Nosaka})\fft{(\lambda-\fft{1}{24})^2}{\lambda^2}\nn\\
		&\quad+\mf^\text{ABJM,(lmf)}(b,\bDelta_\text{Nosaka})\fft{\zeta(3)}{8\pi^2\lambda^2}\,.\label{ABJM:lmf:lambda}
	\end{align}
	Table~\ref{table:ABJM:tHooft} shows that the numerical estimates from the \texttt{LinearModelFit} (\ref{ABJM:lmf:lambda}) are consistent with the analytic coefficients read off from (\ref{ABJM:F0:Nosaka}) and (\ref{ABJM:mfc:Nosaka}). Figure~\ref{fig:ABJM:tHooft} visually illustrates this agreement.
	\begin{table}[t]
		\centering
		\footnotesize
		\renewcommand{\arraystretch}{1.4}
		\begin{tabular}{ |c||c|c| } 
			\hline
			\multirow{4}{*}{$b=\sqrt{7}$,~$\bDelta_\text{Nosaka}=\bDelta_\text{sc}$} & $S_\text{2,lead}^\text{ABJM,(lmf)}(b,\bDelta_\text{Nosaka})$ & $S_\text{2,lead}^\text{ABJM}(b,\bDelta_\text{Nosaka})$  \\
			\cline{2-3}
			& $3.38505366724720$ & $\fft{\pi\sqrt{2}}{3}\fft{Q^2}{4}=3.38505366716828$ \\
			\cline{2-3}
			& $\mf^\text{ABJM,(lmf)}(b,\bDelta_\text{Nosaka})$ & $\mf^\text{ABJM}(b,\bDelta_\text{Nosaka})$ \\
			\cline{2-3}
			& $-0.285715226522448$ & $1-\fft{(b-b^{-1})^2}{4}=-\fft27=-0.285714285714286$ \\
			\hline\hline
			\multirow{4}{*}{$b=\sqrt{5}$,~$\bDelta_\text{Nosaka}=(\fft58,\fft25,\fft35,\fft38)$} & $S_\text{2,lead}^\text{ABJM,(lmf)}(b,\bDelta_\text{Nosaka})$ & $S_\text{2,lead}^\text{ABJM}(b,\bDelta_\text{Nosaka})$  \\
			\cline{2-3}
			& $2.52893330316739$ & $\fft{4\pi\sqrt{2\Delta_1\Delta_2\Delta_3\Delta_4}}{3}\fft{Q^2}{4}=2.52893330317466$ \\
			\cline{2-3}
			& $\mf^\text{ABJM,(lmf)}(b,\bDelta_\text{Nosaka})$ & $\mf^\text{ABJM}(b,\bDelta_\text{Nosaka})$ \\
			\cline{2-3}
			& $0.293303408399649$ & $\fft{1782111}{6076000}=0.293303324555629$ \\
			\hline
		\end{tabular}\caption{Comparison between the numerical fitting (\ref{ABJM:lmf:lambda}) and the corresponding analytic expression (\ref{ABJM:F0:Nosaka}).}\label{table:ABJM:tHooft}%
	\end{table}
	\begin{figure}[t]
		\centering
		\includegraphics[width=0.45\textwidth]{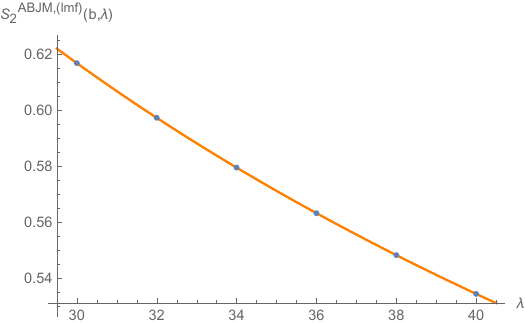}
		\quad
		\includegraphics[width=0.45\textwidth]{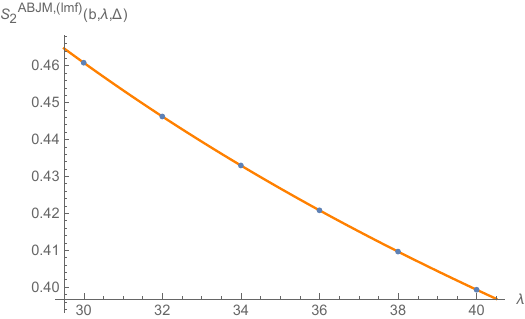}
		\caption{Blue dots represent the numerical values of $S_{2}^\text{ABJM,(lmf)}(b,\lambda,\bDelta_\text{Nosaka})$ in the \texttt{LinearModelFit} (\ref{ABJM:lmf:improve}). Orange lines represent the corresponding analytic expression in~(\ref{ABJM:F0:Nosaka}). The left and right panels are for the i) \& ix) configurations in (\ref{ABJM:data:i}) and (\ref{ABJM:data:ix}), respectively.}\label{fig:ABJM:tHooft}%
	\end{figure}

	\item Table~\ref{table:ABJM:Nosaka} presents selected data that support the analytic expressions (\ref{ABJM:F0:Nosaka}) and (\ref{ABJM:mfc:Nosaka}) by comparing the numerical value of 
	\begin{align}
		\mf^\text{ABJM,(lmf)}(b,\bDelta_\text{Nosaka})&\equiv\fft{8\pi^2\lambda^2}{\zeta(3)}\bigg[S_{2}^\text{ABJM,(lmf)}(b,\lambda,\bDelta_\text{Nosaka})\nn\\
		&\kern5em-\fft{4\pi\sqrt{2\Delta_1\Delta_2\Delta_3\Delta_4}}{3}\fft{Q^2}{4}\fft{(\lambda-\fft{1}{24})^{3/2}}{\lambda^2}N^2\bigg]\,,
	\end{align}
	and the corresponding analytic expression (\ref{ABJM:mfc:Nosaka}).
	\begin{table}[t]
		\centering
		\footnotesize
		\renewcommand{\arraystretch}{1.4}
		\begin{tabular}{ |c||c|c| } 
			\hline
			\text{Data} & $\mf^\text{ABJM,(lmf)}(b,\bDelta_\text{Nosaka})$ & $\mf^\text{ABJM}(b,\bDelta_\text{Nosaka})$  \\
			\hline\hline
			iii) $\bDelta_\text{Nosaka}=(\fft14,\fft14,\fft34,\fft34)$ & $1.16666654026091$ & $\fft76=1.16666666666667$ \\
			\hline
			iv) $\bDelta_\text{Nosaka}=(\fft13,\fft13,\fft23,\fft23)$ & $0.577777833211990$ & $\fft{26}{45}=0.577777777777778$ \\
			\hline
			vi) $\bDelta_\text{Nosaka}=(\fft35,\fft25,\fft35,\fft25)$ & $0.720000000024472$ & $\fft{18}{25}=0.720000000000000$ \\
			\hline
			vii) $\bDelta_\text{Nosaka}=(\fft58,\fft38,\fft58,\fft38)$ & $-0.142856901982585$ & $-\fft17=-0.142857142857143$ \\
			\hline
			iix) $\bDelta_\text{Nosaka}=(\fft35,\fft{5}{12},\fft{7}{12},\fft25)$ & $0.712043477818766$ & $\fft{3344219}{4696650}=0.712043477798005$ \\
			\hline
		\end{tabular}\caption{Numerical data that support the analytic expressions (\ref{ABJM:F0:Nosaka}) and (\ref{ABJM:mfc:Nosaka}).}\label{table:ABJM:Nosaka}%
	\end{table}
\end{itemize}
%

\subsection{ADHM theory}\label{app:data:ADHM}

The following list of $(b,\lambda)$ configurations has been used to deduce the analytic expression (\ref{ADHM:F0}):
\begin{equation}
	\begin{alignedat}{3}
		&\text{i)}&~~ \lambda&=30\,,&\quad b&\in\{1,\sqrt{3}\}\,,\\
		&\text{ii)}&~~ \lambda&\in\{30,32,34,36,38,40\}\,,&\quad b&=\sqrt{5}\,,\label{ADHM:data:ii}\\
		&\text{iii)}&~~ \lambda&\in\{30,35,40\}\,,&\quad b&\in\{\sqrt{7},\sqrt{9},\sqrt{11}\}\,.
	\end{alignedat}
\end{equation}
Below we describe how the numerical values of the $N^2$ leading term $S_2^\text{ADHM,(lmf)}(b,\lambda)$ in the classical action (\ref{ADHM:lmf:improve}) for the above configurations support the analytic expression (\ref{ADHM:F0}) in three distinct ways.
\begin{itemize}
	\item Figure~\ref{fig:ADHM-F0-bsq5} shows that the ii) data above is compatible with the analytic expression (\ref{ADHM:F0}) along with the numerical estimate 
	\begin{equation}
		\mf^\text{ADHM}(b=\sqrt{5})=-0.247365050537827\,.\label{ADHM:mf:bsq=5}
	\end{equation}
	This coefficient (\ref{ADHM:mf:bsq=5}) is determined by fitting the numerical values of $S_2^\text{ADHM,(lmf)}(b,\lambda)$ for the ii) configuration with respect to $\lambda\in\{30,32,\cdots,40\}$, using the two fitting functions 
	\begin{equation}
		\fft{(\lambda-\fft{1}{24}+\fft{b^2+b^{-2}}{3Q^2})^{3/2}}{\lambda^2}\,,\qquad{\rm and}\qquad \fft{1}{\lambda^2}\,,
	\end{equation}
	as in (\ref{ABJM:lmf:lambda}) for the ABJM theory.
	\begin{figure}[t]
		\centering
		\includegraphics[width=0.5\textwidth]{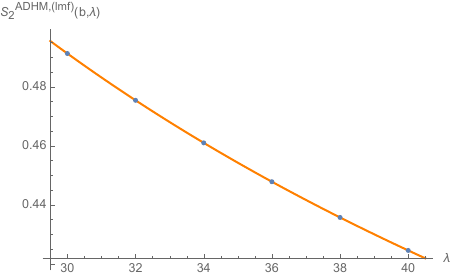}
		\caption{Blue dots represent the numerical values of $S_{2}^\text{ADHM,(lmf)}(b,\lambda)$ in the \texttt{LinearModelFit} (\ref{ADHM:lmf:improve}) for the ii) configuration (\ref{ADHM:data:ii}). The orange line represents the analytic expression in (\ref{ADHM:F0}) with the numerical value (\ref{ADHM:mf:bsq=5}).}\label{fig:ADHM-F0-bsq5}%
	\end{figure}

	\item Table~\ref{table:ADHM:data1} presents the i) data, which support the analytic expression (\ref{ADHM:mfc}) by comparing the numerical values of 
	\begin{align}
		\mfc^\text{ADHM,(lmf)}(b,\lambda)\equiv \lambda^2\bigg[S_{2}^\text{ADHM,(lmf)}(b,\lambda)-\fft{\pi\sqrt{2}}{3}\fft{Q^2}{4}\fft{(\lambda-\fft{1}{24}+\fft{b^2+b^{-2}}{3Q^2})^{3/2}}{\lambda^2}N^2\bigg]\,,\label{ADHM:diff}
	\end{align}
	and the corresponding analytic expression $\mfc^\text{ADHM}(b)$.
	\begin{table}[t]
		\centering
		\footnotesize
		\renewcommand{\arraystretch}{1.4}
		\begin{tabular}{ |c||c|c| } 
			\hline
			$(b,\lambda)$ & $\mfc^\text{ADHM,(lmf)}(b,\lambda)$ & $\mfc^\text{ADHM}(b)$  \\
			\hline\hline
			$(1,30)$ & $-0.071419169040796540447$ & $\fft{\zeta(3)}{8\pi^2}-\fft18\log2=-0.071419169040796528287$ \\
			\hline
			$(\sqrt{3},30)$ & $-0.14250410536682725419$ & $\fft{\zeta(3)}{3\pi^2}-\fft16\log 3=-0.14250410536682725419$ \\
			\hline
		\end{tabular}\caption{Numerical data that support the analytic expression (\ref{ADHM:mfc}).}\label{table:ADHM:data1}%
	\end{table}

	\item Table~\ref{table:ADHM:data2} presents the same quantity (\ref{ADHM:diff}) for the iii) data, highlighting the $\lambda$-independence. This observation strongly supports the structure of the ADHM planar free energy (\ref{ADHM:F0}), while leaving the closed-form expression for $\mfc^\text{ADHM}(b)$ for future work. 
	\begin{table}[t]
		\centering
		\footnotesize
		\renewcommand{\arraystretch}{1.4}
		\begin{tabular}{ |c||c|c|c| } 
			\hline
			$b$ & $\mfc^\text{ADHM,(lmf)}(b,\lambda=30)$ & $\mfc^\text{ADHM,(lmf)}(b,\lambda=35)$ & $\mfc^\text{ADHM,(lmf)}(b,\lambda=40)$  \\
			\hline\hline
			$\sqrt{7}$ & $-0.359318492567472$ & $-0.359318479059387$ & $-0.359318430945854$ \\
			\hline
			$\sqrt{9}$ & $-0.474110276869651$ & $-0.474110155382705$ & $-0.474110482846968$ \\
			\hline
			$\sqrt{11}$ & $-0.590332572015681$ & $-0.590333574547422$ & $-0.590335953236448$ \\
			\hline
		\end{tabular}\caption{Numerical data that support the analytic structure presented in (\ref{ADHM:F0}) for $b\in\{\sqrt{7},\sqrt{9},\sqrt{11}\}$ by exhibiting the $\lambda$-independence of $\mfc^\text{ADHM}(b)$.}\label{table:ADHM:data2}%
	\end{table}
\end{itemize}
%

\section{The small $b$ limit}\label{app:smallb}
\subsection{Saddle point approximation}\label{app:smallb:saddle}
The starting point of the saddle point calculation is the $S^3_b$ partition function of the ADHM model~\eqref{ADHM:Z}. We will use the $b \to 1/b$ symmetry of the matrix model and work with the large $b$ expansion instead of the small $b$ expansion. At the end of the calculation, we will convert the result to the small $b$ expansion. We set $\Delta_m = \frac{Q}{2} \chi_m$ and take $\chi_m$ constant when calculating the large $b$ expansion so that the $\btDelta$ parameters (\ref{ADHM:btDelta}) remain constant.

We first perform the rescaling $\mu_i = 2 \pi b y_i$ so that the saddle point does not move too much when expressed in the variables $y_i$. Next, we take the large $b$ expansion of the integrand using
\begin{equation}
	\sinh \pi b^2 y  \approx \frac{1}{2} \sign(y) e^{\pi b^2 |y|}\,,
\end{equation}
and
\begin{equation}\label{eq:approxsb}
	\log s_b\left( \frac{\ri Q}{2} (1 - \Delta)  + b\ y\right) = {\cal L}(y, \Delta) b^2 + O(b)^{0}\,.
\end{equation}
In the large $b$ expansion (\ref{eq:approxsb}), the function ${\cal L}$ is found to be (see Appendix A of \cite{Hatsuda:2016uqa} for example)
\begin{equation}
	{\cal L}(x, \Delta) = \frac{ \pi \ri }{2}B_2\left(1 -\frac{\Delta}{2}-\ri x\right)-\frac{\ri}{2 \pi } \text{Li}_2\left(e^{- \pi \ri \Delta + 2 \pi x}\right)\,,
\end{equation}
with $B_2$ the Bernoulli polynomial and $\text{Li}_2$ the dilogarithm. The expansion~\eqref{eq:approxsb} is valid if $\Delta$ satisfies the constraint $0 \le \Delta<2$.

Employing the above rescaling and the large $b$ expansion, the integrand of the ADHM partition function \eqref{ADHM:Z} can be approximated in the large $b$ limit as
\begin{align}
	Z^\text{ADHM} &\approx \frac{b^N}{N!} \int d^N y \prod_{i=1}^N e^{-\pi b^2 \chi_m y_i}\prod_{i <j}^N\left[ \sign(y_{ij}) e^{\pi b^2 |y_{ij}|} 2 \sinh(\pi y_{ij})\right] \label{Largeb2}\\
	&\quad\times \prod_{a=1}^3 \prod_{i,j=1}^N \exp \left[ b^2\ {\cal L} \left(- y_{ij}, \Delta_a \right) \right]
	\times \prod_{q=1}^{N_f} \prod_{i=1}^N  \exp \left[  b^2 \left(   {\cal L}  \left( - y_i ,\Delta_{\mu_q} \right) + {\cal L}  \left( y_i , \tDelta_{\mu_q}  \right)\right) \right]\,.\nn
\end{align}
Using the Laplace method to calculate the large $b$ expansion of the integral~\eqref{Largeb2} gives
\begin{equation}\label{eq:FreeEnergyLargeb}
 \log Z^\text{ADHM} \approx \rho\ b^2 + O(b)^{0}\,,
\end{equation}
where
\begin{equation}\label{eq:defrho}
\rho = \underset{y}{\max} f(y)\,,
\end{equation}
with the function $f(y)$ defined as
\begin{equation}
\begin{split}
	f(y) &=  -\sum_{i=1}^N \pi \chi_m y_i+ \sum_{i <j}^N \pi  |y_{ij}| \\
	&\quad + \sum_{a=1}^3 \sum_{i,j=1}^N {\cal L} \left(- y_{ij}, \Delta_a \right) + \sum_{q=1}^{N_f} \sum_{i=1}^N  \Big[    {\cal L}  \left( - y_i ,\Delta_{\mu_q} \right) + {\cal L}  \left( y_i , \tDelta_{\mu_q}  \right) \Big]\,.
\end{split}\label{def:functionf}
\end{equation}
Notice that, importantly, there is no $\log b$ term in the expansion~\eqref{eq:FreeEnergyLargeb}. The small $b$ expansion then reads
\begin{equation}\label{eq:FreeEnergySmallb}
	\log Z^\text{ADHM} \approx \rho\ b^{-2} + O(b)^{0}\,.
\end{equation}
We have checked the expansion~\eqref{eq:FreeEnergySmallb} by computing the left hand side numerically for various values of $N$ and many small values of $b$, and comparing them against the right hand side. Because the $S^3$ partition function of the $V^{5,2}$ theory has the same form as the ADHM theory, the above formulas also apply to the $V^{5,2}$ theory.

\medskip
\noindent\textbf{Numerical example}
\medskip

We illustrate the calculation of the value $\hg^\text{(num)}_0(N_f,\btDelta)$ in Table~\ref{table:evi3:ADHM:sc} for the case $\btDelta=\btDelta_\text{sc}$ and $N_f=1$ based on the saddle point approximation described above. First, we calculate $\rho$ for different values of $N$, see Table \ref{tablerho1}.
\begin{table}[H]
	\centering
	\footnotesize
	\renewcommand{\arraystretch}{1.1}
	$$
	\begin{array}{|c|c|c|}
		\hline
		N & \rho & y_{\star} \\
		\hline\hline
 2& -1.39079 & (-0.132328, 0.132328)\\
3 & -2.37148 & (-0.233632, 0. , 0.233632)\\
\vdots & \vdots & \vdots \\
10 & -12.6388 & ( -0.713068, -0.487065, \ldots, 0.487065, 0.713068 )\\
		\hline
	\end{array}
	$$
	\caption{$\rho$ is defined in equation~\eqref{eq:defrho}, $y_{\star}$ is the point at which the function $f$ defined in~\eqref{def:functionf} is maximized.}\label{tablerho1}
\end{table}
\noindent Then combining the small $b$ expansion \eqref{eq:FreeEnergySmallb} along with \eqref{F:S3b:Airy} and~\eqref{mA:num:smallb}, we obtain  
\begin{equation}
	\hat{g}_0^\text{num} = - \rho - \frac{\pi \alpha}{4}  (N - \beta)^{3/2}\,.
\end{equation} 
Substituting the values of $\rho$ from Table \ref{tablerho1} and the values of $\alpha$ and $\beta$ from Table \ref{table:abc} into the above expression, we estimate the $\hat{g}_0^\text{num}$ coefficient, see Table \ref{tableestgonum}.

\begin{table}[H]
	\centering
	\footnotesize
	\renewcommand{\arraystretch}{1.1}
	$$
	\begin{array}{|c|c|}
		\hline
		N & \hat{g}_0^\text{num} \\
\hline\hline
 2 & -0.18383993271077603 \\
 3 & -0.1838409213600123 \\
 4 & -0.18384101054942192 \\
 5 & -0.18384102137056857 \\
 6 & -0.18384102298546967 \\
 7 & -0.1838410232669867 \\
 8 & -0.18384102332245966 \\
 9 & -0.18384102333451757 \\
 10 & -0.18384102333738106 \\
		\hline
	\end{array}
	$$
	\caption{If $N$ increases, $\hat{g}_0^\text{num}$ approaches the value $-0.1838410233\cdots$, which matches the value reported in Table \ref{table:evi3:ADHM:sc}. }\label{tableestgonum}
\end{table}

\subsection{Evidence for the analytic expressions (\ref{ABJM:g0:f0})}\label{app:smallb:mfc}
Here we support the analytic expressions (\ref{ABJM:g0:f0}) based on results from the literature \cite{Bobev:2022eus,Bobev:2022wem}.

\medskip

We start with the $\hg_0^\text{ABJM}(k,\bDelta)$ coefficient in the superconformal index. We evaluate it numerically for various large values of $k$ (from $N=101\sim351$ in step of 10 with fixed $\lambda=N/k=30$) and for given configurations of $\bDelta$ parameters and find that the results are compatible with the following analytic expression
\begin{align}
	\hat{g}_0^\text{ABJM,(num)}(k,\bDelta)\equiv\fft{1}{2\pi}\Im\mV^\text{ABJM}-\fft{\pi\sqrt{2k\Delta_1\Delta_2\Delta_3\Delta_4}}{3}\bigg(N-\fft{k}{24}+\fft{1}{12k}\sum_{a=1}^4\fft{1}{\Delta_a}\bigg)^{3/2}\,,
\end{align}
where we have used the expression for the ABJM Bethe potential $\mV^\text{ABJM}$ presented in \cite{Bobev:2022wem,Bobev:2024mqw}. We then use \texttt{LinearModelFit} with respect to $k$ to find the numerical values of $\hg_0^\text{ABJM,(num)}$ using the expansion 
\begin{align}
	\hat{g}_0^\text{ABJM,(num)}(k,\bDelta)=\hat{g}_{0,2}^\text{ABJM,(num)}(\bDelta)\fft{\zeta(3)}{8\pi^2}k^2+\sum_{g=0}^{23}\hat{g}_{0,-2g}^\text{ABJM,(num)}(\bDelta)k^{-2g}\,,\label{ABJM:g0:largek}
\end{align}
estimating the $k^2$ leading order coefficient $\hat{g}_{0,2}^\text{ABJM,(num)}(\bDelta)$. In Table~\ref{table:ABJM:g0} we show evidence that this coefficient is consistent with the corresponding analytic expression $\hat{g}_{0,2}^\text{ABJM}(\bDelta)$ in (\ref{ABJM:g0}) for various $\bDelta$ configurations. Note that the choices of parameters we take go beyond the Nosaka configuration (\ref{ABJM:Nosaka}).
\begin{table}[t]
	\centering
	\footnotesize
	\renewcommand*{\arraystretch}{1.3}
	\begin{tabular}{ |c||c|c| } 
		\hline 
		& $\hat g^\text{ABJM,(num)}_{0,2}(\bDelta)$ & $\hat g^\text{ABJM}_{0,2}(\bDelta)$\\
		\hline\hline
		$\bDelta=(\fft37,\fft12,\fft12,\fft47)$ & $-0.244884877027426$ & $-\fft{18719}{76440}=-0.244884877027734$ \\
		\hline
		$\bDelta=(\fft49,\fft59,\fft12,\fft12)$ & $-0.248447234644140$ & $-\fft{52001}{209304}=-0.248447234644345$ \\
		\hline
		$\bDelta=(\fft{5}{11},\fft{5}{11},\fft{6}{11},\fft{6}{11})$ & $-0.243801652892334$ & $-\fft{59}{242}=-0.243801652892562$ \\
		\hline
		$\bDelta=(\fft{5}{12},\fft{7}{12},\fft{5}{12},\fft{7}{12})$ & $-0.243055555554787$ & $-\fft{35}{144}=-0.243055555555556$ \\
		\hline
		$\bDelta=(\fft25,\fft25,\fft25,\fft45)$ & $-0.213333333331287$ & $-\fft{16}{75}=-0.213333333333333$ \\
		\hline
		$\bDelta=(\fft{15}{40},\fft{17}{40},\fft{21}{40},\fft{27}{40})$ & $-0.218610304168689$ & $-\fft{511723}{2340800}=-0.218610304169515$ \\
		\hline
	\end{tabular}\caption{Comparison between $\hat g^\text{ABJM,(num)}_{0,2}(\bDelta)$ estimated from (\ref{ABJM:g0:largek}) and the analytic expression $\hat{g}_{0,2}(\bDelta)$ presented in (\ref{ABJM:g0})}\label{table:ABJM:g0}
\end{table}

\medskip

The analytic expression for the $\hf_0^\text{ABJM}(k,\bDelta)$ coefficient at the $k^2$ leading order was found in \cite{Bobev:2022eus} by analyzing the $S^1\times S^2$ topologically twisted index (TTI) of ABJM theory. The result can be read off from \cite{Bobev:2022eus} by identifying the flavor magnetic fluxes with the corresponding flavor chemical potentials ($\mn_a=\Delta_a$) as follows
\begin{subequations}
	\begin{align}
		\hat{f}_0^\text{ABJM,(from TTI)}(k,\bDelta)&=\overbrace{\bigg[-\sum_{a=1}^4\hat{f}_{0,2,a}(\bDelta)\Delta_a\bigg]}^{=\hat{f}^\text{ABJM,(from TTI)}_{0,2}(\bDelta)}\fft{\zeta(3)}{8\pi^2}k^2+\mO(\log k)\,,\\
		\hat{f}_{0,2,1}(\bDelta)&=\Delta_1+\fft{\Delta_1\Delta_3}{\Delta_1+\Delta_4}+\fft{\Delta_1\Delta_4}{\Delta_1+\Delta_3}+\fft{\Delta_1\Delta_4(\Delta_2+\Delta_3)}{(\Delta_1+\Delta_4)^2}+\fft{\Delta_1\Delta_3(\Delta_2+\Delta_4)}{(\Delta_1+\Delta_3)^2} \nn \\
		&\quad-\fft{2\Delta_3\Delta_4}{(\Delta_1+\Delta_3)(\Delta_1+\Delta_4)}-\fft{\Delta_2^2(\Delta_1-\Delta_2)}{(\Delta_2+\Delta_3)(\Delta_2+\Delta_4)} \nn \\
		&\quad+\fft{\Delta_2\Delta_3(\Delta_1+\Delta_4)}{(\Delta_1+\Delta_3)(\Delta_2+\Delta_3)}+\fft{\Delta_2\Delta_4(\Delta_1+\Delta_3)}{(\Delta_1+\Delta_4)(\Delta_2+\Delta_4)}\,,\\
		\hat f_{0,2,2}(\bDelta)&=\hat f_{0,2,1}(\bDelta)|_{\Delta_1\leftrightarrow\Delta_2} \, , \\
		\hat f_{0,2,3}(\bDelta)&=\hat f_{0,2,1}(\bDelta)|_{(\Delta_1,\Delta_2)\leftrightarrow(\Delta_3,\Delta_4)} \, , \\
		\hat f_{0,2,4}(\bDelta)&=\hat f_{0,2,1}(\bDelta)|_{(\Delta_1,\Delta_2)\leftrightarrow(\Delta_4,\Delta_3)} \, .
	\end{align}\label{ABJM:f0:TTI}%
\end{subequations}
By comparing (\ref{ABJM:f0}) and (\ref{ABJM:f0:TTI}), we have indeed confirmed that
\begin{equation}
	\hat{f}_{0,2}^\text{ABJM}(\bDelta)=\hat{f}_{0,2}^\text{ABJM,(from TTI)}(\bDelta)
\end{equation}
for generic values of $\bDelta$ parameters, including values that go beyond the Nosaka configuration (\ref{ABJM:Nosaka}). Importantly the analytic expression $\hat{f}_{0,2}^\text{ABJM,(from TTI)}(\bDelta)$ derived from the TTI does not rely on any information concerning the factorization formula connecting the $S^3_b$ partition function and the SCI, yet it perfectly matches $\hat{f}_{0,2}^\text{ABJM}(\bDelta)$ obtained using the latter method. We consider this to be a non-trivial check of our calculational methods and the form of the Airy conjecture we propose.

\bibliography{AiryTale}
\bibliographystyle{JHEP}


\end{document}